\documentclass[aps,prx,nofootinbib,amsmath,amssymb,superscriptaddress,twocolumn]{revtex4-2}
\usepackage[utf8]{inputenc}
\usepackage[T1]{fontenc}
\usepackage{graphicx}
\usepackage{xcolor}
\usepackage[english]{babel}
\usepackage{verbatim}
\usepackage{soul}
\usepackage[separate-uncertainty=true]{siunitx}
\usepackage[version=3]{mhchem}
\usepackage{hyperref}
\hypersetup{
    pdfborder = {0 0 0},
    colorlinks,
    citecolor=red,
    filecolor=Darkgreen,
    linkcolor=blue,
    urlcolor=blue
}
\usepackage[full]{textcomp}
\usepackage{mathtools}
\usepackage{tikz}
\usepackage{stix}
\usepackage{makecell}
\usepackage{array}

\usepackage{ragged2e}   
\usepackage{etoolbox}   
\makeatletter
\renewcommand{\@makecaption}[2]{%
  \setbox\@tempboxa\hbox{%
  \textbf{#1}~$\vert$\justifying #2
  }%
  \ifdim \wd\@tempboxa >\hsize
  \textbf{#1}$~\vert~$\justifying #2\par
  \else
  \hbox to\hsize{\hfil\box\@tempboxa\hfil}
  \fi
  \vskip\abovecaptionskip%
}
\patchcmd{\@makecaption}{#1.~#2}{\textbf{#1}$\vert$\justifying #2}{}{}
\makeatother
\AtBeginDocument{\justifying}
\graphicspath{{figures/}}
\newrobustcmd*{\mycircle}[1]{\tikz{\filldraw[draw=#1,fill=#1] (0,0) circle [radius=0.1cm];}}
\newrobustcmd*{\mytriangle}[1]{\tikz{\filldraw[draw=#1,fill=#1] (0,0) --
(0.2cm,0) -- (0.1cm,0.2cm);}}

\newrobustcmd*{\mytriangleUSD}[1]{\tikz{\filldraw[draw=#1,fill=#1] (0,0.2cm) --
(0.2cm,0.2cm) -- (0.1cm,0cm);}}

\newrobustcmd*{\mysquare}[1]{\tikz{\filldraw[draw=#1,fill=#1] (0,0) --
(0,0.2cm) -- (0.2cm,0.2cm) -- (0.2cm,0cm);}}
\newcommand{\COMM}[1]{\iffalse {#1} \fi} 
\newcommand{\highlightingOn}{0}

\if\highlightingOn
\definecolor{hL1}{rgb}{1,0,0}
\definecolor{hL2}{rgb}{0.7,0,0}
\definecolor{hL3}{rgb}{0.45,0,0.8}
\else
\definecolor{hL1}{rgb}{1,0,0}
\definecolor{hL2}{rgb}{0,0,0}
\definecolor{hL3}{rgb}{0,0,0}
\fi

\usepackage{tikz,xcolor}
\definecolor{lime}{HTML}{A6CE39}
\DeclareRobustCommand{\orcidicon}{%
	\begin{tikzpicture}
	\draw[lime, fill=lime] (0,0)
	circle [radius=0.16]
	node[white] {{\fontfamily{qag}\selectfont \tiny ID}};
	\draw[white, fill=white] (-0.0625,0.095)
	circle [radius=0.007];
	\end{tikzpicture}
	\hspace{-2mm}
}
\foreach \x in {A, ..., Z}{%
	\expandafter\xdef\csname orcid\x\endcsname{\noexpand\href{https://orcid.org/\csname orcidauthor\x\endcsname}{\noexpand\orcidicon}}
}

\begin{document}
\renewcommand{\figurename}{\textbf{Fig.}}
\author{Huanying Sun\orcidA{}}
\thanks{These authors contributed equally to this work.}
\affiliation{Beijing Key Laboratory of Fault-Tolerant Quantum Computing, Beijing Academy of Quantum Information Sciences, Beijing 100193, China}

\author{Yanlin Chen}
\thanks{These authors contributed equally to this work.}
\affiliation{Beijing Key Laboratory of Fault-Tolerant Quantum Computing, Beijing Academy of Quantum Information Sciences, Beijing 100193, China}
\affiliation{Beijing National Laboratory for Condensed Matter Physics, Institute of Physics, Chinese Academy of Sciences, Beijing 100190, China}
\affiliation{University of Chinese Academy of Sciences, Beijing 100049, China}

\author{Qichun Liu}
\thanks{These authors contributed equally to this work.}
\affiliation{Beijing Key Laboratory of Fault-Tolerant Quantum Computing, Beijing Academy of Quantum Information Sciences, Beijing 100193, China}

\author{Haihua Wu}
\affiliation{Beijing Key Laboratory of Fault-Tolerant Quantum Computing, Beijing Academy of Quantum Information Sciences, Beijing 100193, China}

\author{Yuqing Wang}
\affiliation{Beijing Key Laboratory of Fault-Tolerant Quantum Computing, Beijing Academy of Quantum Information Sciences, Beijing 100193, China}

\author{Tiefu Li}
\email{litf@tsinghua@edu.cn}
\affiliation{School of Integrated Circuits and Frontier Science Center for Quantum Information, Tsinghua University, Beijing 100084, China}
\affiliation{Beijing Key Laboratory of Fault-Tolerant Quantum Computing, Beijing Academy of Quantum Information Sciences, Beijing 100193, China}

\author{Yulong Liu\orcidB{}}
\email{liuyl@baqis.ac.cn}
\affiliation{Beijing Key Laboratory of Fault-Tolerant Quantum Computing, Beijing Academy of Quantum Information Sciences, Beijing 100193, China}
\affiliation{Beijing National Laboratory for Condensed Matter Physics, Institute of Physics, Chinese Academy of Sciences, Beijing 100190, China}
\affiliation{University of Chinese Academy of Sciences, Beijing 100049, China}

\title{Superior Frequency Stability and Long-Lived State-Swapping in Cubic-SiC Mechanical Mode Pairs}

\date{\today}
\begin{abstract}
The multimode cavity optomechanical system offers versatile applications including state transduction, coherent interconnection, and many-body simulations. In this study, we developed a cavity electromechanical system that integrates a 3C-SiC membrane and a rectangular superconducting cavity to observe the generation of nearly degenerate pairs of mechanical modes. Subsequently, we derive the expression for intrinsic frequency under nonuniform stress and find that this method supports a remarkably resolution for stress analysis in thin films. Experimentally, we perform collective fitting on the measured set of 57 mechanical modes, revealing deviations in biaxial non-uniform stress on the order of MPa. These degeneracy-broken mechanical modes exhibit exceptional quality factors as high as $10^8$ in a thermal bath of 10 mK. Furthermore, Allan deviation indicates that these modes exhibit extremely stable frequencies compared with different types of optomechanical devices. We then performed state-swapping between near-degenerate mode pairs, demonstrating the transfer efficiency exceeding 78\%, attributed to their exceptionally long lifetimes. This study paves the way for the design of compact quantum phononic devices featuring high-quality-factor mechanical multimodes.
\end{abstract}
\maketitle
\noindent\textbf{Introduction}\\
\indent
Phonons have broad applications in thermal management, materials science, and quantum information processing due to their ability to mediate interactions among heat, electricity, and light~\cite{Herring_1954_PR,Sinba_1973_book,Hillenbrand_2002_nature}. Multiple mechanical mode-based optomechanical and electromechanical systems not only hold promising applications in quantum memory \cite{Biswas_2017_PRA, HXTang_2015_NC, Rabl_2012_PRL, Barclay_2021_NC} or quantum transduction \cite{Schuetz_2015_PRL, Hailin_Wang_2010_PRA, Habraken_2012_NJP, Zoller_2017_PRL, Kippenberg_2010_PRL, Burgwal_2023_NC}, but also serve as ideal platforms for studying intriguing physical phenomenon, e.g., quantum entanglements~\cite{Simon_2018_Nature,Mika_2018_Nature,Hujing_2024_PRApplied,Kapil_2014_PRA,Jieqiao_Liao_2022_PRA}, multimode phonon lasing~\cite{John_2018_PRL,HaibinWu_2024_PRL,Ewold_2021_PRL}, and nonreciprocal photon (phonon) transport~\cite{Mika_2020_PRL,Bernier2017,Seif2018,Fang2017,Mathew2020,Ruesink2016,Harris_2019_Nature}. Currently, multimode optomechanical or electromechanical systems are typically developed through two pathways: integrating electronic or optical cavities with spatially separated mechanical resonators~\cite{Albert_2017_PNAS,Vitali_2018_NJP,Kippenberg_2024_Science,Weaver_2017_NC}, or utilizing different-order modes within a single mechanical resonator to couple with one cavity~\cite{Vitali_2021_NJP,SiShiWu_2022_PRL,JGE_2014_PRL,Paul_2019_Optica}. In the latter configuration, mechanical resonators of various geometries have attracted significant attention, including one-dimensional (1D) structures such as nanowires\cite{Hajime_2020_CP,Laure_2021_PRX} and cantilevers~\cite{Doolin_2014_NJP,Vladimir_2011_NL}, as well as phononic crystal beams~\cite{YongjunHuang_2017_SC,Mayer_2023_NC,Painter_2020_Science,Simon_2022_NP,Celand_2013_NP}. Two-dimensional (2D) structures include membrane~\cite{Pitanti_2016_NC,Albert_2017_PNAS,Torba_2020_NP} and phononic bandgap shield resonators~\cite{Connie_2015_SR,Painter_2009_NC,Lawall_2014_PRL,Schliesser_2022_Nc,Simon_2023_NC}. Three-dimensional (3D) structures encompass micro toroidal~\cite{Doolin_2014_NJP}, disk~\cite{Bowen_2022_SA,Ivan_2022_NL,MoLi_2023_NC,YueYu_2022_LSA}, trapped-particles~\cite{Lukas_2021_Nature,Lukas_2023_NP}, and surface or bulk acoustic mode-based devices~\cite{KejieFang_2029_OE,KejieFang_2020_NC,Peter_2019_SA,Rakich_2022_PRApplied,Paulo_2023_NC,YiwenChu_2024_NP,MoLi_2015_Optica}. Among these configurations, 2D structures such as silicon nitride (SiN) square membranes have been widely used in microwave electromechanical and cavity optomechanical systems due to their advantages of high-stress-induced dissipation dilution and high \( Q \) factors~\cite{andrews_bidirectional_2014,PhysRevLett.103.207204,bagci_optical_2014,higginbotham_harnessing_2018,delaney_superconducting-qubit_2022}.
\par
For a typical square-shaped membrane resonator under isotropic stress, the resonance frequencies of out-of-plane mode can be expressed as $f_{mn}=\sqrt{(m^2+n^2)\sigma/(4\rho L^2)}$, where the $m$ and $n$ represent the node number of the mode shape, $\sigma$ is the isotropic stress in the membrane, the $\rho$ is the mass density, and $L$ is the side length of the square~\cite{Tao_2006_book}. In this scenario, modes become degenerate when they share the same mode indices $m$ and $n$, hence resulting in the optomechanical coupling occurs only between a limited number of mechanical modes and cavity fields. For example, in microwave optomechanical devices integrated with SiN films, the coupling electrodes typically interact with only one mode of the degenerate mode pair within the microwave cavity~\cite{Noguchi_2016_NJP,YulongLiu_2023_NPJ,PhysRevLett.127.273603}. In addition, the dark modes that spontaneously form from degenerate mechanical modes will hinder ground state cooling~\cite{PhysRevA.102.011502, PhysRevA.106.013526}, and the preparation of macroscopic squeezed or entangled states~\cite{PhysRevA.108.013516,PhysRevA.106.063506,PhysRevLett.129.063602,huang2023dark}.
\par
It is worth noting that the introduction of asymmetric stress along the x-axis and y-axis in the membrane would break the degenerate mechanical resonance state with equal node numbers~\cite{YulongLiu_2025_NC,Menno_2021_Micromechines}. The altered states may exhibit similar vibration mode shapes with rotational symmetry, potentially leading to a similar coupling strength between the optical mode and all modified mechanical modes~\cite{Tamayo_2023_ACS,Lincoln_2015_NL,Harris_2014_PRL,Vengalattore_2014_PRL,Nielsen_2016_APB}. Although the degeneracy breaking of mechanical modes in square membrane resonators has been observed in some studies, however, measuring the asymmetry of stress in thin films with high precision is challenging for the traditional methods like X-ray diffraction (XRD) and Raman spectroscopy, as their accuracy is restricted to tens of MPa ~\cite{Jakubek_2020_JRS,Odake_2008_ASpe}. Moreover, the performance, including frequency stability and coherence time, of these non-uniform stress-induced degeneracy-broken mechanical modes has not been systematically studied.
\par
In this work, we analyzed the asymmetric stress along the x- and y-axes of the zinc-blende crystal structure SiC membrane by thoroughly examining the resonance characteristics of all 57 mechanical modes. By deriving the membrane resonance frequency under asymmetric stress and fitting the resonance characteristics, we found a strong agreement between theoretical predictions and experimental results. This approach allows us to obtain the exact biaxial stress distribution with a resolution well below MPa, surpassing the capabilities of current commercial membrane stress analysis instruments. Subsequently, we measured the frequency stability of the aforementioned degeneracy-broken mechanical modes. The results from the Allan variance measurements indicate that the dominant oscillator noise is primarily white noise, and no frequency drift was observed over the 40-day measurement period. This represents the highest level of frequency stability reported in the literature for micro- and nano-mechanical oscillators, outperforming even bulk silica glass. Benefiting from exceptionally high frequency stability and quality factors, we demonstrate highly efficient coherent state swapping between a pair of degenerate-broken mechanical modes in their ground state, achieving over 78\% efficiency in a cavity electromechanical system composed of the mechanical resonator and a superconducting microwave circuit cavity.
\bigskip
\\
\noindent\textbf{Results}\\
\noindent\textbf{Stress-Anisotropy induced Non-Degenerated Mode-pairs}\\
\begin{figure}[thbp]
	\centering
	\includegraphics[width=0.49\textwidth]{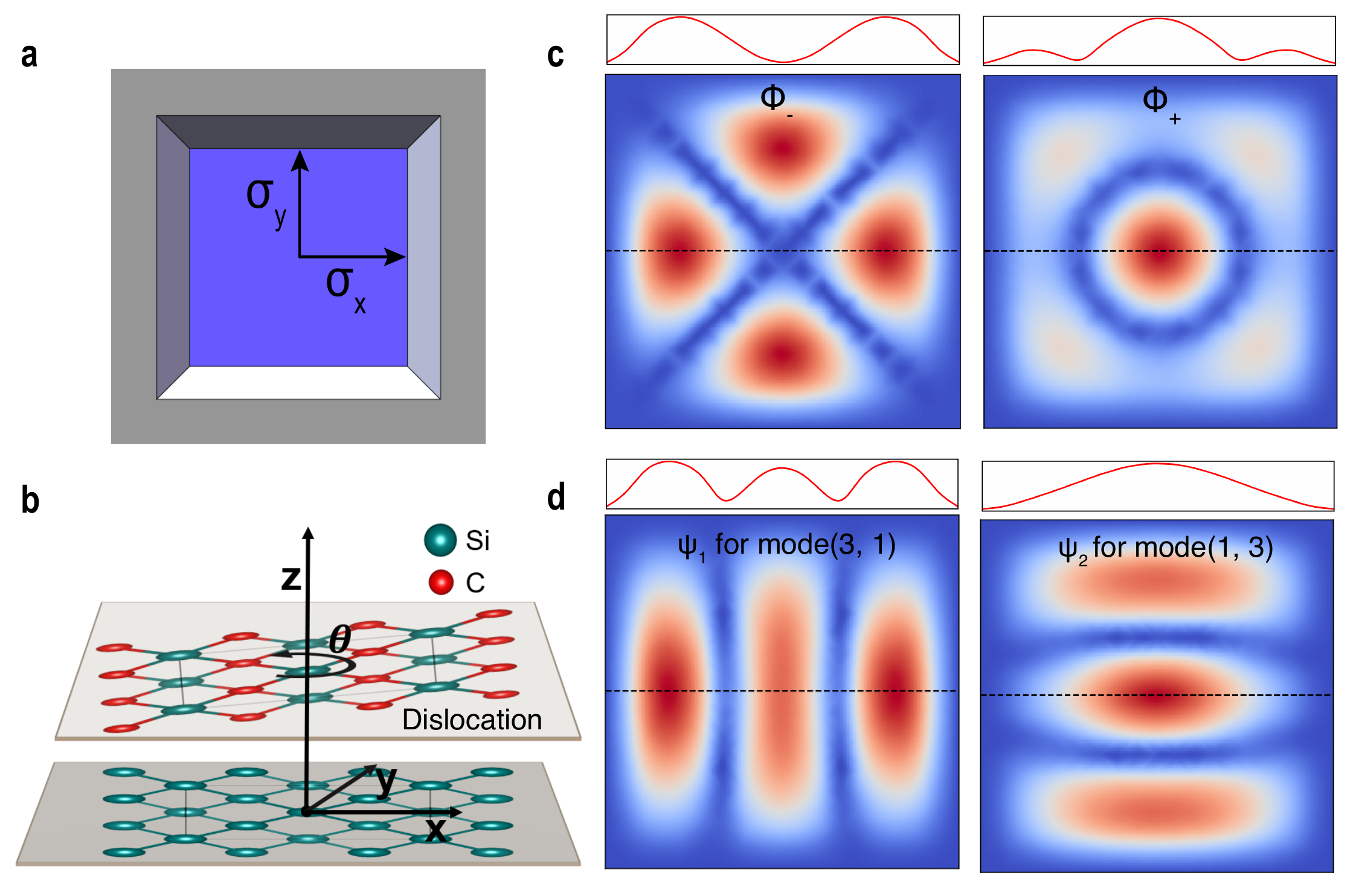}
	\caption{{\bf Schematic of the SiC membrane resonator and out-of-plane mode-shapes with isotropic and anisotropic stress.}
    {\bf a} Schematic of the square SiC membrane resonator, where the grey color represents the silicon frame, and the blue region indicates the SiC membrane window. The quantities $\sigma_x$ and $\sigma_y$ denote the stresses along the x- and y-axis, respectively.
    {\bf b} A simplified model demonstrating the origin of stress biaxial anisotropy, arising from the relative rotation dislocation between the substrate silicon lattice <100> and the overlying silicon carbide lattice (3C phase) during the growth process.
    {\bf c} Surface shapes of two out-of-plane modes, denoted as $\Phi_{-}$ and $\Phi_{+}$, under isotropic stress applied to the square membrane.
    {\bf d} Surface shapes of mode(3,1) and mode(1,3), with the mode indices determined by the number of antinodes along the x- and y-directions.
    The panel at the top of each mode shape presents the vibration amplitude (red-curves) across the center, as indicated by the dashed lines in the mode shape figures.
    }
    \label{fig:stress_ani}
\end{figure}
\indent Considering a square membrane with unequal stress in x and y directions, as shown in Fig.~\hyperref[fig:stress_ani]{\ref{fig:stress_ani}}a, the out-of-plane resonance characteristics of the membrane resonator can be derived from the Euler-Bernoulli equation:
\begin{equation}
	\label{eq:1} \sigma_x\frac{\partial^2w}{\partial^2x}+\sigma_y\frac{\partial^2w}{\partial^2y}=\rho\frac{\partial^2w}{\partial^2t}
\end{equation}
Where the $w$ represents the transverse displacement, $\sigma_x$($\sigma_y$) is the corresponding stress in x(y) direction, $\rho$ is the mass density. The resonance frequencies can be determined through the application of boundary conditions and are expressed as follows:
\begin{equation}
	\label{eq:2}
	\frac{\omega_{mn}}{2\pi}=\frac{1}{2L}\sqrt{\frac{m^2\sigma_x+n^2\sigma_y}{\rho}}
\end{equation}
Where the $L$ is the side length of the square membrane,  $m$($n$) denotes the antinode numbers along the x(y) direction. In the scenario of isotropic stress ($\sigma$ = $\sigma_x$ = $\sigma_y$) on the membrane, the resonance frequencies shorted as the general form $\omega_{m,n}/2\pi=\sqrt{(m^2+n^2)\sigma/(4L^2\rho)}$. The detailed derivation of Eq.~(\ref{eq:2}) is provided in Supplementary Note 1.

In our experiment, the SiC membrane is in 3C-phase and features a 2D square shape. Notably, it experiences unequal stresses along the x-axis and y-axis. Figure~\hyperref[fig:stress_ani]{\ref{fig:stress_ani}}a illustrates a schematic of a standard suspended SiC membrane chip, depicting a square membrane subjected to differential stresses along the x- and y axes. The lattice plane dislocation between the 3C-SiC membrane and the substrate Si <100> illustrates the origin of the anisotropic stress~\cite{Long_1999_JAP,Ernst_1989_JMR,Lavia_2018_MSSP,Massimo_2018_JCG}, as shown in Fig.~\hyperref[fig:stress_ani]{\ref{fig:stress_ani}}b.
\begin{figure*}[thbp]
\centering
\includegraphics[width=0.95\textwidth]{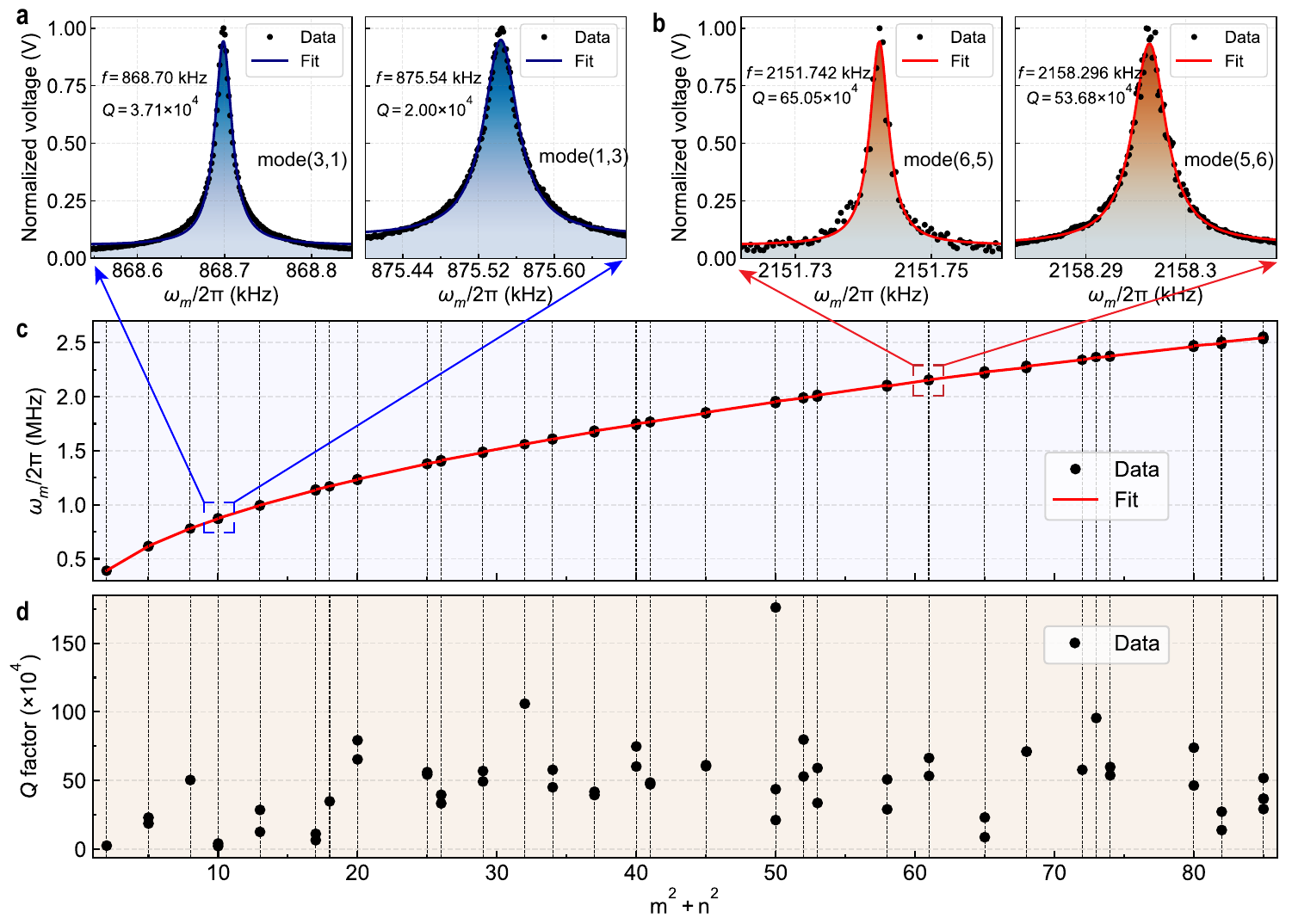}
\caption{Typical vibration spectra of degenerate modes induced by anisotropic stress at low frequencies  {\bf a} and high frequencies  {\bf b} in a square SiC membrane at room temperature. {\bf c} Dependence of the experimental and theoretical resonance frequencies for 57 modes on the sum of the squares of corresponding mode indices $m$ and $n$. Black circular dots represent the experimental results and the red line corresponds to the theoretical fit. The fitted results show the stress in x- and y axes are 242.95 MPa and 245.81 MPa, respectively, with the coefficient of determination ($R^2$) exceeding 99.99\%. {\bf d} The quality factors of all 57 modes are measured at room temperature, from which mode breaking induced by the stress anisotropy is evident as the distinct quality factor values.}
\label{fig:fq_mn}
\end{figure*}

To illustrate the impact of stress anisotropy, figure~\hyperref[fig:stress_ani]{\ref{fig:stress_ani}}c and ~\hyperref[fig:stress_ani]{\ref{fig:stress_ani}}d compare the vibrational mode shapes of a square membrane under isotropic and anisotropic stress, as determined by finite element simulations (FES). In the case of isotropic stress, using the second degenerate mode pair of a square membrane as an example, their vibration amplitude shapes (referred to as mode shapes) exhibit distinct characteristics, as shown in Fig.~\hyperref[fig:stress_ani]{\ref{fig:stress_ani}}c. It demonstrates that the maximum amplitude appears off-center of the membrane for the $\Phi_{-}$ and at the center for $\Phi_{+}$. These two vibration amplitudes $\Phi_{\pm}$ can be further expressed as superpositions of another two orthogonal mode shapes with 90-degree rotational symmetry. Mathematically, we express this as $\Phi_{\pm} = \alpha  \phi_{(3,1)}\pm \beta \phi_{(1,3)}$, where the mode shapes of $\phi_{(3,1)}$ and $\phi_{(1,3)}$ resemble those amplitudes of mode(3,1) and mode(1,3) in Fig.~\hyperref[fig:stress_ani]{\ref{fig:stress_ani}}d. The indices, such as (1,3) or (3,1), represent the number of antinodes of the mode shape along the x- and y-axes, respectively. The $\alpha$ and $\beta$ represent the occupied proportions associated with each mode shape.

As shown in Fig.~\hyperref[fig:stress_ani]{\ref{fig:stress_ani}}d, both mode(3,1) and mode(1,3) exhibit significant vibration amplitude in the center of the membrane. The surface amplitude for a cross-section through the center of the membrane (indicated by the black dashed line in the mode shape) is shown in the top panel of each figure. It reveals that only the $\Phi_+$ mode behaviors a large amplitude at the membrane's center. Under anisotropic stress, however, two modes can exhibit similar mode shapes, related by a $\pi/2$ rotation about the z-axis, as shown in Fig.~\hyperref[fig:stress_ani]{\ref{fig:stress_ani}}d. Later in the text, these modes are designated as $\psi_1$ for $\phi_{(3,1)}$ [mode(3,1)] and $\psi_2$ for $\phi_{(1,3)}$ [mode(1,3)]. It is noted that both $\psi_1$ and $\psi_2$ exhibit significant amplitude at the center of the membrane, which is advantageous for constructing the multimode cavity optomechanical system based on the single cavity mode~\cite{Zeilinger_2006_Nature,Noguchi_2016_NJP,SiShiWu_2022_PRL,HaibinWu_2023_NC,YulongLiu_2023_NPJ}.
\par
\bigskip
\noindent\textbf{Anisotropic stress analysis of the 2D SiC-membrane}\\
\indent
The mechanical vibration characteristics of breaking modes induced by stress anisotropy in the SiC membrane resonator at room temperature were investigated using a laser Doppler vibrometer. The SiC membrane has dimensions of 500 $\mu$m $\times$ 500 $\mu$m $\times$ 50 nm with crystalline of 3C-phase, and its schematic is illustrated in Fig.~\hyperref[fig:stress_ani]{\ref{fig:stress_ani}}a. The mechanical vibration of the resonator was excited by microwaves provided by a lock-in amplifier via a probe positioned above it. The resonant mechanical spectra of the out-of-plane modes were detected by the laser Doppler vibrometer and analyzed using the lock-in amplifier. All measurements were conducted in a vacuum chamber maintained at a pressure of $1\times 10^{-4}$ Pa.

Figure~\hyperref[fig:fq_mn]{\ref{fig:fq_mn}}a shows a typical vibration spectrum of the breaking modes in both low frequencies and high frequencies, resulting from the anisotropic stresses of the membrane along the x- and y-directions. The low-order modes with frequencies of 868.70 kHz and 875.54 kHz correspond to mode(3,1) and mode(1,3), and the high-order modes with frequencies of 2.152 MHz and 2.158 MHz correspond to mode(6,5) and mode(5,6), respectively, with the mode indices determined from FES (see Supplementary Note 2). The independent vibration nature of each mode is manifest due to the high $Q$ factors.

\par
The dependence of the resonance frequencies on the sum of squares of mode indices $m$ and $n$ is shown in Fig.~\hyperref[fig:fq_mn]{\ref{fig:fq_mn}}c. The detailed information about each measured mechanical mode with indices ($m,n$) are displayed in the Supplementary Note 2. The choice of $m^2+n^2$ as variables for the fitting function Eq.~\hyperref[eq:2]{(\ref{eq:2})} is based on the independent and iterative nature of the indices $m$ and $n$, which describe the antinode numbers of the mode shape along different directions. Excellent agreement is observed between measured frequencies and theoretical predictions across the full 390 kHz-2.556 MHz range for all 57 modes. The fitted results reveal that the stress along the x-axis and y-axis are 242.95 MPa and 245.81 MPa, respectively. The discrimination of subtle discrepancies in stress through resonance characterization offers a simple and highly accurate method for evaluating internal anisotropic stress in materials.
\par
The quality factors of all modes are demonstrated in Fig.~\hyperref[fig:fq_mn]{\ref{fig:fq_mn}d}, which displays that the almost modes have the $Q$ factors on the order of hundreds of thousands. This high $Q$-factor is attributed to the membrane's high aspect ratio and significant tensile stresses, both of which contribute to the reduction of mechanical energy through the dilution dissipation mechanism. The dilution factor $D_Q$ of the square membrane resonator can be determined theoretically as follows~\cite{Liu_2023_JP}:
\begin{eqnarray}
	\label{eq:D_Q}
	D_Q&=&\frac{1}{\lambda}\left[1+\frac{\pi^2\lambda(m^2+n^2)}{4}\right]^{-1}\\
    \lambda &=& \sqrt{\frac{4D}{\bar{\sigma} hL^2}}\\
    D &=& \frac{Eh^3}{12(1-\nu^2)}
\end{eqnarray}
where $h$ is the thickness of the membrane, $E$ and $\nu$  denote the Young's modulus and Poisson ratio of the device material, respectively. The dilution factor of the mode(3,1) is dertimined to be approximately  370 with the consideration of $h = 50$ nm, $L = 500~\mu$m, $\bar{\sigma} = 244.38$ MPa, $E = 437$ GPa~\cite{Matsumoto_2001_JMCS,Varshney_2015_JTAP}, and $\nu = 0.268$~\cite{Thomas_1985_PRB}.
\par
In addition to reducing dissipation, high and anisotropic tensile stress will lead to degeneracy breaking, resulting in near-degenerate mode pairs with similar mode shapes and resonance frequencies. Next, we will demonstrate how these modes facilitate the development of multimode optomechanical/electromechanical systems, which consist of a single optical mode coupled to pairs of mechanical modes. 
\begin{figure}[thbp!]
\justifying
\includegraphics[width=0.49\textwidth]{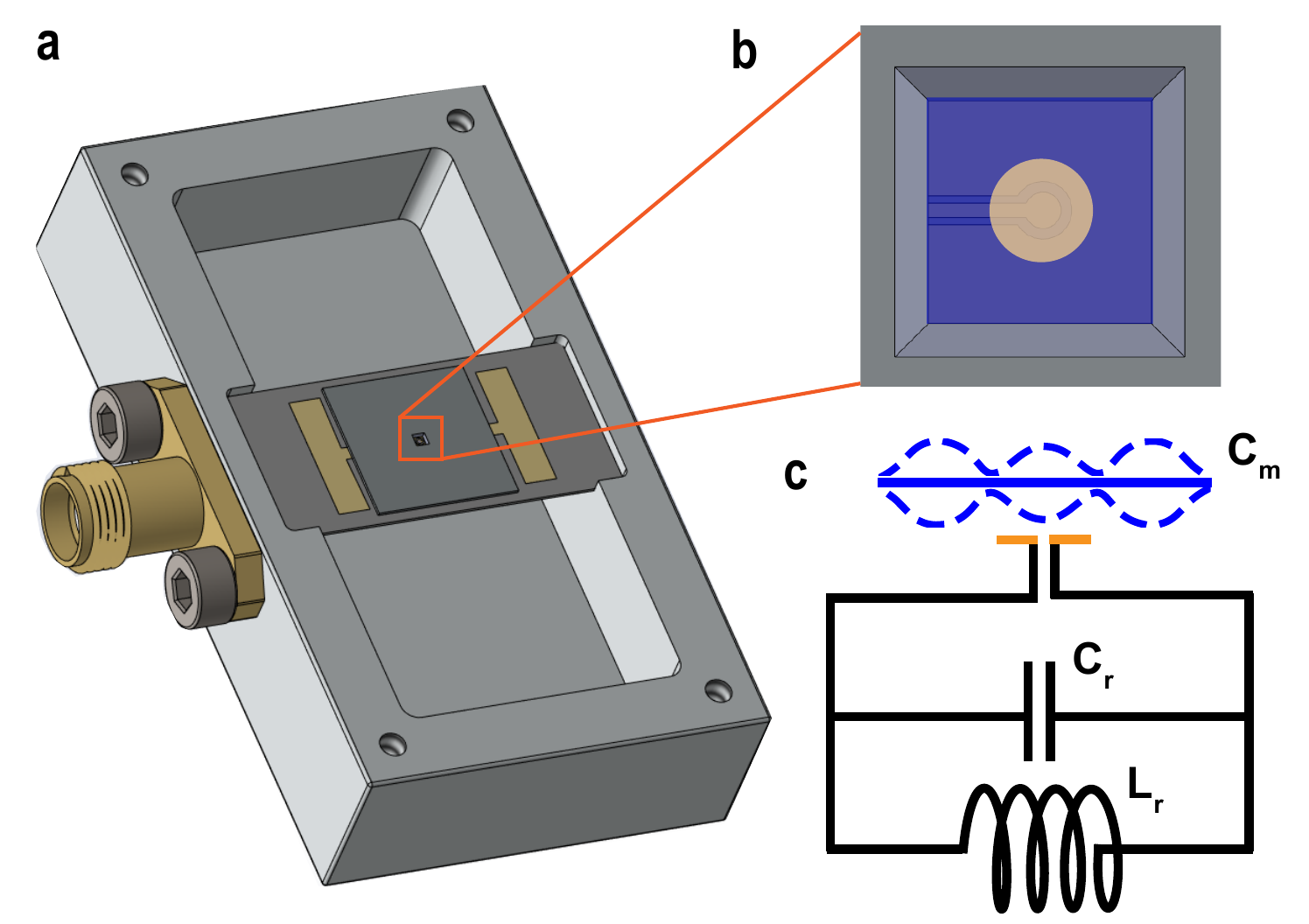}
\caption{{\bf a} The schematic of the cavity electromechanical system, comprising a 3D aluminum rectangular cavity and a SiC membrane. Microwave signal pumping and readout are performed through the SMA port.
{\bf b} A zoom-in of the electromechanical chip. The SiC membrane is flip-chip mounted onto the bottom planar capacitor circuit. The gray, blue, and yellow regions correspond to the silicon frame, the SiC membrane, and the metal-coated area.
{\bf c}  The lumped element mode of the cavity electromechanical system, sketched the mechanical mode(3,1) (blue) coupling to the capacity plate of LC  circuits (orange).  The electromechanical interaction is mediated by changes in mechanical capacitance $C_m$ as the membrane vibrates. $C_r$ and $L_r$ represent the equivalent capacitance and inductance of the 3D metal cavity.
}
\label{fig:Device}
\end{figure}
\\
\bigskip
\\
\noindent\textbf{Electromechanical interface for extremely stable and long-lived mechanical mode pairs}
\begin{figure*}[thbp]
	\centering
	\includegraphics[width=0.75\textwidth]{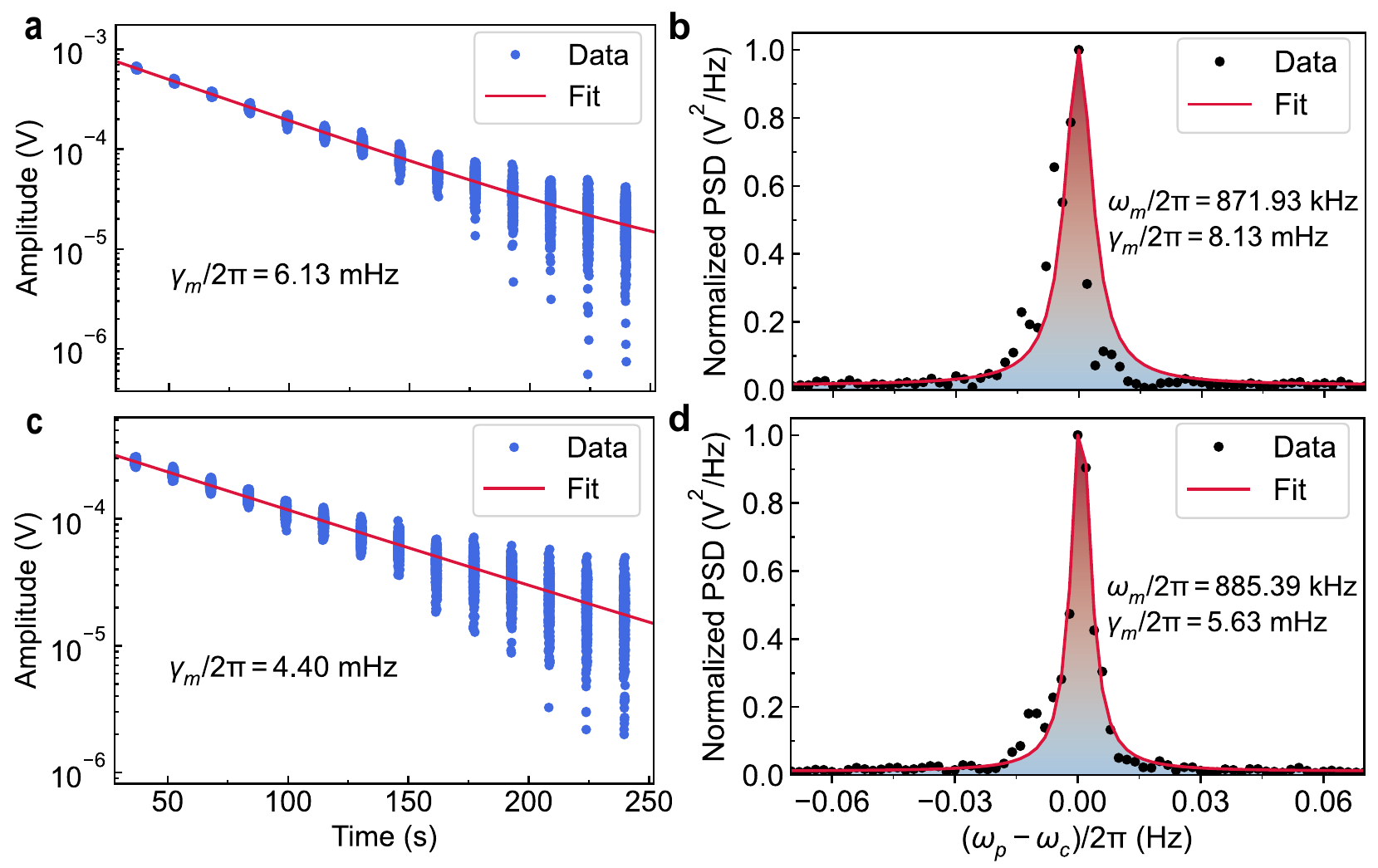}
	\caption{{\bf{Characterization of the mechanical properties of one mode pair at 10 mK using stroboscopic ring-down method and measuring the power spectrum density.}}
    The time dependence of the vibration amplitude for  {\bf a}  mechanical mode(3,1) and  {\bf c}  mode(1,3) measured using the stroboscopic ring-down method.
    The power spectrum density (PSD) of {\bf b} for the mechanical mode(3,1) and {\bf d} for the mode(1,3). The $Q$ factors of the SiC membrane resonator are determined by  $Q=\omega_m/\gamma_m$ with $Q= 1.07\times 10^8$ for mode(3,1) and $Q = 1.57\times 10^8$ for mode(1,3), respectively.}
	\label{fig:Ringdown_Spectrum}
\end{figure*}
\\
\indent
In this section, we demonstrate the state swapping between extremely stable $\psi_1$ and $\psi_2$ based on a cavity electromechanical operations. The detailed schematic of the device is shown in Fig.~\hyperref[fig:Device]{\ref{fig:Device}}a. The system primarily consists of a 3D aluminum rectangular cavity and a SiC membrane that has been metalized to enable electromechanical coupling based on a mechanical capacitor structure. The position-sensitive mechanical capacitor chip is mounted at the center of the 3D cavity, with a single port utilized for applying tones or reading out the system's transmission. Figure~\hyperref[fig:Device]{\ref{fig:Device}}b shows the configuration of the mechanical capacitor, which is composed of the metalized SiC membrane and the bottom planar circuit capacitor. The electromechanical interaction is enabled by the changes in capacitance caused by the membrane's vibration,  and the lumped element model of the electromechanical interaction is demonstrated in Fig.~\hyperref[fig:Device]{\ref{fig:Device}}c. Similar device configurations and measurement techniques for this system have been discussed in our previous works~\cite{YulongLiu_2023_NPJ, YulongLiu_2025_NC}. The mechanical performance of one mode pair $\psi_1$ and $\psi_2$ was first characterized at cryogenic temperatures and then followed by demonstrations of ground-state cooling and coherent-state swapping within this mode pair.
\par
The cavity electromechanical device was located in the mixing chamber stage of a dilution refrigerator with a temperature of 10 mK. The characteristics of the SiC membrane resonator and cavity can be obtained by the spectrum analyzer and the vector network analyzer. It reveals that the microwave cavity has a resonance frequency of $\omega_c/2\pi = 5.39$ GHz and a total linewidth of $\kappa_{\rm tot}/2\pi = 200.25$ kHz, with an external dissipation rate of $\kappa_{\rm ext}/2\pi = 100.25$ kHz and an intrinsic dissipation rate of $\kappa_{\rm int}/2\pi = 100$ kHz, respectively.
\par
The characteristics of the SiC membrane resonator were determined using both the energy stroboscopic ring-down method and vibration spectrum analysis. In the stroboscopic ring-down method, the membrane resonator was first excited by a differential signal composed of tones at the red sideband frequency ($\omega_r = \omega_c - \omega_m$) and the cavity frequency ($\omega_c$), here $\omega_m$ represents the resonance frequency of membrane resonator. Subsequently, the red sideband tones were turned off, and cavity-frequency tones were periodically turned on, in which two sidebands with frequencies at $\omega_r = \omega_c - \omega_m$ and $\omega_b = \omega_c + \omega_m$ will be generated. The free decay of mechanical vibration energy was detected using a spectrum analyzer, with the acquisition frequency set near the red sideband here.

The experimental results obtained from the stroboscopic ring-down method for $\psi_1$ and $\psi_2$ are presented in Fig.~\hyperref[fig:Ringdown_Spectrum]{\ref{fig:Ringdown_Spectrum}}a and \hyperref[fig:Ringdown_Spectrum]{\ref{fig:Ringdown_Spectrum}}c, respectively. It demonstrates an exponential decay in the vibration amplitude of the SiC membrane resonator over time. The periodic disappearance of amplitude is attributed to the intermittent cessation of cavity-frequency tones. Compared with continuous measurement methods,  the stroboscopic approach mitigates the influence of the microwave field on the membrane, thereby improving the accuracy of the measured energy dissipation rate. Theoretical fitting reveals energy dissipation rates of 6.13 mHz and 4.4 mHz for $\psi_1$ and $\psi_2$, respectively.
\begin{figure}[t]
\centering
\includegraphics[width=0.45\textwidth]{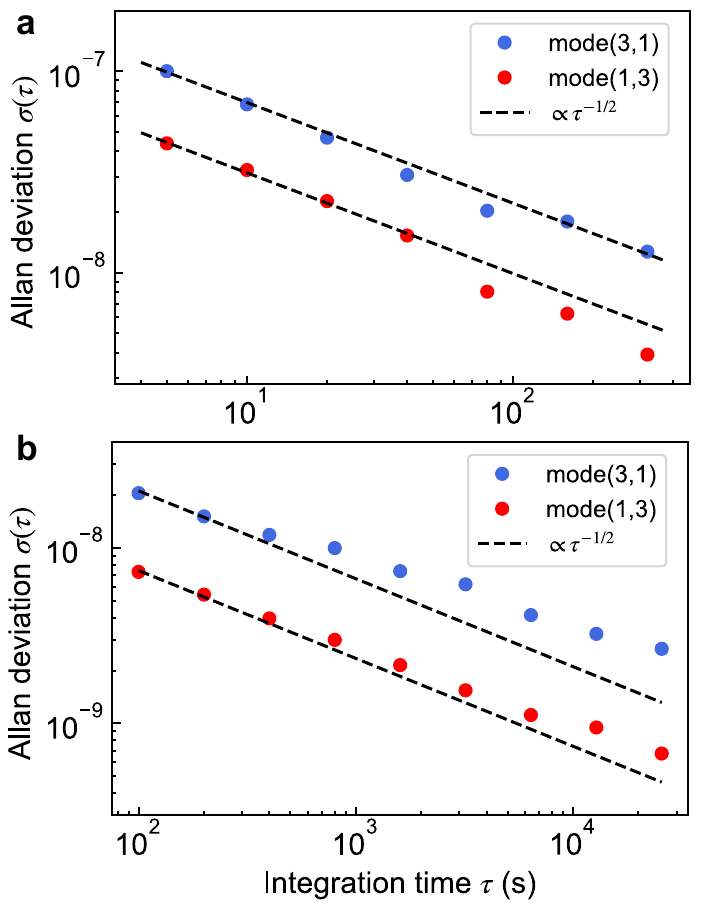}
\caption{Allan deviation $\sigma(\tau)$ versus integration time $\tau$ for modes(3,1) and mode(1,3). Measurements were performed using {\bf a}  5s and {\bf b}  100s sampling intervals. Black dashed lines are the least-squares fits to the $\sigma(\tau) = A\tau^{-1/2}$ noise model, where $A$ is the fitting parameter.}
\label{fig:fre_stablity}
\end{figure}
\par
Additionally, direct detection of the mechanical vibration spectrum was employed to characterize the SiC membrane resonator, where the oscillation of the SiC membrane resonator was excited by environmental thermal fluctuations. By continuously applying the red sideband tone, the characteristics of the mechanical oscillator can be extracted from the upconversion sideband spectrum with frequency at $\omega_c$. The vibration spectrum of the SiC membrane resonator for mode $\psi_1$ and $\psi_2$ are shown in Fig.~\hyperref[fig:Ringdown_Spectrum]{\ref{fig:Ringdown_Spectrum}}b and \hyperref[fig:Ringdown_Spectrum]{\ref{fig:Ringdown_Spectrum}}d, respectively. The Lorentzian fitting reveals that the resonance frequencies and dissipation rates are 871.93 kHz and 8.13 mHz for the $\psi_1$ and 885.39 kHz and 5.63 mHz for $\psi_2$, respectively. It illustrates that this device operates in the resolved-sideband regime ($\kappa_c<\omega_m$). Correspondingly, the $Q$ factors for the $\psi_1$ and  $\psi_2$ are $1.07\times 10^8$ and $1.57\times 10^8$, respectively.
\par
The high $Q$ factors observed at 10 mK are attributed to the exceptional thermal conductivity of silicon carbide, in addition to the dissipation dilution mechanism arising from stress and geometric nonlinearity. This behavior shows that thermoelastic dissipation of 3C-phase SiC decreases with temperature, even at 10 mK, a characteristic unattainable by most semiconductors and amorphous materials. The higher dissipation rate observed here, compared to the results from the stroboscopic method, is due to the fact that it represents the total dissipation rate of the mechanical oscillator, which includes both the thermal decoherence rate and the dephasing rate~\cite{Kippenberg_2023_NP}, rather than only the thermal decoherence rate measured by the stroboscopic ring-down method. Therefore, a pure dephasing rate as low as 1.23 mHz can be extracted for this SiC membrane resonator.
\bigskip
\\
\noindent\textbf{Superior frequency stability at cryogenic temperatures}\\
\indent
The SiC resonator exhibits superior frequency stability owing to its low thermal expansion coefficient, vibration-mode-insensitive volumetric deformation, and the high thermal conductivity that suppresses thermoelastic dissipation by rapidly relaxing thermal gradient-induced stresses. The frequency stability of the resonator can be analyzed by Allan deviation, which can be described mathematically as~\cite{Allan_1966_IEEE,Schimid_2010_PRB}:
\begin{equation}
    \sigma(\tau) = \sqrt{\frac{1}{2(N-1)} \sum_{i=1}^{N-1}
            \left(
                \frac{\bar{f}_{i+1} - \bar{f}_i}{f_0}
            \right)^2}
\end{equation}
Where $N$ is the total number of sampled resonance frequencies ($\bar{f}_1, \bar{f}_2, \cdots, \bar{f}_N$), each averaged over an integration time interval $\tau$.  $f_0$ is the reference resonance frequency of the resonator.
\begin{figure}[t]
\centering
\includegraphics[width=0.52\textwidth]{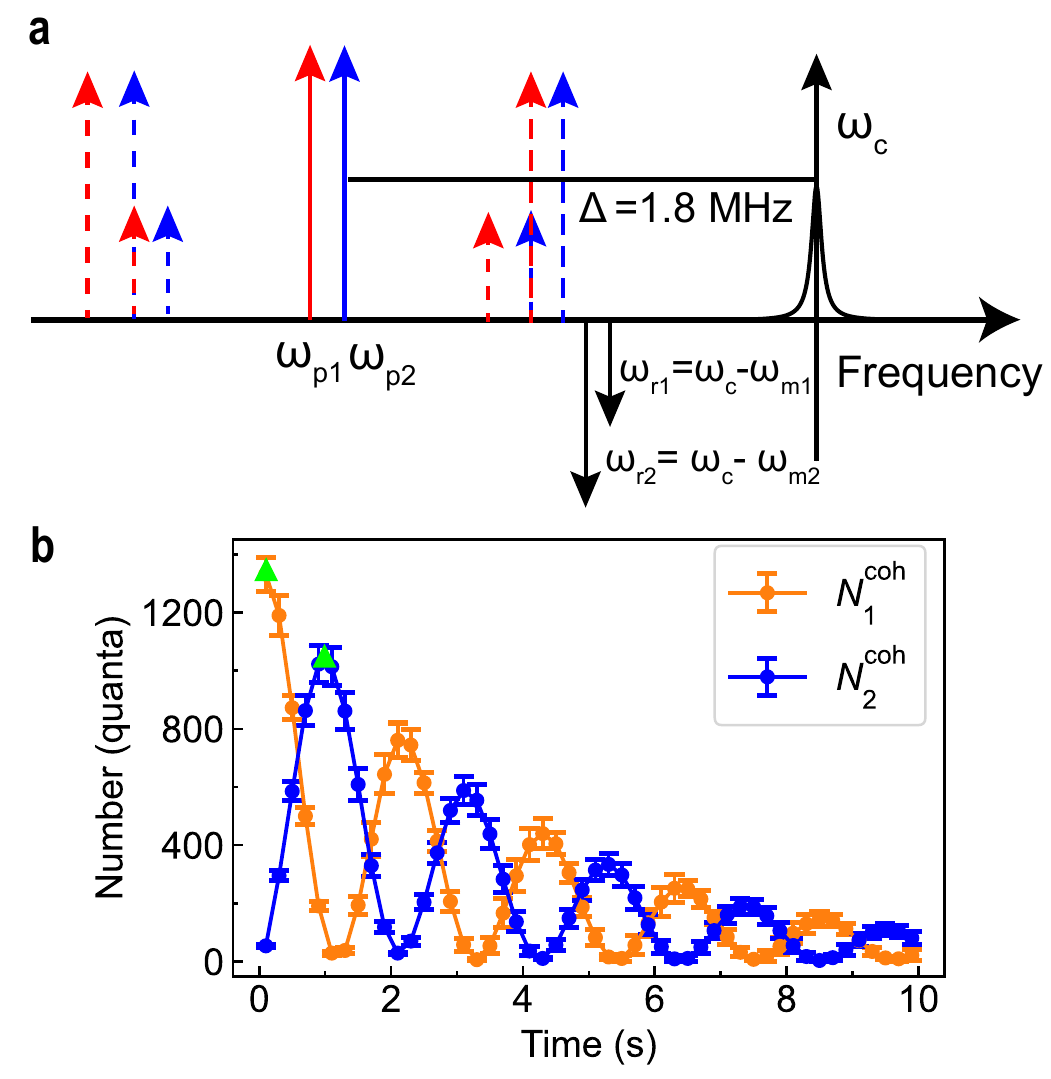}
\caption{{\bf a} The frequencies in the optomechanical STIRAP scheme. The black downward arrows indicate the red sidebands, whose separations from the cavity resonance frequency match the frequencies of the two mechanical modes. The swapping tones are represented by the solid red and blue upward arrows, while the corresponding sidebands are shown as dashed arrows. The swapping between the two mechanical modes is induced due to the overlapping of the sidebands.
{\bf b} Dependence of the coherent occupancies of two mechanical modes on swapping time. }
\label{fig:transfer}
\end{figure}
\par
The integration time dependence of the Allan deviation for the degeneracy-broken pair of mode(3,1) and mode(1,3) is presented in Fig.~\ref{fig:fre_stablity}, with distinct sampling intervals of 5s (Fig.~\ref{fig:fre_stablity} a) and 100s (Fig.~\ref{fig:fre_stablity} b).
These intervals were chosen to characterize noise origins and analyze long-term frequency stability. The minimum sampling time was constrained by both the high $Q$ factors of the resonator and the spectral acquisition methodology. The experimental data (colored circles) are well fitted by the characteristic power-law relationship $A\tau^{-1/2}$, indicative of white frequency noise. 

Figure~\ref{fig:fre_stablity} shows the Allan deviation decreases monotonically across all integration times, regardless of whether they are shorter or longer than the resonator's response time $\tau_r = 1/(\pi\gamma_m)$, and the best $\sigma(\tau) = 6\times 10^{-10}$ are obtained at $\tau=3\times 10^4$s. The observed  $\tau^{-1/2}$ scaling in all integration time identifies white frequency-modulation noise as the dominant noise source, which we attribute primarily to detection noise~\cite{Schmid_2020_PRB}. As summarized in Table~\ref{tab:allan-deviation}, the measured frequency stability of our degeneracy-broken mode pair demonstrates superior performance compared with previously reported resonators fabricated from alternative materials.
\begin{table*}[t]
\caption{\label{tab:allan-deviation} Comparison of frequency stability of resonator fabricated from different materials.}
\begin{tabular}{@{} c @{\extracolsep{\fill}} c @{\extracolsep{\fill}} c @{\extracolsep{\fill}} c @{\extracolsep{\fill}} c @{\extracolsep{\fill}} c @{\extracolsep{\fill}} c @{\extracolsep{\fill}} c @{\extracolsep{\fill}} c @{\extracolsep{\fill}} c @{}}
  \hline \hline \noalign{\vspace*{0.1cm}}
 \parbox{0.8cm} {\centering Material} & \parbox{2cm} {\centering Structure} & \parbox{1.5cm} {\centering Mode } & \parbox{1.2cm} {\centering $f$ (MHz) } & \parbox{1.5cm} {\centering $Q~(\times 10^3) $ } & \parbox{1.8cm} {\centering Freq. stability (ppm) } & \parbox{1.1cm} {\centering Duration } &\parbox{1cm} {\centering T (K)} & \parbox{3cm} {\centering Best $\sigma(\tau)$} & \parbox{1cm} {\centering Ref.}\\ \noalign{\vspace*{0.1cm}}
   \hline \noalign{\vspace*{0.1cm}}

    Si  & \parbox{2.5cm} {Center-anchored} &Flexural &0.15 & 33 & $\pm 5$ &2500 hours & Room &$8\times 10^{-6}~(\tau \approx 10^5$ s) &~\cite{Kim_2005_Conference}\\

     Si &Clamped-clamped	 &Flexural  &0.21	&20	&$\pm$9	&6000 s	&323 &$3\times 10^{-7}~(\tau \approx 10$~s)	&\cite{Ashwin_2021_JMS}\\

     Si&	Clamped–clamped& Flexural &	0.068	&- &	$\pm$87.7	&600 s	& Room&	$1.2\times 10^{-8}~(\tau \approx 11$~s) & \cite{Antonio_2012_NC}\\	

     Si&	Cantilever &Flexural&	0.589	&8	&0.4 &	60 s& Room &$5\times 10^{-9}~(\tau \approx 100$~ms)& \cite{Selim_2014_PNAS}\\

     Si nanowire &	Clamped–clamped	&Flexural &	187.9&	-&	0.15	&1 s&	25  & $6\times 10^{-8}~(\tau \approx 8$~s) & \cite{XL_Feng_2007_Confernce}	\\

     Silica glass	&Twin-microbottle &	Breathing &	56.7	&3.5	&-	&-	&Room &$1\times 10^{-10}~(\tau \approx 30$~ms)&	\cite{Asano_2024_Arxiv}\\

     Si/SiO$_2$ &Center-anchored &Extensional 	& 6.759 &41	&-3.6	&40 hours& 20 &	$2.7\times 10^{-9}~(\tau \approx 1$~s) &\cite{XL_Feng_2025_JMS}\\

     Carbon nanotube & Clamped-clamped &Flexural &1865&	-	&3&	30 s & 5.5& $2\times 10^{-6}~(\tau \approx 30$~s) & 	~\cite{Chaste_2012_NN}\\

      Carbon	 nanotube& Clamped-clamped &Flexural&28.9	&-&	$\pm$52	&60 s	&4.2 &$3\times 10^{-5}~(\tau \approx 30$~s)	&\cite{Adrian_2018_APL}\\

     Carbon	 nanotube& Clampled-clampled &	Flexural &	55.6	& 480 & $\pm$4.5	&500 s	& 0.03 &$4\times 10^{-7}~(\tau \approx 100$~s)&\cite{Moser_2014_NN}\\

    AlN/Si&Clamped–clamped & Flexural &	65	&4230	&-	&- &Room &$4\times 10^{-9}~(\tau \approx 1$~s)	&\cite{XL_Feng_2023_IEEE}\\

    AlN/Si$_3$N$_4$&	Cantilever &	Flexural	&0.088&	0.098	&-	&-	&Room & $2\times 10^{-8}~(\tau \approx 10$~s)&\cite{Song_2017_JMM}\\

     Mo/Si &Center-anchored &Lam\'{e}& 18.284&	309.9	&-	&- &Room &$5\times 10^{-9}~(\tau \approx 100$~ms)& \cite{Guoqiang_Wu_2023_APL}\\

    Si$_3$N$_4$ &Membrane	&Flexural&	0.09	&1180&-	&- &Room &$4\times 10^{-9}~(\tau \approx 0.1$~s)	&\cite{ChangZhang_2023_APL}	\\

    Si$_3$N$_4$ 	&Membrane	&Flexural&	0.076 &	2533 &	$\pm$2.6	&300 s & Room &	$9\times 10^{-9}~(\tau \approx 0.2$~s)	&\cite{Raphael_2022_PRApplied}\\

    Si$_3$N$_4$ & Membranes&	Flexural&	0.188	&30	&-	&- &Room &$3\times 10^{-7}~(\tau \approx 1$~s)	&\cite{Peter_20232_PRApplied}\\

    Si$_3$N$_4$ & Membranes&	Flexural&	0.082	&97	&-	&-	&Room & $8\times 10^{-7}~(\tau \approx 1$~s) &\cite{Silvan_2023_PRApplied}\\

    Si$_3$N$_4$ &Clamped-clamped & Flexural &2.844 &378.7 &0.7	&10 s & Room &	$2\times 10^{-8}~(\tau \approx 1$~s) &	\cite{Gavartin_2013_NC}\\

    SiN  string &Clamped–clamped & Flexural &	0.264	&300	&- &- &Room &$7\times 10^{-8}~(\tau \approx 20$~s)
    &\cite{Silvan_2020_PRB}\\

    SiC&Clamped-clamped& Flexural&	428.2& 2.5&	2&500 s	& 22K &-	&\cite{Feng_2008_NN}	\\
   SiC & Membrane & Flexural & 0.885 & $1.57 \times 10^5$ & $\pm$ 0.02 & 214 hours & 0.01 & $6 \times 10^{-10}$ ($\tau \approx 3 \times 10^4$ s) & Our work \\
\noalign{\vspace*{0.1cm}}
  \hline \hline
\end{tabular}
\end{table*}
\bigskip
\\
\noindent\textbf{Coherent state swapping within the mode pair}
\\
\indent
Benefiting from an extremely small dephasing rate and superior frequency stability, these mode pairs are promising candidates for constructing long-lived phononic memories. The stimulated Raman adiabatic passage (STIRAP) in optomechanics exhibits significant potential for transferring states with high fidelity, attributed to the intermediary dark state involved in this process. Here, we employ the optomechanical STIRAP scheme to carry out the state swapping between mode(3,1) and mode(1,3) by modulating inter-mode coupling at their difference frequency.

The pumping tones and the corresponding generated sidebands of our STIRAP scheme are illustrated in Fig.~\hyperref[fig:transfer]{\ref{fig:transfer}}a. The downward-pointing black arrows labeled $\omega_{r1}$ ($\omega_{r2}$) indicate the red sideband frequencies of modes $\psi_1$ and $\psi_2$. The shifts relative to the cavity resonance frequency $\omega_c$ match the resonance frequencies of modes $\psi_1$ and $\psi_2$, respectively, which demonstrates the resolved sideband condition. The solid red (blue) upward-pointing arrows denote the pumping tones at frequencies $\omega_{\rm p1}$ ($\omega_{\rm p2}$), while the dashed arrows in corresponding colors represent their associated sidebands. One pumping tone applied to two mechanical modes will generate four sidebands, resulting in a total of eight sidebands with two pumping tones. To mitigate the effects of cavity damping and unmatched motional sidebands, we set the pumping tones with a lager detuning from the cavity resonance at $\Delta/2\pi = 1.8$ MHz. The separation between the pumping tones matches the frequency difference of the two mechanical modes~\cite{Dirk_2021_PRL}. The swapping of the states between $\psi_1$ and $\psi_2$ occurs as a result of the overlap of the sidebands, as illustrated in Fig.~\hyperref[fig:transfer]{\ref{fig:transfer}}a.
\par
The complete state-swapping process involves four primary phases: cooling, excitation, swapping, and readout. Initially, two mechanical modes are cooled to near their quantum ground states using their individual red sideband tones simultaneously. Next, one mode, here is the $\psi_1$, is excited by the difference frequency signal through concurrent injection of tones at frequencies of cavity $\omega_c$ and red sideband $\omega_c-\omega_{\rm m1}$. After turning off the excitation tones, the swapping tones are injected to facilitate the swapping processes. Once the swapping tones are turned off, the mechanical state is read out via blue sideband amplification. Further details regarding the characterization in cooling and readout can be found in our previous works~\cite{YulongLiu_2023_NPJ,YulongLiu_2025_NC}.
\par
By repeating the aforementioned four processes and varying the swapping time in each experiment, we can characterize the swapping dynamics. The dependence of coherent phonons occupation of two mechanical modes on swapping time is shown in Fig.~\hyperref[fig:transfer]{\ref{fig:transfer}}b, illustrating that mechanical excitations oscillate between the two modes as the swapping time increases, with maximum occupancies alternately exchanging. The swapping period is 2.1 seconds. As the swapping time increases, the coherent occupancies of the mechanical modes decrease and eventually reach thermal equilibrium. The transfer efficiency is 78.18\%, defined as the ratio of the number of phonons in $\psi_2$ at the end of the first swapping to the number of phonons in $\psi_1$ at the beginning of the swapping~\cite{Dirk_2021_PRL}, as indicated by the green triangles.
\bigskip
\\
\noindent\textbf{Discussion}\\
\indent
Asymmetric stress induced degenerancy-broken offers a pathway to develop multimode opto-/electro-mechanical systems utilizing multiple phononic mode pairs within a single mechanical oscillator. By performing collective fitting on a measured set of 57 mechanical modes, we uncover deviations in biaxial nonuniform stress on the order of MPa, supports a remarkably high resolution for stress analysis in thin films. In contrast, detecting stress anisotropy necessitates additional adjustments, such as geometric configurations~\cite{Luca_1994_JAP,Olabi_2012_MD}, polarization tuning of light~\cite{Song_2017_JMM,Gries_2008_JAP}, and signal enhancement techniques~\cite{Katrin_2007_PT,Langer_2020_ACSNano}. Our method, which combines dynamic resonance measurements with theoretical analysis, provides a simpler detection process and more accurate stress values for characterizing stress asymmetry in thin films.
\par
The results from the Allan deviation measurements highlight the exceptional frequency stability of the degenerate-broken mechanical mode pairs. The dominant oscillator noise is primarily white noise, and importantly, no frequency drift was observed over the 40-day measurement period. This level of frequency stability represents the highest reported in the literature for micro- and nano-mechanical oscillators, even surpassing that of bulk silica glass. By integrating a superconducting resonant cavity with the SiC membrane-based electromechanical chip to form a cavity electromechanical system, we experimentally demonstrated high-fidelity state exchange between a degeneracy-breaking mode pair. The experimental results show that the transfer efficiency exceeds 78\%, which is attributed to the high quality factors of the mechanical modes and the strong coupling between the two mechanical modes and the shared cavity mode. These findings hold significant potential for advancing quantum information storage, scalable quantum computing, distributed quantum networks, and quantum simulation research based on multiple phonon modes via stress engineering~\cite{Habraken_2012_NJP,HailinWang_2018_PRX,YiwenChu_2018_Nature,YiwenChu_2024_NP}.
\bigskip
\\
\noindent{\textbf {Method}}\\
{\bf Device fabrication}\\
\indent The device measured at room temperature utilizes a commercial 3C-phase silicon carbide (Norcada PSCX5050A). The SiC membrane features a window with the dimensions of $500~\mu$m $~\times~500~\mu$m and a thickness of 50 nm, supported by a 400 $\mu$m thick silicon frame.
\par
At cryogenic temperature,  the cavity electromechanical device is composed of a 3D aluminum (Al) superconducting cavity and a mechanical capacitor chip. The 3D Al rectangular cavity is precision-machined and polished to ensure low microwave absorption and scattering. The mechanical capacitor chip consists of two parallel plates: the upper plate is the metallized SiC membrane and the bottom plate is a planar metal circuit capacitor. The metallization of the SiC membrane is achieved through the deposition of Al via electron beam evaporation (JEB-4, Adnanotek). The bottom planar capacitor features an H-shaped electrode and a concentric niobium (Nb) arc and cicle as its center. The planar capacitor electrode is fabricated from a 120 nm thick niobium (Nb) film, deposited on a high-resistivity silicon substrate using physical vapor deposition (Syskey Technology, SP-lC4-A06). The pattern is defined by laser direct-writing using DWL 66+ laser lithography tool (Heidelberg Instruments), and the unwanted Nb film is removed via reactive ion etching (RIE-10NR, Samco inc.). A commercial ACC$\mu$RA flip-chip bonder is employed to align the center of the upper plate with the center of the bottom capacitor electrode.
\bigskip
\\
{\bf Measurement methods}\\
\indent
The mechanical spectrum was employed to characterize the properties of the SiC membrane resonator. At room temperature, the mechanical modes were excited by microwaves via a probe positioned above it, and the resonant mechanical spectrum is acquired by a laser Doppler vibrometer (ONOSOKKI LV-1800) and analyzed by a lock-in amplifier (Zurich Instruments UHFLI, 600 MHz). All measurements were conducted in a vacuum chamber maintained at a pressure of $1\times 10^{-4}$ Pa. The resonant frequencies and quality factors were extracted by fitting the mechanical spectrum to the Lorentzian function.
\par
At cryogenic temperatures, the cavity electromechanical device was detected at the 10 mK stage of a dilution refrigerator (BlueFors LD400). The detection was preformed via reflection measurements using a circulator. In order to avoid the saturation of high electron mobility transistor (HEMT), a canceling tone and the output tone were integrated through a directional coupler, the reflected pump tone can be destructive interference by fine-tuning the amplitude and phase of the cancelling tone. An arbitrary waveform generator (Tektronix AWG5014C) was combined with a mixer to generate pulse tones, including sideband pump tone ($\omega_c-\omega_{m1,m2}$, $\omega_{p1}$, $\omega_{p2}$), amplified tone ($\omega_c+\omega_{m1,m2}$), and signal tone ($\omega_c$). These tones were combined using a broadband power splitter at room temperature and transmitted to the device through low-loss coaxial lines. High-isolation microwave switches were positioned after each mixer to ensure precise on-off control of individual tones, with the switch control signals and mixer trigger signals generated simultaneously via the marker channel and analog channel of the AWG, respectively. The real-time signal was acquired using a spectrum analyzer (Tektronix RSA5126B). For S21 parameter measurements of the device, the PNA (Rohde \& Schwarz ZNB20) probe tone was integrated into the input line. All instruments were phase-locked to a 10 MHz rubidium frequency standard.
\bigskip
\\
\noindent\textbf{Data availability}
The data that support the findings of this study are available from the corresponding authors upon reasonable request.
\bibliography{references}

\begin{thebibliography}{131}%
\makeatletter
\providecommand \@ifxundefined [1]{%
 \@ifx{#1\undefined}
}%
\providecommand \@ifnum [1]{%
 \ifnum #1\expandafter \@firstoftwo
 \else \expandafter \@secondoftwo
 \fi
}%
\providecommand \@ifx [1]{%
 \ifx #1\expandafter \@firstoftwo
 \else \expandafter \@secondoftwo
 \fi
}%
\providecommand \natexlab [1]{#1}%
\providecommand \enquote  [1]{``#1''}%
\providecommand \bibnamefont  [1]{#1}%
\providecommand \bibfnamefont [1]{#1}%
\providecommand \citenamefont [1]{#1}%
\providecommand \href@noop [0]{\@secondoftwo}%
\providecommand \href [0]{\begingroup \@sanitize@url \@href}%
\providecommand \@href[1]{\@@startlink{#1}\@@href}%
\providecommand \@@href[1]{\endgroup#1\@@endlink}%
\providecommand \@sanitize@url [0]{\catcode `\\12\catcode `\$12\catcode
  `\&12\catcode `\#12\catcode `\^12\catcode `\_12\catcode `\%12\relax}%
\providecommand \@@startlink[1]{}%
\providecommand \@@endlink[0]{}%
\providecommand \url  [0]{\begingroup\@sanitize@url \@url }%
\providecommand \@url [1]{\endgroup\@href {#1}{\urlprefix }}%
\providecommand \urlprefix  [0]{URL }%
\providecommand \Eprint [0]{\href }%
\providecommand \doibase [0]{https://doi.org/}%
\providecommand \selectlanguage [0]{\@gobble}%
\providecommand \bibinfo  [0]{\@secondoftwo}%
\providecommand \bibfield  [0]{\@secondoftwo}%
\providecommand \translation [1]{[#1]}%
\providecommand \BibitemOpen [0]{}%
\providecommand \bibitemStop [0]{}%
\providecommand \bibitemNoStop [0]{.\EOS\space}%
\providecommand \EOS [0]{\spacefactor3000\relax}%
\providecommand \BibitemShut  [1]{\csname bibitem#1\endcsname}%
\let\auto@bib@innerbib\@empty
\bibitem [{\citenamefont {Herring}(1954)}]{Herring_1954_PR}%
  \BibitemOpen
  \bibfield  {author} {\bibinfo {author} {\bibfnamefont {C.}~\bibnamefont
  {Herring}},\ }\bibfield  {title} {\bibinfo {title} {Role of low-energy
  phonons in thermal conduction},\ }\href
  {https://doi.org/10.1103/PhysRev.95.954} {\bibfield  {journal} {\bibinfo
  {journal} {Phys. Rev.}\ }\textbf {\bibinfo {volume} {95}},\ \bibinfo {pages}
  {954} (\bibinfo {year} {1954})}\BibitemShut {NoStop}%
\bibitem [{\citenamefont {Sinba}(1973)}]{Sinba_1973_book}%
  \BibitemOpen
  \bibfield  {author} {\bibinfo {author} {\bibfnamefont {S.~K.}\ \bibnamefont
  {Sinba}},\ }\bibfield  {title} {\bibinfo {title} {Phonons in
  semiconductors},\ }\href {https://doi.org/10.1080/10408437308244867}
  {\bibfield  {journal} {\bibinfo  {journal} {Critical Reviews in Solid State
  and Material Sciences}\ }\textbf {\bibinfo {volume} {3}},\ \bibinfo {pages}
  {273} (\bibinfo {year} {1973})}\BibitemShut {NoStop}%
\bibitem [{\citenamefont {Hillenbrand}\ \emph {et~al.}(2002)\citenamefont
  {Hillenbrand}, \citenamefont {Taubner},\ and\ \citenamefont
  {Keilmann}}]{Hillenbrand_2002_nature}%
  \BibitemOpen
  \bibfield  {author} {\bibinfo {author} {\bibfnamefont {R.}~\bibnamefont
  {Hillenbrand}}, \bibinfo {author} {\bibfnamefont {T.}~\bibnamefont
  {Taubner}},\ and\ \bibinfo {author} {\bibfnamefont {F.}~\bibnamefont
  {Keilmann}},\ }\bibfield  {title} {\bibinfo {title} {Phonon-enhanced
  light--matter interaction at the nanometre scale},\ }\href
  {https://doi.org/10.1038/nature00899} {\bibfield  {journal} {\bibinfo
  {journal} {Nature}\ }\textbf {\bibinfo {volume} {418}},\ \bibinfo {pages}
  {159} (\bibinfo {year} {2002})}\BibitemShut {NoStop}%
\bibitem [{\citenamefont {Garg}\ \emph {et~al.}(2017)\citenamefont {Garg},
  \citenamefont {Chauhan},\ and\ \citenamefont {Biswas}}]{Biswas_2017_PRA}%
  \BibitemOpen
  \bibfield  {author} {\bibinfo {author} {\bibfnamefont {D.}~\bibnamefont
  {Garg}}, \bibinfo {author} {\bibfnamefont {A.~K.}\ \bibnamefont {Chauhan}},\
  and\ \bibinfo {author} {\bibfnamefont {A.}~\bibnamefont {Biswas}},\
  }\bibfield  {title} {\bibinfo {title} {Adiabatic transfer of energy
  fluctuations between membranes inside an optical cavity},\ }\href
  {https://doi.org/10.1103/PhysRevA.96.023837} {\bibfield  {journal} {\bibinfo
  {journal} {Phys. Rev. A}\ }\textbf {\bibinfo {volume} {96}},\ \bibinfo
  {pages} {023837} (\bibinfo {year} {2017})}\BibitemShut {NoStop}%
\bibitem [{\citenamefont {Fan}\ \emph {et~al.}(2015)\citenamefont {Fan},
  \citenamefont {Fong}, \citenamefont {Poot},\ and\ \citenamefont
  {Tang}}]{HXTang_2015_NC}%
  \BibitemOpen
  \bibfield  {author} {\bibinfo {author} {\bibfnamefont {L.}~\bibnamefont
  {Fan}}, \bibinfo {author} {\bibfnamefont {K.~Y.}\ \bibnamefont {Fong}},
  \bibinfo {author} {\bibfnamefont {M.}~\bibnamefont {Poot}},\ and\ \bibinfo
  {author} {\bibfnamefont {H.~X.}\ \bibnamefont {Tang}},\ }\bibfield  {title}
  {\bibinfo {title} {Cascaded optical transparency in multimode-cavity
  optomechanical systems},\ }\href {https://doi.org/10.1038/ncomms6850}
  {\bibfield  {journal} {\bibinfo  {journal} {Nature Communications}\ }\textbf
  {\bibinfo {volume} {6}},\ \bibinfo {pages} {5850} (\bibinfo {year}
  {2015})}\BibitemShut {NoStop}%
\bibitem [{\citenamefont {Stannigel}\ \emph {et~al.}(2012)\citenamefont
  {Stannigel}, \citenamefont {Komar}, \citenamefont {Habraken}, \citenamefont
  {Bennett}, \citenamefont {Lukin}, \citenamefont {Zoller},\ and\ \citenamefont
  {Rabl}}]{Rabl_2012_PRL}%
  \BibitemOpen
  \bibfield  {author} {\bibinfo {author} {\bibfnamefont {K.}~\bibnamefont
  {Stannigel}}, \bibinfo {author} {\bibfnamefont {P.}~\bibnamefont {Komar}},
  \bibinfo {author} {\bibfnamefont {S.~J.~M.}\ \bibnamefont {Habraken}},
  \bibinfo {author} {\bibfnamefont {S.~D.}\ \bibnamefont {Bennett}}, \bibinfo
  {author} {\bibfnamefont {M.~D.}\ \bibnamefont {Lukin}}, \bibinfo {author}
  {\bibfnamefont {P.}~\bibnamefont {Zoller}},\ and\ \bibinfo {author}
  {\bibfnamefont {P.}~\bibnamefont {Rabl}},\ }\bibfield  {title} {\bibinfo
  {title} {Optomechanical quantum information processing with photons and
  phonons},\ }\href {https://doi.org/10.1103/PhysRevLett.109.013603} {\bibfield
   {journal} {\bibinfo  {journal} {Phys. Rev. Lett.}\ }\textbf {\bibinfo
  {volume} {109}},\ \bibinfo {pages} {013603} (\bibinfo {year}
  {2012})}\BibitemShut {NoStop}%
\bibitem [{\citenamefont {Lake}\ \emph {et~al.}(2021)\citenamefont {Lake},
  \citenamefont {Mitchell}, \citenamefont {Sukachev},\ and\ \citenamefont
  {Barclay}}]{Barclay_2021_NC}%
  \BibitemOpen
  \bibfield  {author} {\bibinfo {author} {\bibfnamefont {D.~P.}\ \bibnamefont
  {Lake}}, \bibinfo {author} {\bibfnamefont {M.}~\bibnamefont {Mitchell}},
  \bibinfo {author} {\bibfnamefont {D.~D.}\ \bibnamefont {Sukachev}},\ and\
  \bibinfo {author} {\bibfnamefont {P.~E.}\ \bibnamefont {Barclay}},\
  }\bibfield  {title} {\bibinfo {title} {Processing light with an optically
  tunable mechanical memory},\ }\href
  {https://doi.org/10.1038/s41467-021-20899-w} {\bibfield  {journal} {\bibinfo
  {journal} {Nature Communications}\ }\textbf {\bibinfo {volume} {12}},\
  \bibinfo {pages} {663} (\bibinfo {year} {2021})}\BibitemShut {NoStop}%
\bibitem [{\citenamefont {Schuetz}\ \emph {et~al.}(2015)\citenamefont
  {Schuetz}, \citenamefont {Kessler}, \citenamefont {Giedke}, \citenamefont
  {Vandersypen}, \citenamefont {Lukin},\ and\ \citenamefont
  {Cirac}}]{Schuetz_2015_PRL}%
  \BibitemOpen
  \bibfield  {author} {\bibinfo {author} {\bibfnamefont {M.~J.~A.}\
  \bibnamefont {Schuetz}}, \bibinfo {author} {\bibfnamefont {E.~M.}\
  \bibnamefont {Kessler}}, \bibinfo {author} {\bibfnamefont {G.}~\bibnamefont
  {Giedke}}, \bibinfo {author} {\bibfnamefont {L.~M.~K.}\ \bibnamefont
  {Vandersypen}}, \bibinfo {author} {\bibfnamefont {M.~D.}\ \bibnamefont
  {Lukin}},\ and\ \bibinfo {author} {\bibfnamefont {J.~I.}\ \bibnamefont
  {Cirac}},\ }\bibfield  {title} {\bibinfo {title} {Universal quantum
  transducers based on surface acoustic waves},\ }\href
  {https://doi.org/10.1103/PhysRevX.5.031031} {\bibfield  {journal} {\bibinfo
  {journal} {Phys. Rev. X}\ }\textbf {\bibinfo {volume} {5}},\ \bibinfo {pages}
  {031031} (\bibinfo {year} {2015})}\BibitemShut {NoStop}%
\bibitem [{\citenamefont {Tian}\ and\ \citenamefont
  {Wang}(2010)}]{Hailin_Wang_2010_PRA}%
  \BibitemOpen
  \bibfield  {author} {\bibinfo {author} {\bibfnamefont {L.}~\bibnamefont
  {Tian}}\ and\ \bibinfo {author} {\bibfnamefont {H.}~\bibnamefont {Wang}},\
  }\bibfield  {title} {\bibinfo {title} {Optical wavelength conversion of
  quantum states with optomechanics},\ }\href
  {https://doi.org/10.1103/PhysRevA.82.053806} {\bibfield  {journal} {\bibinfo
  {journal} {Phys. Rev. A}\ }\textbf {\bibinfo {volume} {82}},\ \bibinfo
  {pages} {053806} (\bibinfo {year} {2010})}\BibitemShut {NoStop}%
\bibitem [{\citenamefont {Habraken}\ \emph {et~al.}(2012)\citenamefont
  {Habraken}, \citenamefont {Stannigel}, \citenamefont {Lukin}, \citenamefont
  {Zoller},\ and\ \citenamefont {Rabl}}]{Habraken_2012_NJP}%
  \BibitemOpen
  \bibfield  {author} {\bibinfo {author} {\bibfnamefont {S.~J.~M.}\
  \bibnamefont {Habraken}}, \bibinfo {author} {\bibfnamefont {K.}~\bibnamefont
  {Stannigel}}, \bibinfo {author} {\bibfnamefont {M.~D.}\ \bibnamefont
  {Lukin}}, \bibinfo {author} {\bibfnamefont {P.}~\bibnamefont {Zoller}},\ and\
  \bibinfo {author} {\bibfnamefont {P.}~\bibnamefont {Rabl}},\ }\bibfield
  {title} {\bibinfo {title} {Continuous mode cooling and phonon routers for
  phononic quantum networks},\ }\href
  {https://doi.org/10.1088/1367-2630/14/11/115004} {\bibfield  {journal}
  {\bibinfo  {journal} {New Journal of Physics}\ }\textbf {\bibinfo {volume}
  {14}},\ \bibinfo {pages} {115004} (\bibinfo {year} {2012})}\BibitemShut
  {NoStop}%
\bibitem [{\citenamefont {Vermersch}\ \emph {et~al.}(2017)\citenamefont
  {Vermersch}, \citenamefont {Guimond}, \citenamefont {Pichler},\ and\
  \citenamefont {Zoller}}]{Zoller_2017_PRL}%
  \BibitemOpen
  \bibfield  {author} {\bibinfo {author} {\bibfnamefont {B.}~\bibnamefont
  {Vermersch}}, \bibinfo {author} {\bibfnamefont {P.-O.}\ \bibnamefont
  {Guimond}}, \bibinfo {author} {\bibfnamefont {H.}~\bibnamefont {Pichler}},\
  and\ \bibinfo {author} {\bibfnamefont {P.}~\bibnamefont {Zoller}},\
  }\bibfield  {title} {\bibinfo {title} {Quantum state transfer via noisy
  photonic and phononic waveguides},\ }\href
  {https://doi.org/10.1103/PhysRevLett.118.133601} {\bibfield  {journal}
  {\bibinfo  {journal} {Phys. Rev. Lett.}\ }\textbf {\bibinfo {volume} {118}},\
  \bibinfo {pages} {133601} (\bibinfo {year} {2017})}\BibitemShut {NoStop}%
\bibitem [{\citenamefont {Dobrindt}\ and\ \citenamefont
  {Kippenberg}(2010)}]{Kippenberg_2010_PRL}%
  \BibitemOpen
  \bibfield  {author} {\bibinfo {author} {\bibfnamefont {J.~M.}\ \bibnamefont
  {Dobrindt}}\ and\ \bibinfo {author} {\bibfnamefont {T.~J.}\ \bibnamefont
  {Kippenberg}},\ }\bibfield  {title} {\bibinfo {title} {Theoretical analysis
  of mechanical displacement measurement using a multiple cavity mode
  transducer},\ }\href {https://doi.org/10.1103/PhysRevLett.104.033901}
  {\bibfield  {journal} {\bibinfo  {journal} {Phys. Rev. Lett.}\ }\textbf
  {\bibinfo {volume} {104}},\ \bibinfo {pages} {033901} (\bibinfo {year}
  {2010})}\BibitemShut {NoStop}%
\bibitem [{\citenamefont {Burgwal}\ and\ \citenamefont
  {Verhagen}(2023)}]{Burgwal_2023_NC}%
  \BibitemOpen
  \bibfield  {author} {\bibinfo {author} {\bibfnamefont {R.}~\bibnamefont
  {Burgwal}}\ and\ \bibinfo {author} {\bibfnamefont {E.}~\bibnamefont
  {Verhagen}},\ }\bibfield  {title} {\bibinfo {title} {Enhanced nonlinear
  optomechanics in a coupled-mode photonic crystal device},\ }\href
  {https://doi.org/10.1038/s41467-023-37138-z} {\bibfield  {journal} {\bibinfo
  {journal} {Nature Communications}\ }\textbf {\bibinfo {volume} {14}},\
  \bibinfo {pages} {1526} (\bibinfo {year} {2023})}\BibitemShut {NoStop}%
\bibitem [{\citenamefont {Riedinger}\ \emph {et~al.}(2018)\citenamefont
  {Riedinger}, \citenamefont {Wallucks}, \citenamefont {Marinkovi{\'{c}}},
  \citenamefont {L{\"o}schnauer}, \citenamefont {Aspelmeyer}, \citenamefont
  {Hong},\ and\ \citenamefont {Gr{\"o}blacher}}]{Simon_2018_Nature}%
  \BibitemOpen
  \bibfield  {author} {\bibinfo {author} {\bibfnamefont {R.}~\bibnamefont
  {Riedinger}}, \bibinfo {author} {\bibfnamefont {A.}~\bibnamefont {Wallucks}},
  \bibinfo {author} {\bibfnamefont {I.}~\bibnamefont {Marinkovi{\'{c}}}},
  \bibinfo {author} {\bibfnamefont {C.}~\bibnamefont {L{\"o}schnauer}},
  \bibinfo {author} {\bibfnamefont {M.}~\bibnamefont {Aspelmeyer}}, \bibinfo
  {author} {\bibfnamefont {S.}~\bibnamefont {Hong}},\ and\ \bibinfo {author}
  {\bibfnamefont {S.}~\bibnamefont {Gr{\"o}blacher}},\ }\bibfield  {title}
  {\bibinfo {title} {Remote quantum entanglement between two micromechanical
  oscillators},\ }\href {https://doi.org/10.1038/s41586-018-0036-z} {\bibfield
  {journal} {\bibinfo  {journal} {Nature}\ }\textbf {\bibinfo {volume} {556}},\
  \bibinfo {pages} {473} (\bibinfo {year} {2018})}\BibitemShut {NoStop}%
\bibitem [{\citenamefont {Ockeloen-Korppi}\ \emph {et~al.}(2018)\citenamefont
  {Ockeloen-Korppi}, \citenamefont {Damsk{\"a}gg}, \citenamefont
  {Pirkkalainen}, \citenamefont {Asjad}, \citenamefont {Clerk}, \citenamefont
  {Massel}, \citenamefont {Woolley},\ and\ \citenamefont
  {Sillanp{\"a}{\"a}}}]{Mika_2018_Nature}%
  \BibitemOpen
  \bibfield  {author} {\bibinfo {author} {\bibfnamefont {C.~F.}\ \bibnamefont
  {Ockeloen-Korppi}}, \bibinfo {author} {\bibfnamefont {E.}~\bibnamefont
  {Damsk{\"a}gg}}, \bibinfo {author} {\bibfnamefont {J.-M.}\ \bibnamefont
  {Pirkkalainen}}, \bibinfo {author} {\bibfnamefont {M.}~\bibnamefont {Asjad}},
  \bibinfo {author} {\bibfnamefont {A.~A.}\ \bibnamefont {Clerk}}, \bibinfo
  {author} {\bibfnamefont {F.}~\bibnamefont {Massel}}, \bibinfo {author}
  {\bibfnamefont {M.~J.}\ \bibnamefont {Woolley}},\ and\ \bibinfo {author}
  {\bibfnamefont {M.~A.}\ \bibnamefont {Sillanp{\"a}{\"a}}},\ }\bibfield
  {title} {\bibinfo {title} {Stabilized entanglement of massive mechanical
  oscillators},\ }\href {https://doi.org/10.1038/s41586-018-0038-x} {\bibfield
  {journal} {\bibinfo  {journal} {Nature}\ }\textbf {\bibinfo {volume} {556}},\
  \bibinfo {pages} {478} (\bibinfo {year} {2018})}\BibitemShut {NoStop}%
\bibitem [{\citenamefont {Lu}\ \emph {et~al.}(2024)\citenamefont {Lu},
  \citenamefont {Li}, \citenamefont {Wang}, \citenamefont {Wang}, \citenamefont
  {Xiao},\ and\ \citenamefont {Jing}}]{Hujing_2024_PRApplied}%
  \BibitemOpen
  \bibfield  {author} {\bibinfo {author} {\bibfnamefont {T.-X.}\ \bibnamefont
  {Lu}}, \bibinfo {author} {\bibfnamefont {B.}~\bibnamefont {Li}}, \bibinfo
  {author} {\bibfnamefont {Y.}~\bibnamefont {Wang}}, \bibinfo {author}
  {\bibfnamefont {D.-Y.}\ \bibnamefont {Wang}}, \bibinfo {author}
  {\bibfnamefont {X.}~\bibnamefont {Xiao}},\ and\ \bibinfo {author}
  {\bibfnamefont {H.}~\bibnamefont {Jing}},\ }\bibfield  {title} {\bibinfo
  {title} {Directional quantum-squeezing-enabled nonreciprocal enhancement of
  entanglement},\ }\href {https://doi.org/10.1103/PhysRevApplied.22.064001}
  {\bibfield  {journal} {\bibinfo  {journal} {Phys. Rev. Appl.}\ }\textbf
  {\bibinfo {volume} {22}},\ \bibinfo {pages} {064001} (\bibinfo {year}
  {2024})}\BibitemShut {NoStop}%
\bibitem [{\citenamefont {Ghosh}\ \emph {et~al.}(2024)\citenamefont {Ghosh},
  \citenamefont {Mondal}, \citenamefont {Varshney},\ and\ \citenamefont
  {Debnath}}]{Kapil_2014_PRA}%
  \BibitemOpen
  \bibfield  {author} {\bibinfo {author} {\bibfnamefont {J.}~\bibnamefont
  {Ghosh}}, \bibinfo {author} {\bibfnamefont {S.}~\bibnamefont {Mondal}},
  \bibinfo {author} {\bibfnamefont {S.~K.}\ \bibnamefont {Varshney}},\ and\
  \bibinfo {author} {\bibfnamefont {K.}~\bibnamefont {Debnath}},\ }\bibfield
  {title} {\bibinfo {title} {Simultaneous control of quantum phase
  synchronization and entanglement dynamics in a gain-loss optomechanical
  cavity system},\ }\href {https://doi.org/10.1103/PhysRevA.109.023512}
  {\bibfield  {journal} {\bibinfo  {journal} {Phys. Rev. A}\ }\textbf {\bibinfo
  {volume} {109}},\ \bibinfo {pages} {023512} (\bibinfo {year}
  {2024})}\BibitemShut {NoStop}%
\bibitem [{\citenamefont {Huang}\ \emph
  {et~al.}(2022{\natexlab{a}})\citenamefont {Huang}, \citenamefont {Lai},\ and\
  \citenamefont {Liao}}]{Jieqiao_Liao_2022_PRA}%
  \BibitemOpen
  \bibfield  {author} {\bibinfo {author} {\bibfnamefont {J.}~\bibnamefont
  {Huang}}, \bibinfo {author} {\bibfnamefont {D.-G.}\ \bibnamefont {Lai}},\
  and\ \bibinfo {author} {\bibfnamefont {J.-Q.}\ \bibnamefont {Liao}},\
  }\bibfield  {title} {\bibinfo {title} {Thermal-noise-resistant optomechanical
  entanglement via general dark-mode control},\ }\href
  {https://doi.org/10.1103/PhysRevA.106.063506} {\bibfield  {journal} {\bibinfo
   {journal} {Phys. Rev. A}\ }\textbf {\bibinfo {volume} {106}},\ \bibinfo
  {pages} {063506} (\bibinfo {year} {2022}{\natexlab{a}})}\BibitemShut
  {NoStop}%
\bibitem [{\citenamefont {Kemiktarak}\ \emph
  {et~al.}(2014{\natexlab{a}})\citenamefont {Kemiktarak}, \citenamefont
  {Durand}, \citenamefont {Metcalfe},\ and\ \citenamefont
  {Lawall}}]{John_2018_PRL}%
  \BibitemOpen
  \bibfield  {author} {\bibinfo {author} {\bibfnamefont {U.}~\bibnamefont
  {Kemiktarak}}, \bibinfo {author} {\bibfnamefont {M.}~\bibnamefont {Durand}},
  \bibinfo {author} {\bibfnamefont {M.}~\bibnamefont {Metcalfe}},\ and\
  \bibinfo {author} {\bibfnamefont {J.}~\bibnamefont {Lawall}},\ }\bibfield
  {title} {\bibinfo {title} {Mode competition and anomalous cooling in a
  multimode phonon laser},\ }\href
  {https://doi.org/10.1103/PhysRevLett.113.030802} {\bibfield  {journal}
  {\bibinfo  {journal} {Phys. Rev. Lett.}\ }\textbf {\bibinfo {volume} {113}},\
  \bibinfo {pages} {030802} (\bibinfo {year} {2014}{\natexlab{a}})}\BibitemShut
  {NoStop}%
\bibitem [{\citenamefont {Sheng}\ \emph {et~al.}(2020)\citenamefont {Sheng},
  \citenamefont {Wei}, \citenamefont {Yang},\ and\ \citenamefont
  {Wu}}]{HaibinWu_2024_PRL}%
  \BibitemOpen
  \bibfield  {author} {\bibinfo {author} {\bibfnamefont {J.}~\bibnamefont
  {Sheng}}, \bibinfo {author} {\bibfnamefont {X.}~\bibnamefont {Wei}}, \bibinfo
  {author} {\bibfnamefont {C.}~\bibnamefont {Yang}},\ and\ \bibinfo {author}
  {\bibfnamefont {H.}~\bibnamefont {Wu}},\ }\bibfield  {title} {\bibinfo
  {title} {Self-organized synchronization of phonon lasers},\ }\href
  {https://doi.org/10.1103/PhysRevLett.124.053604} {\bibfield  {journal}
  {\bibinfo  {journal} {Phys. Rev. Lett.}\ }\textbf {\bibinfo {volume} {124}},\
  \bibinfo {pages} {053604} (\bibinfo {year} {2020})}\BibitemShut {NoStop}%
\bibitem [{\citenamefont {Mercad\'e}\ \emph {et~al.}(2021)\citenamefont
  {Mercad\'e}, \citenamefont {Pelka}, \citenamefont {Burgwal}, \citenamefont
  {Xuereb}, \citenamefont {Mart\'{\i}nez},\ and\ \citenamefont
  {Verhagen}}]{Ewold_2021_PRL}%
  \BibitemOpen
  \bibfield  {author} {\bibinfo {author} {\bibfnamefont {L.}~\bibnamefont
  {Mercad\'e}}, \bibinfo {author} {\bibfnamefont {K.}~\bibnamefont {Pelka}},
  \bibinfo {author} {\bibfnamefont {R.}~\bibnamefont {Burgwal}}, \bibinfo
  {author} {\bibfnamefont {A.}~\bibnamefont {Xuereb}}, \bibinfo {author}
  {\bibfnamefont {A.}~\bibnamefont {Mart\'{\i}nez}},\ and\ \bibinfo {author}
  {\bibfnamefont {E.}~\bibnamefont {Verhagen}},\ }\bibfield  {title} {\bibinfo
  {title} {Floquet phonon lasing in multimode optomechanical systems},\ }\href
  {https://doi.org/10.1103/PhysRevLett.127.073601} {\bibfield  {journal}
  {\bibinfo  {journal} {Phys. Rev. Lett.}\ }\textbf {\bibinfo {volume} {127}},\
  \bibinfo {pages} {073601} (\bibinfo {year} {2021})}\BibitemShut {NoStop}%
\bibitem [{\citenamefont {Mercier~de L\'epinay}\ \emph
  {et~al.}(2020)\citenamefont {Mercier~de L\'epinay}, \citenamefont
  {Ockeloen-Korppi}, \citenamefont {Malz},\ and\ \citenamefont
  {Sillanp\"a\"a}}]{Mika_2020_PRL}%
  \BibitemOpen
  \bibfield  {author} {\bibinfo {author} {\bibfnamefont {L.}~\bibnamefont
  {Mercier~de L\'epinay}}, \bibinfo {author} {\bibfnamefont {C.~F.}\
  \bibnamefont {Ockeloen-Korppi}}, \bibinfo {author} {\bibfnamefont
  {D.}~\bibnamefont {Malz}},\ and\ \bibinfo {author} {\bibfnamefont {M.~A.}\
  \bibnamefont {Sillanp\"a\"a}},\ }\bibfield  {title} {\bibinfo {title}
  {Nonreciprocal transport based on cavity floquet modes in optomechanics},\
  }\href {https://doi.org/10.1103/PhysRevLett.125.023603} {\bibfield  {journal}
  {\bibinfo  {journal} {Phys. Rev. Lett.}\ }\textbf {\bibinfo {volume} {125}},\
  \bibinfo {pages} {023603} (\bibinfo {year} {2020})}\BibitemShut {NoStop}%
\bibitem [{\citenamefont {Bernier}\ \emph {et~al.}(2017)\citenamefont
  {Bernier}, \citenamefont {T{\'o}th}, \citenamefont {Koottandavida},
  \citenamefont {Ioannou}, \citenamefont {Malz}, \citenamefont {Nunnenkamp},
  \citenamefont {Feofanov},\ and\ \citenamefont {Kippenberg}}]{Bernier2017}%
  \BibitemOpen
  \bibfield  {author} {\bibinfo {author} {\bibfnamefont {N.~R.}\ \bibnamefont
  {Bernier}}, \bibinfo {author} {\bibfnamefont {L.~D.}\ \bibnamefont
  {T{\'o}th}}, \bibinfo {author} {\bibfnamefont {A.}~\bibnamefont
  {Koottandavida}}, \bibinfo {author} {\bibfnamefont {M.~A.}\ \bibnamefont
  {Ioannou}}, \bibinfo {author} {\bibfnamefont {D.}~\bibnamefont {Malz}},
  \bibinfo {author} {\bibfnamefont {A.}~\bibnamefont {Nunnenkamp}}, \bibinfo
  {author} {\bibfnamefont {A.~K.}\ \bibnamefont {Feofanov}},\ and\ \bibinfo
  {author} {\bibfnamefont {T.~J.}\ \bibnamefont {Kippenberg}},\ }\bibfield
  {title} {\bibinfo {title} {Nonreciprocal reconfigurable microwave
  optomechanical circuit},\ }\href {https://doi.org/10.1038/s41467-017-00447-1}
  {\bibfield  {journal} {\bibinfo  {journal} {Nature Communications}\ }\textbf
  {\bibinfo {volume} {8}},\ \bibinfo {pages} {604} (\bibinfo {year}
  {2017})}\BibitemShut {NoStop}%
\bibitem [{\citenamefont {Seif}\ \emph {et~al.}(2018)\citenamefont {Seif},
  \citenamefont {DeGottardi}, \citenamefont {Esfarjani},\ and\ \citenamefont
  {Hafezi}}]{Seif2018}%
  \BibitemOpen
  \bibfield  {author} {\bibinfo {author} {\bibfnamefont {A.}~\bibnamefont
  {Seif}}, \bibinfo {author} {\bibfnamefont {W.}~\bibnamefont {DeGottardi}},
  \bibinfo {author} {\bibfnamefont {K.}~\bibnamefont {Esfarjani}},\ and\
  \bibinfo {author} {\bibfnamefont {M.}~\bibnamefont {Hafezi}},\ }\bibfield
  {title} {\bibinfo {title} {Thermal management and non-reciprocal control of
  phonon flow via optomechanics},\ }\href
  {https://doi.org/10.1038/s41467-018-03624-y} {\bibfield  {journal} {\bibinfo
  {journal} {Nature Communications}\ }\textbf {\bibinfo {volume} {9}},\
  \bibinfo {pages} {1207} (\bibinfo {year} {2018})}\BibitemShut {NoStop}%
\bibitem [{\citenamefont {Fang}\ \emph {et~al.}(2017)\citenamefont {Fang},
  \citenamefont {Luo}, \citenamefont {Metelmann}, \citenamefont {Matheny},
  \citenamefont {Marquardt}, \citenamefont {Clerk},\ and\ \citenamefont
  {Painter}}]{Fang2017}%
  \BibitemOpen
  \bibfield  {author} {\bibinfo {author} {\bibfnamefont {K.}~\bibnamefont
  {Fang}}, \bibinfo {author} {\bibfnamefont {J.}~\bibnamefont {Luo}}, \bibinfo
  {author} {\bibfnamefont {A.}~\bibnamefont {Metelmann}}, \bibinfo {author}
  {\bibfnamefont {M.~H.}\ \bibnamefont {Matheny}}, \bibinfo {author}
  {\bibfnamefont {F.}~\bibnamefont {Marquardt}}, \bibinfo {author}
  {\bibfnamefont {A.~A.}\ \bibnamefont {Clerk}},\ and\ \bibinfo {author}
  {\bibfnamefont {O.}~\bibnamefont {Painter}},\ }\bibfield  {title} {\bibinfo
  {title} {Generalized non-reciprocity in an optomechanical circuit via
  synthetic magnetism and reservoir engineering},\ }\href
  {https://doi.org/10.1038/nphys4009} {\bibfield  {journal} {\bibinfo
  {journal} {Nature Physics}\ }\textbf {\bibinfo {volume} {13}},\ \bibinfo
  {pages} {465} (\bibinfo {year} {2017})}\BibitemShut {NoStop}%
\bibitem [{\citenamefont {Mathew}\ \emph {et~al.}(2020)\citenamefont {Mathew},
  \citenamefont {Pino},\ and\ \citenamefont {Verhagen}}]{Mathew2020}%
  \BibitemOpen
  \bibfield  {author} {\bibinfo {author} {\bibfnamefont {J.~P.}\ \bibnamefont
  {Mathew}}, \bibinfo {author} {\bibfnamefont {J.~d.}\ \bibnamefont {Pino}},\
  and\ \bibinfo {author} {\bibfnamefont {E.}~\bibnamefont {Verhagen}},\
  }\bibfield  {title} {\bibinfo {title} {Synthetic gauge fields for phonon
  transport in a nano-optomechanical system},\ }\href
  {https://doi.org/10.1038/s41565-019-0630-8} {\bibfield  {journal} {\bibinfo
  {journal} {Nature Nanotechnology}\ }\textbf {\bibinfo {volume} {15}},\
  \bibinfo {pages} {198} (\bibinfo {year} {2020})}\BibitemShut {NoStop}%
\bibitem [{\citenamefont {Ruesink}\ \emph {et~al.}(2016)\citenamefont
  {Ruesink}, \citenamefont {Miri}, \citenamefont {Al{\`u}},\ and\ \citenamefont
  {Verhagen}}]{Ruesink2016}%
  \BibitemOpen
  \bibfield  {author} {\bibinfo {author} {\bibfnamefont {F.}~\bibnamefont
  {Ruesink}}, \bibinfo {author} {\bibfnamefont {M.-A.}\ \bibnamefont {Miri}},
  \bibinfo {author} {\bibfnamefont {A.}~\bibnamefont {Al{\`u}}},\ and\ \bibinfo
  {author} {\bibfnamefont {E.}~\bibnamefont {Verhagen}},\ }\bibfield  {title}
  {\bibinfo {title} {Nonreciprocity and magnetic-free isolation based on
  optomechanical interactions},\ }\href {https://doi.org/10.1038/ncomms13662}
  {\bibfield  {journal} {\bibinfo  {journal} {Nature Communications}\ }\textbf
  {\bibinfo {volume} {7}},\ \bibinfo {pages} {13662} (\bibinfo {year}
  {2016})}\BibitemShut {NoStop}%
\bibitem [{\citenamefont {Xu}\ \emph {et~al.}(2019)\citenamefont {Xu},
  \citenamefont {Jiang}, \citenamefont {Clerk},\ and\ \citenamefont
  {Harris}}]{Harris_2019_Nature}%
  \BibitemOpen
  \bibfield  {author} {\bibinfo {author} {\bibfnamefont {H.}~\bibnamefont
  {Xu}}, \bibinfo {author} {\bibfnamefont {L.}~\bibnamefont {Jiang}}, \bibinfo
  {author} {\bibfnamefont {A.~A.}\ \bibnamefont {Clerk}},\ and\ \bibinfo
  {author} {\bibfnamefont {J.~G.~E.}\ \bibnamefont {Harris}},\ }\bibfield
  {title} {\bibinfo {title} {Nonreciprocal control and cooling of phonon modes
  in an optomechanical system},\ }\href
  {https://doi.org/10.1038/s41586-019-1061-2} {\bibfield  {journal} {\bibinfo
  {journal} {Nature}\ }\textbf {\bibinfo {volume} {568}},\ \bibinfo {pages}
  {65} (\bibinfo {year} {2019})}\BibitemShut {NoStop}%
\bibitem [{\citenamefont {Nielsen}\ \emph {et~al.}(2017)\citenamefont
  {Nielsen}, \citenamefont {Tsaturyan}, \citenamefont {Møller}, \citenamefont
  {Polzik},\ and\ \citenamefont {Schliesser}}]{Albert_2017_PNAS}%
  \BibitemOpen
  \bibfield  {author} {\bibinfo {author} {\bibfnamefont {W.~H.~P.}\
  \bibnamefont {Nielsen}}, \bibinfo {author} {\bibfnamefont {Y.}~\bibnamefont
  {Tsaturyan}}, \bibinfo {author} {\bibfnamefont {C.~B.}\ \bibnamefont
  {Møller}}, \bibinfo {author} {\bibfnamefont {E.~S.}\ \bibnamefont
  {Polzik}},\ and\ \bibinfo {author} {\bibfnamefont {A.}~\bibnamefont
  {Schliesser}},\ }\bibfield  {title} {\bibinfo {title} {Multimode
  optomechanical system in the quantum regime},\ }\href
  {https://doi.org/10.1073/pnas.1608412114} {\bibfield  {journal} {\bibinfo
  {journal} {Proceedings of the National Academy of Sciences}\ }\textbf
  {\bibinfo {volume} {114}},\ \bibinfo {pages} {62} (\bibinfo {year}
  {2017})}\BibitemShut {NoStop}%
\bibitem [{\citenamefont {Piergentili}\ \emph {et~al.}(2018)\citenamefont
  {Piergentili}, \citenamefont {Catalini}, \citenamefont {Bawaj}, \citenamefont
  {Zippilli}, \citenamefont {Malossi}, \citenamefont {Natali}, \citenamefont
  {Vitali},\ and\ \citenamefont {Giuseppe}}]{Vitali_2018_NJP}%
  \BibitemOpen
  \bibfield  {author} {\bibinfo {author} {\bibfnamefont {P.}~\bibnamefont
  {Piergentili}}, \bibinfo {author} {\bibfnamefont {L.}~\bibnamefont
  {Catalini}}, \bibinfo {author} {\bibfnamefont {M.}~\bibnamefont {Bawaj}},
  \bibinfo {author} {\bibfnamefont {S.}~\bibnamefont {Zippilli}}, \bibinfo
  {author} {\bibfnamefont {N.}~\bibnamefont {Malossi}}, \bibinfo {author}
  {\bibfnamefont {R.}~\bibnamefont {Natali}}, \bibinfo {author} {\bibfnamefont
  {D.}~\bibnamefont {Vitali}},\ and\ \bibinfo {author} {\bibfnamefont {G.~D.}\
  \bibnamefont {Giuseppe}},\ }\bibfield  {title} {\bibinfo {title}
  {Two-membrane cavity optomechanics},\ }\href
  {https://doi.org/10.1088/1367-2630/aad85f} {\bibfield  {journal} {\bibinfo
  {journal} {New Journal of Physics}\ }\textbf {\bibinfo {volume} {20}},\
  \bibinfo {pages} {083024} (\bibinfo {year} {2018})}\BibitemShut {NoStop}%
\bibitem [{\citenamefont {Chegnizadeh}\ \emph {et~al.}(2024)\citenamefont
  {Chegnizadeh}, \citenamefont {Scigliuzzo}, \citenamefont {Youssefi},
  \citenamefont {Kono}, \citenamefont {Guzovskii},\ and\ \citenamefont
  {Kippenberg}}]{Kippenberg_2024_Science}%
  \BibitemOpen
  \bibfield  {author} {\bibinfo {author} {\bibfnamefont {M.}~\bibnamefont
  {Chegnizadeh}}, \bibinfo {author} {\bibfnamefont {M.}~\bibnamefont
  {Scigliuzzo}}, \bibinfo {author} {\bibfnamefont {A.}~\bibnamefont
  {Youssefi}}, \bibinfo {author} {\bibfnamefont {S.}~\bibnamefont {Kono}},
  \bibinfo {author} {\bibfnamefont {E.}~\bibnamefont {Guzovskii}},\ and\
  \bibinfo {author} {\bibfnamefont {T.~J.}\ \bibnamefont {Kippenberg}},\
  }\bibfield  {title} {\bibinfo {title} {Quantum collective motion of
  macroscopic mechanical oscillators},\ }\href
  {https://doi.org/10.1126/science.adr8187} {\bibfield  {journal} {\bibinfo
  {journal} {Science}\ }\textbf {\bibinfo {volume} {386}},\ \bibinfo {pages}
  {1383} (\bibinfo {year} {2024})}\BibitemShut {NoStop}%
\bibitem [{\citenamefont {Weaver}\ \emph {et~al.}(2017)\citenamefont {Weaver},
  \citenamefont {Buters}, \citenamefont {Luna}, \citenamefont {Eerkens},
  \citenamefont {Heeck}, \citenamefont {de~Man},\ and\ \citenamefont
  {Bouwmeester}}]{Weaver_2017_NC}%
  \BibitemOpen
  \bibfield  {author} {\bibinfo {author} {\bibfnamefont {M.~J.}\ \bibnamefont
  {Weaver}}, \bibinfo {author} {\bibfnamefont {F.}~\bibnamefont {Buters}},
  \bibinfo {author} {\bibfnamefont {F.}~\bibnamefont {Luna}}, \bibinfo {author}
  {\bibfnamefont {H.}~\bibnamefont {Eerkens}}, \bibinfo {author} {\bibfnamefont
  {K.}~\bibnamefont {Heeck}}, \bibinfo {author} {\bibfnamefont
  {S.}~\bibnamefont {de~Man}},\ and\ \bibinfo {author} {\bibfnamefont
  {D.}~\bibnamefont {Bouwmeester}},\ }\bibfield  {title} {\bibinfo {title}
  {Coherent optomechanical state transfer between disparate mechanical
  resonators},\ }\href {https://doi.org/10.1038/s41467-017-00968-9} {\bibfield
  {journal} {\bibinfo  {journal} {Nature Communications}\ }\textbf {\bibinfo
  {volume} {8}},\ \bibinfo {pages} {824} (\bibinfo {year} {2017})}\BibitemShut
  {NoStop}%
\bibitem [{\citenamefont {Piergentili}\ \emph {et~al.}(2021)\citenamefont
  {Piergentili}, \citenamefont {Li}, \citenamefont {Natali}, \citenamefont
  {Malossi}, \citenamefont {Vitali},\ and\ \citenamefont
  {Giuseppe}}]{Vitali_2021_NJP}%
  \BibitemOpen
  \bibfield  {author} {\bibinfo {author} {\bibfnamefont {P.}~\bibnamefont
  {Piergentili}}, \bibinfo {author} {\bibfnamefont {W.}~\bibnamefont {Li}},
  \bibinfo {author} {\bibfnamefont {R.}~\bibnamefont {Natali}}, \bibinfo
  {author} {\bibfnamefont {N.}~\bibnamefont {Malossi}}, \bibinfo {author}
  {\bibfnamefont {D.}~\bibnamefont {Vitali}},\ and\ \bibinfo {author}
  {\bibfnamefont {G.~D.}\ \bibnamefont {Giuseppe}},\ }\bibfield  {title}
  {\bibinfo {title} {Two-membrane cavity optomechanics: non-linear dynamics},\
  }\href {https://doi.org/10.1088/1367-2630/abdd6a} {\bibfield  {journal}
  {\bibinfo  {journal} {New Journal of Physics}\ }\textbf {\bibinfo {volume}
  {23}},\ \bibinfo {pages} {073013} (\bibinfo {year} {2021})}\BibitemShut
  {NoStop}%
\bibitem [{\citenamefont {Wu}\ \emph {et~al.}(2022)\citenamefont {Wu},
  \citenamefont {Liu}, \citenamefont {Liu}, \citenamefont {Wang}, \citenamefont
  {Chen},\ and\ \citenamefont {Li}}]{SiShiWu_2022_PRL}%
  \BibitemOpen
  \bibfield  {author} {\bibinfo {author} {\bibfnamefont {S.}~\bibnamefont
  {Wu}}, \bibinfo {author} {\bibfnamefont {Y.}~\bibnamefont {Liu}}, \bibinfo
  {author} {\bibfnamefont {Q.}~\bibnamefont {Liu}}, \bibinfo {author}
  {\bibfnamefont {S.-P.}\ \bibnamefont {Wang}}, \bibinfo {author}
  {\bibfnamefont {Z.}~\bibnamefont {Chen}},\ and\ \bibinfo {author}
  {\bibfnamefont {T.}~\bibnamefont {Li}},\ }\bibfield  {title} {\bibinfo
  {title} {Hybridized frequency combs in multimode cavity electromechanical
  system},\ }\href {https://doi.org/10.1103/PhysRevLett.128.153901} {\bibfield
  {journal} {\bibinfo  {journal} {Phys. Rev. Lett.}\ }\textbf {\bibinfo
  {volume} {128}},\ \bibinfo {pages} {153901} (\bibinfo {year}
  {2022})}\BibitemShut {NoStop}%
\bibitem [{\citenamefont {Shkarin}\ \emph
  {et~al.}(2014{\natexlab{a}})\citenamefont {Shkarin}, \citenamefont
  {Flowers-Jacobs}, \citenamefont {Hoch}, \citenamefont {Kashkanova},
  \citenamefont {Deutsch}, \citenamefont {Reichel},\ and\ \citenamefont
  {Harris}}]{JGE_2014_PRL}%
  \BibitemOpen
  \bibfield  {author} {\bibinfo {author} {\bibfnamefont {A.~B.}\ \bibnamefont
  {Shkarin}}, \bibinfo {author} {\bibfnamefont {N.~E.}\ \bibnamefont
  {Flowers-Jacobs}}, \bibinfo {author} {\bibfnamefont {S.~W.}\ \bibnamefont
  {Hoch}}, \bibinfo {author} {\bibfnamefont {A.~D.}\ \bibnamefont
  {Kashkanova}}, \bibinfo {author} {\bibfnamefont {C.}~\bibnamefont {Deutsch}},
  \bibinfo {author} {\bibfnamefont {J.}~\bibnamefont {Reichel}},\ and\ \bibinfo
  {author} {\bibfnamefont {J.~G.~E.}\ \bibnamefont {Harris}},\ }\bibfield
  {title} {\bibinfo {title} {Optically mediated hybridization between two
  mechanical modes},\ }\href {https://doi.org/10.1103/PhysRevLett.112.013602}
  {\bibfield  {journal} {\bibinfo  {journal} {Phys. Rev. Lett.}\ }\textbf
  {\bibinfo {volume} {112}},\ \bibinfo {pages} {013602} (\bibinfo {year}
  {2014}{\natexlab{a}})}\BibitemShut {NoStop}%
\bibitem [{\citenamefont {Schneider}\ \emph {et~al.}(2019)\citenamefont
  {Schneider}, \citenamefont {Baumgartner}, \citenamefont {H\"{o}nl},
  \citenamefont {Welter}, \citenamefont {Hahn}, \citenamefont {Wilson},
  \citenamefont {Czornomaz},\ and\ \citenamefont {Seidler}}]{Paul_2019_Optica}%
  \BibitemOpen
  \bibfield  {author} {\bibinfo {author} {\bibfnamefont {K.}~\bibnamefont
  {Schneider}}, \bibinfo {author} {\bibfnamefont {Y.}~\bibnamefont
  {Baumgartner}}, \bibinfo {author} {\bibfnamefont {S.}~\bibnamefont
  {H\"{o}nl}}, \bibinfo {author} {\bibfnamefont {P.}~\bibnamefont {Welter}},
  \bibinfo {author} {\bibfnamefont {H.}~\bibnamefont {Hahn}}, \bibinfo {author}
  {\bibfnamefont {D.~J.}\ \bibnamefont {Wilson}}, \bibinfo {author}
  {\bibfnamefont {L.}~\bibnamefont {Czornomaz}},\ and\ \bibinfo {author}
  {\bibfnamefont {P.}~\bibnamefont {Seidler}},\ }\bibfield  {title} {\bibinfo
  {title} {Optomechanics with one-dimensional gallium phosphide photonic
  crystal cavities},\ }\href {https://doi.org/10.1364/OPTICA.6.000577}
  {\bibfield  {journal} {\bibinfo  {journal} {Optica}\ }\textbf {\bibinfo
  {volume} {6}},\ \bibinfo {pages} {577} (\bibinfo {year} {2019})}\BibitemShut
  {NoStop}%
\bibitem [{\citenamefont {Asano}\ \emph {et~al.}(2020)\citenamefont {Asano},
  \citenamefont {Zhang}, \citenamefont {Tawara}, \citenamefont {Yamaguchi},\
  and\ \citenamefont {Okamoto}}]{Hajime_2020_CP}%
  \BibitemOpen
  \bibfield  {author} {\bibinfo {author} {\bibfnamefont {M.}~\bibnamefont
  {Asano}}, \bibinfo {author} {\bibfnamefont {G.}~\bibnamefont {Zhang}},
  \bibinfo {author} {\bibfnamefont {T.}~\bibnamefont {Tawara}}, \bibinfo
  {author} {\bibfnamefont {H.}~\bibnamefont {Yamaguchi}},\ and\ \bibinfo
  {author} {\bibfnamefont {H.}~\bibnamefont {Okamoto}},\ }\bibfield  {title}
  {\bibinfo {title} {Near-field cavity optomechanical coupling in a compound
  semiconductor nanowire},\ }\href {https://doi.org/10.1038/s42005-020-00498-9}
  {\bibfield  {journal} {\bibinfo  {journal} {Communications Physics}\ }\textbf
  {\bibinfo {volume} {3}},\ \bibinfo {pages} {230} (\bibinfo {year}
  {2020})}\BibitemShut {NoStop}%
\bibitem [{\citenamefont {Fogliano}\ \emph {et~al.}(2021)\citenamefont
  {Fogliano}, \citenamefont {Besga}, \citenamefont {Reigue}, \citenamefont
  {Heringlake}, \citenamefont {Mercier~de L\'epinay}, \citenamefont {Vaneph},
  \citenamefont {Reichel}, \citenamefont {Pigeau},\ and\ \citenamefont
  {Arcizet}}]{Laure_2021_PRX}%
  \BibitemOpen
  \bibfield  {author} {\bibinfo {author} {\bibfnamefont {F.}~\bibnamefont
  {Fogliano}}, \bibinfo {author} {\bibfnamefont {B.}~\bibnamefont {Besga}},
  \bibinfo {author} {\bibfnamefont {A.}~\bibnamefont {Reigue}}, \bibinfo
  {author} {\bibfnamefont {P.}~\bibnamefont {Heringlake}}, \bibinfo {author}
  {\bibfnamefont {L.}~\bibnamefont {Mercier~de L\'epinay}}, \bibinfo {author}
  {\bibfnamefont {C.}~\bibnamefont {Vaneph}}, \bibinfo {author} {\bibfnamefont
  {J.}~\bibnamefont {Reichel}}, \bibinfo {author} {\bibfnamefont
  {B.}~\bibnamefont {Pigeau}},\ and\ \bibinfo {author} {\bibfnamefont
  {O.}~\bibnamefont {Arcizet}},\ }\bibfield  {title} {\bibinfo {title} {Mapping
  the cavity optomechanical interaction with subwavelength-sized ultrasensitive
  nanomechanical force sensors},\ }\href
  {https://doi.org/10.1103/PhysRevX.11.021009} {\bibfield  {journal} {\bibinfo
  {journal} {Phys. Rev. X}\ }\textbf {\bibinfo {volume} {11}},\ \bibinfo
  {pages} {021009} (\bibinfo {year} {2021})}\BibitemShut {NoStop}%
\bibitem [{\citenamefont {Doolin}\ \emph {et~al.}(2014)\citenamefont {Doolin},
  \citenamefont {Kim}, \citenamefont {Hauer}, \citenamefont {MacDonald},\ and\
  \citenamefont {Davis}}]{Doolin_2014_NJP}%
  \BibitemOpen
  \bibfield  {author} {\bibinfo {author} {\bibfnamefont {C.}~\bibnamefont
  {Doolin}}, \bibinfo {author} {\bibfnamefont {P.~H.}\ \bibnamefont {Kim}},
  \bibinfo {author} {\bibfnamefont {B.~D.}\ \bibnamefont {Hauer}}, \bibinfo
  {author} {\bibfnamefont {A.~J.~R.}\ \bibnamefont {MacDonald}},\ and\ \bibinfo
  {author} {\bibfnamefont {J.~P.}\ \bibnamefont {Davis}},\ }\bibfield  {title}
  {\bibinfo {title} {Multidimensional optomechanical cantilevers for
  high-frequency force sensing},\ }\href
  {https://doi.org/10.1088/1367-2630/16/3/035001} {\bibfield  {journal}
  {\bibinfo  {journal} {New Journal of Physics}\ }\textbf {\bibinfo {volume}
  {16}},\ \bibinfo {pages} {035001} (\bibinfo {year} {2014})}\BibitemShut
  {NoStop}%
\bibitem [{\citenamefont {Srinivasan}\ \emph {et~al.}(2011)\citenamefont
  {Srinivasan}, \citenamefont {Miao}, \citenamefont {Rakher}, \citenamefont
  {Davan{\c{c}}o},\ and\ \citenamefont {Aksyuk}}]{Vladimir_2011_NL}%
  \BibitemOpen
  \bibfield  {author} {\bibinfo {author} {\bibfnamefont {K.}~\bibnamefont
  {Srinivasan}}, \bibinfo {author} {\bibfnamefont {H.}~\bibnamefont {Miao}},
  \bibinfo {author} {\bibfnamefont {M.~T.}\ \bibnamefont {Rakher}}, \bibinfo
  {author} {\bibfnamefont {M.}~\bibnamefont {Davan{\c{c}}o}},\ and\ \bibinfo
  {author} {\bibfnamefont {V.}~\bibnamefont {Aksyuk}},\ }\bibfield  {title}
  {\bibinfo {title} {Optomechanical transduction of an integrated silicon
  cantilever probe using a microdisk resonator},\ }\href
  {https://doi.org/10.1021/nl104018r} {\bibfield  {journal} {\bibinfo
  {journal} {Nano Letters}\ }\textbf {\bibinfo {volume} {11}},\ \bibinfo
  {pages} {791} (\bibinfo {year} {2011})}\BibitemShut {NoStop}%
\bibitem [{\citenamefont {Huang}\ \emph {et~al.}(2017)\citenamefont {Huang},
  \citenamefont {Flores}, \citenamefont {Cai}, \citenamefont {Yu},
  \citenamefont {Kwong}, \citenamefont {Wen}, \citenamefont {Churchill},\ and\
  \citenamefont {Wong}}]{YongjunHuang_2017_SC}%
  \BibitemOpen
  \bibfield  {author} {\bibinfo {author} {\bibfnamefont {Y.}~\bibnamefont
  {Huang}}, \bibinfo {author} {\bibfnamefont {J.~G.~F.}\ \bibnamefont
  {Flores}}, \bibinfo {author} {\bibfnamefont {Z.}~\bibnamefont {Cai}},
  \bibinfo {author} {\bibfnamefont {M.}~\bibnamefont {Yu}}, \bibinfo {author}
  {\bibfnamefont {D.-L.}\ \bibnamefont {Kwong}}, \bibinfo {author}
  {\bibfnamefont {G.}~\bibnamefont {Wen}}, \bibinfo {author} {\bibfnamefont
  {L.}~\bibnamefont {Churchill}},\ and\ \bibinfo {author} {\bibfnamefont
  {C.~W.}\ \bibnamefont {Wong}},\ }\bibfield  {title} {\bibinfo {title} {A
  low-frequency chip-scale optomechanical oscillator with 58 {kHz} mechanical
  stiffening and more than 100th-order stable harmonics},\ }\href
  {https://doi.org/10.1038/s41598-017-04882-4} {\bibfield  {journal} {\bibinfo
  {journal} {Scientific Reports}\ }\textbf {\bibinfo {volume} {7}},\ \bibinfo
  {pages} {4383} (\bibinfo {year} {2017})}\BibitemShut {NoStop}%
\bibitem [{\citenamefont {Primo}\ \emph {et~al.}(2023)\citenamefont {Primo},
  \citenamefont {Pinho}, \citenamefont {Benevides}, \citenamefont
  {Gr{\"o}blacher}, \citenamefont {Wiederhecker},\ and\ \citenamefont
  {Alegre}}]{Mayer_2023_NC}%
  \BibitemOpen
  \bibfield  {author} {\bibinfo {author} {\bibfnamefont {A.~G.}\ \bibnamefont
  {Primo}}, \bibinfo {author} {\bibfnamefont {P.~V.}\ \bibnamefont {Pinho}},
  \bibinfo {author} {\bibfnamefont {R.}~\bibnamefont {Benevides}}, \bibinfo
  {author} {\bibfnamefont {S.}~\bibnamefont {Gr{\"o}blacher}}, \bibinfo
  {author} {\bibfnamefont {G.~S.}\ \bibnamefont {Wiederhecker}},\ and\ \bibinfo
  {author} {\bibfnamefont {T.~P.~M.}\ \bibnamefont {Alegre}},\ }\bibfield
  {title} {\bibinfo {title} {Dissipative optomechanics in high-frequency
  nanomechanical resonators},\ }\href
  {https://doi.org/10.1038/s41467-023-41127-7} {\bibfield  {journal} {\bibinfo
  {journal} {Nature Communications}\ }\textbf {\bibinfo {volume} {14}},\
  \bibinfo {pages} {5793} (\bibinfo {year} {2023})}\BibitemShut {NoStop}%
\bibitem [{\citenamefont {MacCabe}\ \emph {et~al.}(2020)\citenamefont
  {MacCabe}, \citenamefont {Ren}, \citenamefont {Luo}, \citenamefont {Cohen},
  \citenamefont {Zhou}, \citenamefont {Sipahigil}, \citenamefont
  {Mirhosseini},\ and\ \citenamefont {Painter}}]{Painter_2020_Science}%
  \BibitemOpen
  \bibfield  {author} {\bibinfo {author} {\bibfnamefont {G.~S.}\ \bibnamefont
  {MacCabe}}, \bibinfo {author} {\bibfnamefont {H.}~\bibnamefont {Ren}},
  \bibinfo {author} {\bibfnamefont {J.}~\bibnamefont {Luo}}, \bibinfo {author}
  {\bibfnamefont {J.~D.}\ \bibnamefont {Cohen}}, \bibinfo {author}
  {\bibfnamefont {H.}~\bibnamefont {Zhou}}, \bibinfo {author} {\bibfnamefont
  {A.}~\bibnamefont {Sipahigil}}, \bibinfo {author} {\bibfnamefont
  {M.}~\bibnamefont {Mirhosseini}},\ and\ \bibinfo {author} {\bibfnamefont
  {O.}~\bibnamefont {Painter}},\ }\bibfield  {title} {\bibinfo {title}
  {Nano-acoustic resonator with ultralong phonon lifetime},\ }\href
  {https://doi.org/10.1126/science.abc7312} {\bibfield  {journal} {\bibinfo
  {journal} {Science}\ }\textbf {\bibinfo {volume} {370}},\ \bibinfo {pages}
  {840} (\bibinfo {year} {2020})}\BibitemShut {NoStop}%
\bibitem [{\citenamefont {Zivari}\ \emph {et~al.}(2022)\citenamefont {Zivari},
  \citenamefont {Stockill}, \citenamefont {Fiaschi},\ and\ \citenamefont
  {Gr{\"o}blacher}}]{Simon_2022_NP}%
  \BibitemOpen
  \bibfield  {author} {\bibinfo {author} {\bibfnamefont {A.}~\bibnamefont
  {Zivari}}, \bibinfo {author} {\bibfnamefont {R.}~\bibnamefont {Stockill}},
  \bibinfo {author} {\bibfnamefont {N.}~\bibnamefont {Fiaschi}},\ and\ \bibinfo
  {author} {\bibfnamefont {S.}~\bibnamefont {Gr{\"o}blacher}},\ }\bibfield
  {title} {\bibinfo {title} {Non-classical mechanical states guided in a
  phononic waveguide},\ }\href {https://doi.org/10.1038/s41567-022-01612-0}
  {\bibfield  {journal} {\bibinfo  {journal} {Nature Physics}\ }\textbf
  {\bibinfo {volume} {18}},\ \bibinfo {pages} {789} (\bibinfo {year}
  {2022})}\BibitemShut {NoStop}%
\bibitem [{\citenamefont {Bochmann}\ \emph {et~al.}(2013)\citenamefont
  {Bochmann}, \citenamefont {Vainsencher}, \citenamefont {Awschalom},\ and\
  \citenamefont {Cleland}}]{Celand_2013_NP}%
  \BibitemOpen
  \bibfield  {author} {\bibinfo {author} {\bibfnamefont {J.}~\bibnamefont
  {Bochmann}}, \bibinfo {author} {\bibfnamefont {A.}~\bibnamefont
  {Vainsencher}}, \bibinfo {author} {\bibfnamefont {D.~D.}\ \bibnamefont
  {Awschalom}},\ and\ \bibinfo {author} {\bibfnamefont {A.~N.}\ \bibnamefont
  {Cleland}},\ }\bibfield  {title} {\bibinfo {title} {Nanomechanical coupling
  between microwave and optical photons},\ }\href
  {https://doi.org/10.1038/nphys2748} {\bibfield  {journal} {\bibinfo
  {journal} {Nature Physics}\ }\textbf {\bibinfo {volume} {9}},\ \bibinfo
  {pages} {712} (\bibinfo {year} {2013})}\BibitemShut {NoStop}%
\bibitem [{\citenamefont {Fink}\ \emph {et~al.}(2016)\citenamefont {Fink},
  \citenamefont {Kalaee}, \citenamefont {Pitanti}, \citenamefont {Norte},
  \citenamefont {Heinzle}, \citenamefont {Davan{\c{c}}o}, \citenamefont
  {Srinivasan},\ and\ \citenamefont {Painter}}]{Pitanti_2016_NC}%
  \BibitemOpen
  \bibfield  {author} {\bibinfo {author} {\bibfnamefont {J.~M.}\ \bibnamefont
  {Fink}}, \bibinfo {author} {\bibfnamefont {M.}~\bibnamefont {Kalaee}},
  \bibinfo {author} {\bibfnamefont {A.}~\bibnamefont {Pitanti}}, \bibinfo
  {author} {\bibfnamefont {R.}~\bibnamefont {Norte}}, \bibinfo {author}
  {\bibfnamefont {L.}~\bibnamefont {Heinzle}}, \bibinfo {author} {\bibfnamefont
  {M.}~\bibnamefont {Davan{\c{c}}o}}, \bibinfo {author} {\bibfnamefont
  {K.}~\bibnamefont {Srinivasan}},\ and\ \bibinfo {author} {\bibfnamefont
  {O.}~\bibnamefont {Painter}},\ }\bibfield  {title} {\bibinfo {title} {Quantum
  electromechanics on silicon nitride nanomembranes},\ }\href
  {https://doi.org/10.1038/ncomms12396} {\bibfield  {journal} {\bibinfo
  {journal} {Nature Communications}\ }\textbf {\bibinfo {volume} {7}},\
  \bibinfo {pages} {12396} (\bibinfo {year} {2016})}\BibitemShut {NoStop}%
\bibitem [{\citenamefont {Pate}\ \emph {et~al.}(2020)\citenamefont {Pate},
  \citenamefont {Goryachev}, \citenamefont {Chiao}, \citenamefont {Sharping},\
  and\ \citenamefont {Tobar}}]{Torba_2020_NP}%
  \BibitemOpen
  \bibfield  {author} {\bibinfo {author} {\bibfnamefont {J.~M.}\ \bibnamefont
  {Pate}}, \bibinfo {author} {\bibfnamefont {M.}~\bibnamefont {Goryachev}},
  \bibinfo {author} {\bibfnamefont {R.~Y.}\ \bibnamefont {Chiao}}, \bibinfo
  {author} {\bibfnamefont {J.~E.}\ \bibnamefont {Sharping}},\ and\ \bibinfo
  {author} {\bibfnamefont {M.~E.}\ \bibnamefont {Tobar}},\ }\bibfield  {title}
  {\bibinfo {title} {Casimir spring and dilution in macroscopic cavity
  optomechanics},\ }\href {https://doi.org/10.1038/s41567-020-0975-9}
  {\bibfield  {journal} {\bibinfo  {journal} {Nature Physics}\ }\textbf
  {\bibinfo {volume} {16}},\ \bibinfo {pages} {1117} (\bibinfo {year}
  {2020})}\BibitemShut {NoStop}%
\bibitem [{\citenamefont {Yang}\ \emph {et~al.}(2015)\citenamefont {Yang},
  \citenamefont {Gerke}, \citenamefont {Ng}, \citenamefont {Rao}, \citenamefont
  {Chase},\ and\ \citenamefont {Chang-Hasnain}}]{Connie_2015_SR}%
  \BibitemOpen
  \bibfield  {author} {\bibinfo {author} {\bibfnamefont {W.}~\bibnamefont
  {Yang}}, \bibinfo {author} {\bibfnamefont {S.~A.}\ \bibnamefont {Gerke}},
  \bibinfo {author} {\bibfnamefont {K.~W.}\ \bibnamefont {Ng}}, \bibinfo
  {author} {\bibfnamefont {Y.}~\bibnamefont {Rao}}, \bibinfo {author}
  {\bibfnamefont {C.}~\bibnamefont {Chase}},\ and\ \bibinfo {author}
  {\bibfnamefont {C.~J.}\ \bibnamefont {Chang-Hasnain}},\ }\bibfield  {title}
  {\bibinfo {title} {Laser optomechanics},\ }\href
  {https://doi.org/10.1038/srep13700} {\bibfield  {journal} {\bibinfo
  {journal} {Scientific Reports}\ }\textbf {\bibinfo {volume} {5}},\ \bibinfo
  {pages} {13700} (\bibinfo {year} {2015})}\BibitemShut {NoStop}%
\bibitem [{\citenamefont {Eichenfield}\ \emph {et~al.}(2009)\citenamefont
  {Eichenfield}, \citenamefont {Chan}, \citenamefont {Camacho}, \citenamefont
  {Vahala},\ and\ \citenamefont {Painter}}]{Painter_2009_NC}%
  \BibitemOpen
  \bibfield  {author} {\bibinfo {author} {\bibfnamefont {M.}~\bibnamefont
  {Eichenfield}}, \bibinfo {author} {\bibfnamefont {J.}~\bibnamefont {Chan}},
  \bibinfo {author} {\bibfnamefont {R.~M.}\ \bibnamefont {Camacho}}, \bibinfo
  {author} {\bibfnamefont {K.~J.}\ \bibnamefont {Vahala}},\ and\ \bibinfo
  {author} {\bibfnamefont {O.}~\bibnamefont {Painter}},\ }\bibfield  {title}
  {\bibinfo {title} {Optomechanical crystals},\ }\href
  {https://doi.org/10.1038/nature08524} {\bibfield  {journal} {\bibinfo
  {journal} {Nature}\ }\textbf {\bibinfo {volume} {462}},\ \bibinfo {pages}
  {78} (\bibinfo {year} {2009})}\BibitemShut {NoStop}%
\bibitem [{\citenamefont {Kemiktarak}\ \emph
  {et~al.}(2014{\natexlab{b}})\citenamefont {Kemiktarak}, \citenamefont
  {Durand}, \citenamefont {Metcalfe},\ and\ \citenamefont
  {Lawall}}]{Lawall_2014_PRL}%
  \BibitemOpen
  \bibfield  {author} {\bibinfo {author} {\bibfnamefont {U.}~\bibnamefont
  {Kemiktarak}}, \bibinfo {author} {\bibfnamefont {M.}~\bibnamefont {Durand}},
  \bibinfo {author} {\bibfnamefont {M.}~\bibnamefont {Metcalfe}},\ and\
  \bibinfo {author} {\bibfnamefont {J.}~\bibnamefont {Lawall}},\ }\bibfield
  {title} {\bibinfo {title} {Mode competition and anomalous cooling in a
  multimode phonon laser},\ }\href
  {https://doi.org/10.1103/PhysRevLett.113.030802} {\bibfield  {journal}
  {\bibinfo  {journal} {Phys. Rev. Lett.}\ }\textbf {\bibinfo {volume} {113}},\
  \bibinfo {pages} {030802} (\bibinfo {year} {2014}{\natexlab{b}})}\BibitemShut
  {NoStop}%
\bibitem [{\citenamefont {Seis}\ \emph {et~al.}(2022)\citenamefont {Seis},
  \citenamefont {Capelle}, \citenamefont {Langman}, \citenamefont {Saarinen},
  \citenamefont {Planz},\ and\ \citenamefont
  {Schliesser}}]{Schliesser_2022_Nc}%
  \BibitemOpen
  \bibfield  {author} {\bibinfo {author} {\bibfnamefont {Y.}~\bibnamefont
  {Seis}}, \bibinfo {author} {\bibfnamefont {T.}~\bibnamefont {Capelle}},
  \bibinfo {author} {\bibfnamefont {E.}~\bibnamefont {Langman}}, \bibinfo
  {author} {\bibfnamefont {S.}~\bibnamefont {Saarinen}}, \bibinfo {author}
  {\bibfnamefont {E.}~\bibnamefont {Planz}},\ and\ \bibinfo {author}
  {\bibfnamefont {A.}~\bibnamefont {Schliesser}},\ }\bibfield  {title}
  {\bibinfo {title} {Ground state cooling of an ultracoherent electromechanical
  system},\ }\href {https://doi.org/10.1038/s41467-022-29115-9} {\bibfield
  {journal} {\bibinfo  {journal} {Nature Communications}\ }\textbf {\bibinfo
  {volume} {13}},\ \bibinfo {pages} {1507} (\bibinfo {year}
  {2022})}\BibitemShut {NoStop}%
\bibitem [{\citenamefont {Guo}\ \emph {et~al.}(2023)\citenamefont {Guo},
  \citenamefont {Chang}, \citenamefont {Yao},\ and\ \citenamefont
  {Gr{\"o}blacher}}]{Simon_2023_NC}%
  \BibitemOpen
  \bibfield  {author} {\bibinfo {author} {\bibfnamefont {J.}~\bibnamefont
  {Guo}}, \bibinfo {author} {\bibfnamefont {J.}~\bibnamefont {Chang}}, \bibinfo
  {author} {\bibfnamefont {X.}~\bibnamefont {Yao}},\ and\ \bibinfo {author}
  {\bibfnamefont {S.}~\bibnamefont {Gr{\"o}blacher}},\ }\bibfield  {title}
  {\bibinfo {title} {Active-feedback quantum control of an integrated
  low-frequency mechanical resonator},\ }\href
  {https://doi.org/10.1038/s41467-023-40442-3} {\bibfield  {journal} {\bibinfo
  {journal} {Nature Communications}\ }\textbf {\bibinfo {volume} {14}},\
  \bibinfo {pages} {4721} (\bibinfo {year} {2023})}\BibitemShut {NoStop}%
\bibitem [{\citenamefont {Meng}\ \emph {et~al.}(2022)\citenamefont {Meng},
  \citenamefont {Brawley}, \citenamefont {Khademi}, \citenamefont {Bridge},
  \citenamefont {Bennett},\ and\ \citenamefont {Bowen}}]{Bowen_2022_SA}%
  \BibitemOpen
  \bibfield  {author} {\bibinfo {author} {\bibfnamefont {C.}~\bibnamefont
  {Meng}}, \bibinfo {author} {\bibfnamefont {G.~A.}\ \bibnamefont {Brawley}},
  \bibinfo {author} {\bibfnamefont {S.}~\bibnamefont {Khademi}}, \bibinfo
  {author} {\bibfnamefont {E.~M.}\ \bibnamefont {Bridge}}, \bibinfo {author}
  {\bibfnamefont {J.~S.}\ \bibnamefont {Bennett}},\ and\ \bibinfo {author}
  {\bibfnamefont {W.~P.}\ \bibnamefont {Bowen}},\ }\bibfield  {title} {\bibinfo
  {title} {Measurement-based preparation of multimode mechanical states},\
  }\href {https://doi.org/10.1126/sciadv.abm7585} {\bibfield  {journal}
  {\bibinfo  {journal} {Science Advances}\ }\textbf {\bibinfo {volume} {8}},\
  \bibinfo {pages} {eabm7585} (\bibinfo {year} {2022})}\BibitemShut {NoStop}%
\bibitem [{\citenamefont {Sbarra}\ \emph {et~al.}(2022)\citenamefont {Sbarra},
  \citenamefont {Waquier}, \citenamefont {Suffit}, \citenamefont {Lemaître},\
  and\ \citenamefont {Favero}}]{Ivan_2022_NL}%
  \BibitemOpen
  \bibfield  {author} {\bibinfo {author} {\bibfnamefont {S.}~\bibnamefont
  {Sbarra}}, \bibinfo {author} {\bibfnamefont {L.}~\bibnamefont {Waquier}},
  \bibinfo {author} {\bibfnamefont {S.}~\bibnamefont {Suffit}}, \bibinfo
  {author} {\bibfnamefont {A.}~\bibnamefont {Lemaître}},\ and\ \bibinfo
  {author} {\bibfnamefont {I.}~\bibnamefont {Favero}},\ }\bibfield  {title}
  {\bibinfo {title} {Multimode optomechanical weighting of a single
  nanoparticle},\ }\href {https://doi.org/10.1021/acs.nanolett.1c03890}
  {\bibfield  {journal} {\bibinfo  {journal} {Nano Letters}\ }\textbf {\bibinfo
  {volume} {22}},\ \bibinfo {pages} {710} (\bibinfo {year} {2022})}\BibitemShut
  {NoStop}%
\bibitem [{\citenamefont {Chen}\ \emph {et~al.}(2023)\citenamefont {Chen},
  \citenamefont {Li}, \citenamefont {Lee}, \citenamefont {Chakravarthi},
  \citenamefont {Fu},\ and\ \citenamefont {Li}}]{MoLi_2023_NC}%
  \BibitemOpen
  \bibfield  {author} {\bibinfo {author} {\bibfnamefont {I.-T.}\ \bibnamefont
  {Chen}}, \bibinfo {author} {\bibfnamefont {B.}~\bibnamefont {Li}}, \bibinfo
  {author} {\bibfnamefont {S.}~\bibnamefont {Lee}}, \bibinfo {author}
  {\bibfnamefont {S.}~\bibnamefont {Chakravarthi}}, \bibinfo {author}
  {\bibfnamefont {K.-M.}\ \bibnamefont {Fu}},\ and\ \bibinfo {author}
  {\bibfnamefont {M.}~\bibnamefont {Li}},\ }\bibfield  {title} {\bibinfo
  {title} {Optomechanical ring resonator for efficient microwave-optical
  frequency conversion},\ }\href {https://doi.org/10.1038/s41467-023-43393-x}
  {\bibfield  {journal} {\bibinfo  {journal} {Nature Communications}\ }\textbf
  {\bibinfo {volume} {14}},\ \bibinfo {pages} {7594} (\bibinfo {year}
  {2023})}\BibitemShut {NoStop}%
\bibitem [{\citenamefont {Yu}\ \emph {et~al.}(2022)\citenamefont {Yu},
  \citenamefont {Xi},\ and\ \citenamefont {Sun}}]{YueYu_2022_LSA}%
  \BibitemOpen
  \bibfield  {author} {\bibinfo {author} {\bibfnamefont {Y.}~\bibnamefont
  {Yu}}, \bibinfo {author} {\bibfnamefont {X.}~\bibnamefont {Xi}},\ and\
  \bibinfo {author} {\bibfnamefont {X.}~\bibnamefont {Sun}},\ }\bibfield
  {title} {\bibinfo {title} {Observation of mechanical bound states in the
  continuum in an optomechanical microresonator},\ }\href
  {https://doi.org/10.1038/s41377-022-00971-w} {\bibfield  {journal} {\bibinfo
  {journal} {Light: Science {\&} Applications}\ }\textbf {\bibinfo {volume}
  {11}},\ \bibinfo {pages} {328} (\bibinfo {year} {2022})}\BibitemShut
  {NoStop}%
\bibitem [{\citenamefont {Tebbenjohanns}\ \emph {et~al.}(2021)\citenamefont
  {Tebbenjohanns}, \citenamefont {Mattana}, \citenamefont {Rossi},
  \citenamefont {Frimmer},\ and\ \citenamefont {Novotny}}]{Lukas_2021_Nature}%
  \BibitemOpen
  \bibfield  {author} {\bibinfo {author} {\bibfnamefont {F.}~\bibnamefont
  {Tebbenjohanns}}, \bibinfo {author} {\bibfnamefont {M.~L.}\ \bibnamefont
  {Mattana}}, \bibinfo {author} {\bibfnamefont {M.}~\bibnamefont {Rossi}},
  \bibinfo {author} {\bibfnamefont {M.}~\bibnamefont {Frimmer}},\ and\ \bibinfo
  {author} {\bibfnamefont {L.}~\bibnamefont {Novotny}},\ }\bibfield  {title}
  {\bibinfo {title} {Quantum control of a nanoparticle optically levitated in
  cryogenic free space},\ }\href {https://doi.org/10.1038/s41586-021-03617-w}
  {\bibfield  {journal} {\bibinfo  {journal} {Nature}\ }\textbf {\bibinfo
  {volume} {595}},\ \bibinfo {pages} {378} (\bibinfo {year}
  {2021})}\BibitemShut {NoStop}%
\bibitem [{\citenamefont {Piotrowski}\ \emph {et~al.}(2023)\citenamefont
  {Piotrowski}, \citenamefont {Windey}, \citenamefont {Vijayan}, \citenamefont
  {Gonzalez-Ballestero}, \citenamefont {de~los R{\'i}os~Sommer}, \citenamefont
  {Meyer}, \citenamefont {Quidant}, \citenamefont {Romero-Isart}, \citenamefont
  {Reimann},\ and\ \citenamefont {Novotny}}]{Lukas_2023_NP}%
  \BibitemOpen
  \bibfield  {author} {\bibinfo {author} {\bibfnamefont {J.}~\bibnamefont
  {Piotrowski}}, \bibinfo {author} {\bibfnamefont {D.}~\bibnamefont {Windey}},
  \bibinfo {author} {\bibfnamefont {J.}~\bibnamefont {Vijayan}}, \bibinfo
  {author} {\bibfnamefont {C.}~\bibnamefont {Gonzalez-Ballestero}}, \bibinfo
  {author} {\bibfnamefont {A.}~\bibnamefont {de~los R{\'i}os~Sommer}}, \bibinfo
  {author} {\bibfnamefont {N.}~\bibnamefont {Meyer}}, \bibinfo {author}
  {\bibfnamefont {R.}~\bibnamefont {Quidant}}, \bibinfo {author} {\bibfnamefont
  {O.}~\bibnamefont {Romero-Isart}}, \bibinfo {author} {\bibfnamefont
  {R.}~\bibnamefont {Reimann}},\ and\ \bibinfo {author} {\bibfnamefont
  {L.}~\bibnamefont {Novotny}},\ }\bibfield  {title} {\bibinfo {title}
  {Simultaneous ground-state cooling of two mechanical modes of a levitated
  nanoparticle},\ }\href {https://doi.org/10.1038/s41567-023-01956-1}
  {\bibfield  {journal} {\bibinfo  {journal} {Nature Physics}\ }\textbf
  {\bibinfo {volume} {19}},\ \bibinfo {pages} {1009} (\bibinfo {year}
  {2023})}\BibitemShut {NoStop}%
\bibitem [{\citenamefont {Zhao}\ and\ \citenamefont
  {Fang}(2019)}]{KejieFang_2029_OE}%
  \BibitemOpen
  \bibfield  {author} {\bibinfo {author} {\bibfnamefont {M.}~\bibnamefont
  {Zhao}}\ and\ \bibinfo {author} {\bibfnamefont {K.}~\bibnamefont {Fang}},\
  }\bibfield  {title} {\bibinfo {title} {Mechanical bound states in the
  continuum for macroscopic optomechanics},\ }\href
  {https://doi.org/10.1364/OE.27.010138} {\bibfield  {journal} {\bibinfo
  {journal} {Opt. Express}\ }\textbf {\bibinfo {volume} {27}},\ \bibinfo
  {pages} {10138} (\bibinfo {year} {2019})}\BibitemShut {NoStop}%
\bibitem [{\citenamefont {Tong}\ \emph {et~al.}(2020)\citenamefont {Tong},
  \citenamefont {Liu}, \citenamefont {Zhao},\ and\ \citenamefont
  {Fang}}]{KejieFang_2020_NC}%
  \BibitemOpen
  \bibfield  {author} {\bibinfo {author} {\bibfnamefont {H.}~\bibnamefont
  {Tong}}, \bibinfo {author} {\bibfnamefont {S.}~\bibnamefont {Liu}}, \bibinfo
  {author} {\bibfnamefont {M.}~\bibnamefont {Zhao}},\ and\ \bibinfo {author}
  {\bibfnamefont {K.}~\bibnamefont {Fang}},\ }\bibfield  {title} {\bibinfo
  {title} {Observation of phonon trapping in the continuum with topological
  charges},\ }\href {https://doi.org/10.1038/s41467-020-19091-3} {\bibfield
  {journal} {\bibinfo  {journal} {Nature Communications}\ }\textbf {\bibinfo
  {volume} {11}},\ \bibinfo {pages} {5216} (\bibinfo {year}
  {2020})}\BibitemShut {NoStop}%
\bibitem [{\citenamefont {Kharel}\ \emph {et~al.}(2019)\citenamefont {Kharel},
  \citenamefont {Harris}, \citenamefont {Kittlaus}, \citenamefont {Renninger},
  \citenamefont {Otterstrom}, \citenamefont {Harris},\ and\ \citenamefont
  {Rakich}}]{Peter_2019_SA}%
  \BibitemOpen
  \bibfield  {author} {\bibinfo {author} {\bibfnamefont {P.}~\bibnamefont
  {Kharel}}, \bibinfo {author} {\bibfnamefont {G.~I.}\ \bibnamefont {Harris}},
  \bibinfo {author} {\bibfnamefont {E.~A.}\ \bibnamefont {Kittlaus}}, \bibinfo
  {author} {\bibfnamefont {W.~H.}\ \bibnamefont {Renninger}}, \bibinfo {author}
  {\bibfnamefont {N.~T.}\ \bibnamefont {Otterstrom}}, \bibinfo {author}
  {\bibfnamefont {J.~G.~E.}\ \bibnamefont {Harris}},\ and\ \bibinfo {author}
  {\bibfnamefont {P.~T.}\ \bibnamefont {Rakich}},\ }\bibfield  {title}
  {\bibinfo {title} {High-frequency cavity optomechanics using bulk acoustic
  phonons},\ }\href {https://doi.org/10.1126/sciadv.aav0582} {\bibfield
  {journal} {\bibinfo  {journal} {Science Advances}\ }\textbf {\bibinfo
  {volume} {5}},\ \bibinfo {pages} {eaav0582} (\bibinfo {year}
  {2019})}\BibitemShut {NoStop}%
\bibitem [{\citenamefont {Kharel}\ \emph {et~al.}(2022)\citenamefont {Kharel},
  \citenamefont {Chu}, \citenamefont {Mason}, \citenamefont {Kittlaus},
  \citenamefont {Otterstrom}, \citenamefont {Gertler},\ and\ \citenamefont
  {Rakich}}]{Rakich_2022_PRApplied}%
  \BibitemOpen
  \bibfield  {author} {\bibinfo {author} {\bibfnamefont {P.}~\bibnamefont
  {Kharel}}, \bibinfo {author} {\bibfnamefont {Y.}~\bibnamefont {Chu}},
  \bibinfo {author} {\bibfnamefont {D.}~\bibnamefont {Mason}}, \bibinfo
  {author} {\bibfnamefont {E.~A.}\ \bibnamefont {Kittlaus}}, \bibinfo {author}
  {\bibfnamefont {N.~T.}\ \bibnamefont {Otterstrom}}, \bibinfo {author}
  {\bibfnamefont {S.}~\bibnamefont {Gertler}},\ and\ \bibinfo {author}
  {\bibfnamefont {P.~T.}\ \bibnamefont {Rakich}},\ }\bibfield  {title}
  {\bibinfo {title} {Multimode strong coupling in cavity optomechanics},\
  }\href {https://doi.org/10.1103/PhysRevApplied.18.024054} {\bibfield
  {journal} {\bibinfo  {journal} {Phys. Rev. Appl.}\ }\textbf {\bibinfo
  {volume} {18}},\ \bibinfo {pages} {024054} (\bibinfo {year}
  {2022})}\BibitemShut {NoStop}%
\bibitem [{\citenamefont {Kuznetsov}\ \emph {et~al.}(2023)\citenamefont
  {Kuznetsov}, \citenamefont {Biermann}, \citenamefont {Reynoso}, \citenamefont
  {Fainstein},\ and\ \citenamefont {Santos}}]{Paulo_2023_NC}%
  \BibitemOpen
  \bibfield  {author} {\bibinfo {author} {\bibfnamefont {A.~S.}\ \bibnamefont
  {Kuznetsov}}, \bibinfo {author} {\bibfnamefont {K.}~\bibnamefont {Biermann}},
  \bibinfo {author} {\bibfnamefont {A.~A.}\ \bibnamefont {Reynoso}}, \bibinfo
  {author} {\bibfnamefont {A.}~\bibnamefont {Fainstein}},\ and\ \bibinfo
  {author} {\bibfnamefont {P.~V.}\ \bibnamefont {Santos}},\ }\bibfield  {title}
  {\bibinfo {title} {Microcavity phonoritons -- a coherent optical-to-microwave
  interface},\ }\href {https://doi.org/10.1038/s41467-023-40894-7} {\bibfield
  {journal} {\bibinfo  {journal} {Nature Communications}\ }\textbf {\bibinfo
  {volume} {14}},\ \bibinfo {pages} {5470} (\bibinfo {year}
  {2023})}\BibitemShut {NoStop}%
\bibitem [{\citenamefont {von L{\"u}pke}\ \emph {et~al.}(2024)\citenamefont
  {von L{\"u}pke}, \citenamefont {Rodrigues}, \citenamefont {Yang},
  \citenamefont {Fadel},\ and\ \citenamefont {Chu}}]{YiwenChu_2024_NP}%
  \BibitemOpen
  \bibfield  {author} {\bibinfo {author} {\bibfnamefont {U.}~\bibnamefont {von
  L{\"u}pke}}, \bibinfo {author} {\bibfnamefont {I.~C.}\ \bibnamefont
  {Rodrigues}}, \bibinfo {author} {\bibfnamefont {Y.}~\bibnamefont {Yang}},
  \bibinfo {author} {\bibfnamefont {M.}~\bibnamefont {Fadel}},\ and\ \bibinfo
  {author} {\bibfnamefont {Y.}~\bibnamefont {Chu}},\ }\bibfield  {title}
  {\bibinfo {title} {Engineering multimode interactions in circuit quantum
  acoustodynamics},\ }\href {https://doi.org/10.1038/s41567-023-02377-w}
  {\bibfield  {journal} {\bibinfo  {journal} {Nature Physics}\ }\textbf
  {\bibinfo {volume} {20}},\ \bibinfo {pages} {564} (\bibinfo {year}
  {2024})}\BibitemShut {NoStop}%
\bibitem [{\citenamefont {Li}\ \emph {et~al.}(2015)\citenamefont {Li},
  \citenamefont {Tadesse}, \citenamefont {Liu},\ and\ \citenamefont
  {Li}}]{MoLi_2015_Optica}%
  \BibitemOpen
  \bibfield  {author} {\bibinfo {author} {\bibfnamefont {H.}~\bibnamefont
  {Li}}, \bibinfo {author} {\bibfnamefont {S.~A.}\ \bibnamefont {Tadesse}},
  \bibinfo {author} {\bibfnamefont {Q.}~\bibnamefont {Liu}},\ and\ \bibinfo
  {author} {\bibfnamefont {M.}~\bibnamefont {Li}},\ }\bibfield  {title}
  {\bibinfo {title} {Nanophotonic cavity optomechanics with propagating
  acoustic waves at frequencies up to 12 {GHz}},\ }\href
  {https://doi.org/10.1364/OPTICA.2.000826} {\bibfield  {journal} {\bibinfo
  {journal} {Optica}\ }\textbf {\bibinfo {volume} {2}},\ \bibinfo {pages} {826}
  (\bibinfo {year} {2015})}\BibitemShut {NoStop}%
\bibitem [{\citenamefont {Andrews}\ \emph {et~al.}(2014)\citenamefont
  {Andrews}, \citenamefont {Peterson}, \citenamefont {Purdy}, \citenamefont
  {Cicak}, \citenamefont {Simmonds}, \citenamefont {Regal},\ and\ \citenamefont
  {Lehnert}}]{andrews_bidirectional_2014}%
  \BibitemOpen
  \bibfield  {author} {\bibinfo {author} {\bibfnamefont {R.~W.}\ \bibnamefont
  {Andrews}}, \bibinfo {author} {\bibfnamefont {R.~W.}\ \bibnamefont
  {Peterson}}, \bibinfo {author} {\bibfnamefont {T.~P.}\ \bibnamefont {Purdy}},
  \bibinfo {author} {\bibfnamefont {K.}~\bibnamefont {Cicak}}, \bibinfo
  {author} {\bibfnamefont {R.~W.}\ \bibnamefont {Simmonds}}, \bibinfo {author}
  {\bibfnamefont {C.~A.}\ \bibnamefont {Regal}},\ and\ \bibinfo {author}
  {\bibfnamefont {K.~W.}\ \bibnamefont {Lehnert}},\ }\bibfield  {title}
  {\bibinfo {title} {Bidirectional and efficient conversion between microwave
  and optical light},\ }\href {https://doi.org/10.1038/nphys2911} {\bibfield
  {journal} {\bibinfo  {journal} {Nature Physics}\ }\textbf {\bibinfo {volume}
  {10}},\ \bibinfo {pages} {321} (\bibinfo {year} {2014})}\BibitemShut
  {NoStop}%
\bibitem [{\citenamefont {Wilson}\ \emph {et~al.}(2009)\citenamefont {Wilson},
  \citenamefont {Regal}, \citenamefont {Papp},\ and\ \citenamefont
  {Kimble}}]{PhysRevLett.103.207204}%
  \BibitemOpen
  \bibfield  {author} {\bibinfo {author} {\bibfnamefont {D.~J.}\ \bibnamefont
  {Wilson}}, \bibinfo {author} {\bibfnamefont {C.~A.}\ \bibnamefont {Regal}},
  \bibinfo {author} {\bibfnamefont {S.~B.}\ \bibnamefont {Papp}},\ and\
  \bibinfo {author} {\bibfnamefont {H.~J.}\ \bibnamefont {Kimble}},\ }\bibfield
   {title} {\bibinfo {title} {Cavity optomechanics with stoichiometric {SiN}
  films},\ }\href {https://doi.org/10.1103/PhysRevLett.103.207204} {\bibfield
  {journal} {\bibinfo  {journal} {Phys. Rev. Lett.}\ }\textbf {\bibinfo
  {volume} {103}},\ \bibinfo {pages} {207204} (\bibinfo {year}
  {2009})}\BibitemShut {NoStop}%
\bibitem [{\citenamefont {Bagci}\ \emph {et~al.}(2014)\citenamefont {Bagci},
  \citenamefont {Simonsen}, \citenamefont {Schmid}, \citenamefont {Villanueva},
  \citenamefont {Zeuthen}, \citenamefont {Appel}, \citenamefont {Taylor},
  \citenamefont {S{\o}rensen}, \citenamefont {Usami}, \citenamefont
  {Schliesser},\ and\ \citenamefont {Polzik}}]{bagci_optical_2014}%
  \BibitemOpen
  \bibfield  {author} {\bibinfo {author} {\bibfnamefont {T.}~\bibnamefont
  {Bagci}}, \bibinfo {author} {\bibfnamefont {A.}~\bibnamefont {Simonsen}},
  \bibinfo {author} {\bibfnamefont {S.}~\bibnamefont {Schmid}}, \bibinfo
  {author} {\bibfnamefont {L.~G.}\ \bibnamefont {Villanueva}}, \bibinfo
  {author} {\bibfnamefont {E.}~\bibnamefont {Zeuthen}}, \bibinfo {author}
  {\bibfnamefont {J.}~\bibnamefont {Appel}}, \bibinfo {author} {\bibfnamefont
  {J.~M.}\ \bibnamefont {Taylor}}, \bibinfo {author} {\bibfnamefont
  {A.}~\bibnamefont {S{\o}rensen}}, \bibinfo {author} {\bibfnamefont
  {K.}~\bibnamefont {Usami}}, \bibinfo {author} {\bibfnamefont
  {A.}~\bibnamefont {Schliesser}},\ and\ \bibinfo {author} {\bibfnamefont
  {E.~S.}\ \bibnamefont {Polzik}},\ }\bibfield  {title} {\bibinfo {title}
  {Optical detection of radio waves through a nanomechanical transducer},\
  }\href {https://doi.org/10.1038/nature13029} {\bibfield  {journal} {\bibinfo
  {journal} {Nature}\ }\textbf {\bibinfo {volume} {507}},\ \bibinfo {pages}
  {81} (\bibinfo {year} {2014})}\BibitemShut {NoStop}%
\bibitem [{\citenamefont {Higginbotham}\ \emph {et~al.}(2018)\citenamefont
  {Higginbotham}, \citenamefont {Burns}, \citenamefont {Urmey}, \citenamefont
  {Peterson}, \citenamefont {Kampel}, \citenamefont {Brubaker}, \citenamefont
  {Smith}, \citenamefont {Lehnert},\ and\ \citenamefont
  {Regal}}]{higginbotham_harnessing_2018}%
  \BibitemOpen
  \bibfield  {author} {\bibinfo {author} {\bibfnamefont {A.~P.}\ \bibnamefont
  {Higginbotham}}, \bibinfo {author} {\bibfnamefont {P.~S.}\ \bibnamefont
  {Burns}}, \bibinfo {author} {\bibfnamefont {M.~D.}\ \bibnamefont {Urmey}},
  \bibinfo {author} {\bibfnamefont {R.~W.}\ \bibnamefont {Peterson}}, \bibinfo
  {author} {\bibfnamefont {N.~S.}\ \bibnamefont {Kampel}}, \bibinfo {author}
  {\bibfnamefont {B.~M.}\ \bibnamefont {Brubaker}}, \bibinfo {author}
  {\bibfnamefont {G.}~\bibnamefont {Smith}}, \bibinfo {author} {\bibfnamefont
  {K.~W.}\ \bibnamefont {Lehnert}},\ and\ \bibinfo {author} {\bibfnamefont
  {C.~A.}\ \bibnamefont {Regal}},\ }\bibfield  {title} {\bibinfo {title}
  {Harnessing electro-optic correlations in an efficient mechanical
  converter},\ }\href {https://doi.org/10.1038/s41567-018-0210-0} {\bibfield
  {journal} {\bibinfo  {journal} {Nature Physics}\ }\textbf {\bibinfo {volume}
  {14}},\ \bibinfo {pages} {1038} (\bibinfo {year} {2018})}\BibitemShut
  {NoStop}%
\bibitem [{\citenamefont {Delaney}\ \emph {et~al.}(2022)\citenamefont
  {Delaney}, \citenamefont {Urmey}, \citenamefont {Mittal}, \citenamefont
  {Brubaker}, \citenamefont {Kindem}, \citenamefont {Burns}, \citenamefont
  {Regal},\ and\ \citenamefont {Lehnert}}]{delaney_superconducting-qubit_2022}%
  \BibitemOpen
  \bibfield  {author} {\bibinfo {author} {\bibfnamefont {R.~D.}\ \bibnamefont
  {Delaney}}, \bibinfo {author} {\bibfnamefont {M.~D.}\ \bibnamefont {Urmey}},
  \bibinfo {author} {\bibfnamefont {S.}~\bibnamefont {Mittal}}, \bibinfo
  {author} {\bibfnamefont {B.~M.}\ \bibnamefont {Brubaker}}, \bibinfo {author}
  {\bibfnamefont {J.~M.}\ \bibnamefont {Kindem}}, \bibinfo {author}
  {\bibfnamefont {P.~S.}\ \bibnamefont {Burns}}, \bibinfo {author}
  {\bibfnamefont {C.~A.}\ \bibnamefont {Regal}},\ and\ \bibinfo {author}
  {\bibfnamefont {K.~W.}\ \bibnamefont {Lehnert}},\ }\bibfield  {title}
  {\bibinfo {title} {Superconducting-qubit readout via low-backaction
  electro-optic transduction},\ }\href
  {https://doi.org/10.1038/s41586-022-04720-2} {\bibfield  {journal} {\bibinfo
  {journal} {Nature}\ }\textbf {\bibinfo {volume} {606}},\ \bibinfo {pages}
  {489} (\bibinfo {year} {2022})}\BibitemShut {NoStop}%
\bibitem [{\citenamefont {Rao}(2006)}]{Tao_2006_book}%
  \BibitemOpen
  \bibfield  {author} {\bibinfo {author} {\bibfnamefont {S.~S.}\ \bibnamefont
  {Rao}},\ }\bibinfo {title} {Vibration of membranes},\ in\ \href
  {https://doi.org/https://doi.org/10.1002/9780470117866.ch13} {\emph {\bibinfo
  {booktitle} {Vibration of Continuous Systems}}}\ (\bibinfo  {publisher} {John
  Wiley \& Sons, Ltd},\ \bibinfo {year} {2006})\ Chap.~\bibinfo {chapter} {13},
  pp.\ \bibinfo {pages} {420--456}\BibitemShut {NoStop}%
\bibitem [{\citenamefont {Noguchi}\ \emph {et~al.}(2016)\citenamefont
  {Noguchi}, \citenamefont {Yamazaki}, \citenamefont {Ataka}, \citenamefont
  {Fujita}, \citenamefont {Tabuchi}, \citenamefont {Ishikawa}, \citenamefont
  {Usami},\ and\ \citenamefont {Nakamura}}]{Noguchi_2016_NJP}%
  \BibitemOpen
  \bibfield  {author} {\bibinfo {author} {\bibfnamefont {A.}~\bibnamefont
  {Noguchi}}, \bibinfo {author} {\bibfnamefont {R.}~\bibnamefont {Yamazaki}},
  \bibinfo {author} {\bibfnamefont {M.}~\bibnamefont {Ataka}}, \bibinfo
  {author} {\bibfnamefont {H.}~\bibnamefont {Fujita}}, \bibinfo {author}
  {\bibfnamefont {Y.}~\bibnamefont {Tabuchi}}, \bibinfo {author} {\bibfnamefont
  {T.}~\bibnamefont {Ishikawa}}, \bibinfo {author} {\bibfnamefont
  {K.}~\bibnamefont {Usami}},\ and\ \bibinfo {author} {\bibfnamefont
  {Y.}~\bibnamefont {Nakamura}},\ }\bibfield  {title} {\bibinfo {title} {Ground
  state cooling of a quantum electromechanical system with a silicon nitride
  membrane in a {3D} loop-gap cavity},\ }\href
  {https://doi.org/10.1088/1367-2630/18/10/103036} {\bibfield  {journal}
  {\bibinfo  {journal} {New Journal of Physics}\ }\textbf {\bibinfo {volume}
  {18}},\ \bibinfo {pages} {103036} (\bibinfo {year} {2016})}\BibitemShut
  {NoStop}%
\bibitem [{\citenamefont {Liu}\ \emph {et~al.}(2023{\natexlab{a}})\citenamefont
  {Liu}, \citenamefont {Liu}, \citenamefont {Sun}, \citenamefont {Chen},
  \citenamefont {Wang},\ and\ \citenamefont {Li}}]{YulongLiu_2023_NPJ}%
  \BibitemOpen
  \bibfield  {author} {\bibinfo {author} {\bibfnamefont {Y.}~\bibnamefont
  {Liu}}, \bibinfo {author} {\bibfnamefont {Q.}~\bibnamefont {Liu}}, \bibinfo
  {author} {\bibfnamefont {H.}~\bibnamefont {Sun}}, \bibinfo {author}
  {\bibfnamefont {M.}~\bibnamefont {Chen}}, \bibinfo {author} {\bibfnamefont
  {S.}~\bibnamefont {Wang}},\ and\ \bibinfo {author} {\bibfnamefont
  {T.}~\bibnamefont {Li}},\ }\bibfield  {title} {\bibinfo {title} {Coherent
  memory for microwave photons based on long-lived mechanical excitations},\
  }\href {https://doi.org/10.1038/s41534-023-00749-x} {\bibfield  {journal}
  {\bibinfo  {journal} {npj Quantum Information}\ }\textbf {\bibinfo {volume}
  {9}},\ \bibinfo {pages} {80} (\bibinfo {year}
  {2023}{\natexlab{a}})}\BibitemShut {NoStop}%
\bibitem [{\citenamefont {Liu}\ \emph {et~al.}(2021)\citenamefont {Liu},
  \citenamefont {Liu}, \citenamefont {Wang}, \citenamefont {Chen},
  \citenamefont {Sillanp\"a\"a},\ and\ \citenamefont
  {Li}}]{PhysRevLett.127.273603}%
  \BibitemOpen
  \bibfield  {author} {\bibinfo {author} {\bibfnamefont {Y.}~\bibnamefont
  {Liu}}, \bibinfo {author} {\bibfnamefont {Q.}~\bibnamefont {Liu}}, \bibinfo
  {author} {\bibfnamefont {S.}~\bibnamefont {Wang}}, \bibinfo {author}
  {\bibfnamefont {Z.}~\bibnamefont {Chen}}, \bibinfo {author} {\bibfnamefont
  {M.~A.}\ \bibnamefont {Sillanp\"a\"a}},\ and\ \bibinfo {author}
  {\bibfnamefont {T.}~\bibnamefont {Li}},\ }\bibfield  {title} {\bibinfo
  {title} {Optomechanical anti-lasing with infinite group delay at a phase
  singularity},\ }\href {https://doi.org/10.1103/PhysRevLett.127.273603}
  {\bibfield  {journal} {\bibinfo  {journal} {Phys. Rev. Lett.}\ }\textbf
  {\bibinfo {volume} {127}},\ \bibinfo {pages} {273603} (\bibinfo {year}
  {2021})}\BibitemShut {NoStop}%
\bibitem [{\citenamefont {Lai}\ \emph {et~al.}(2020)\citenamefont {Lai},
  \citenamefont {Huang}, \citenamefont {Yin}, \citenamefont {Hou},
  \citenamefont {Li}, \citenamefont {Vitali}, \citenamefont {Nori},\ and\
  \citenamefont {Liao}}]{PhysRevA.102.011502}%
  \BibitemOpen
  \bibfield  {author} {\bibinfo {author} {\bibfnamefont {D.-G.}\ \bibnamefont
  {Lai}}, \bibinfo {author} {\bibfnamefont {J.-F.}\ \bibnamefont {Huang}},
  \bibinfo {author} {\bibfnamefont {X.-L.}\ \bibnamefont {Yin}}, \bibinfo
  {author} {\bibfnamefont {B.-P.}\ \bibnamefont {Hou}}, \bibinfo {author}
  {\bibfnamefont {W.}~\bibnamefont {Li}}, \bibinfo {author} {\bibfnamefont
  {D.}~\bibnamefont {Vitali}}, \bibinfo {author} {\bibfnamefont
  {F.}~\bibnamefont {Nori}},\ and\ \bibinfo {author} {\bibfnamefont {J.-Q.}\
  \bibnamefont {Liao}},\ }\bibfield  {title} {\bibinfo {title} {Nonreciprocal
  ground-state cooling of multiple mechanical resonators},\ }\href
  {https://doi.org/10.1103/PhysRevA.102.011502} {\bibfield  {journal} {\bibinfo
   {journal} {Phys. Rev. A}\ }\textbf {\bibinfo {volume} {102}},\ \bibinfo
  {pages} {011502} (\bibinfo {year} {2020})}\BibitemShut {NoStop}%
\bibitem [{\citenamefont {Huang}\ \emph
  {et~al.}(2022{\natexlab{b}})\citenamefont {Huang}, \citenamefont {Lai},
  \citenamefont {Liu}, \citenamefont {Huang}, \citenamefont {Nori},\ and\
  \citenamefont {Liao}}]{PhysRevA.106.013526}%
  \BibitemOpen
  \bibfield  {author} {\bibinfo {author} {\bibfnamefont {J.}~\bibnamefont
  {Huang}}, \bibinfo {author} {\bibfnamefont {D.-G.}\ \bibnamefont {Lai}},
  \bibinfo {author} {\bibfnamefont {C.}~\bibnamefont {Liu}}, \bibinfo {author}
  {\bibfnamefont {J.-F.}\ \bibnamefont {Huang}}, \bibinfo {author}
  {\bibfnamefont {F.}~\bibnamefont {Nori}},\ and\ \bibinfo {author}
  {\bibfnamefont {J.-Q.}\ \bibnamefont {Liao}},\ }\bibfield  {title} {\bibinfo
  {title} {Multimode optomechanical cooling via general dark-mode control},\
  }\href {https://doi.org/10.1103/PhysRevA.106.013526} {\bibfield  {journal}
  {\bibinfo  {journal} {Phys. Rev. A}\ }\textbf {\bibinfo {volume} {106}},\
  \bibinfo {pages} {013526} (\bibinfo {year} {2022}{\natexlab{b}})}\BibitemShut
  {NoStop}%
\bibitem [{\citenamefont {Huang}\ \emph
  {et~al.}(2023{\natexlab{a}})\citenamefont {Huang}, \citenamefont {Lai},\ and\
  \citenamefont {Liao}}]{PhysRevA.108.013516}%
  \BibitemOpen
  \bibfield  {author} {\bibinfo {author} {\bibfnamefont {J.}~\bibnamefont
  {Huang}}, \bibinfo {author} {\bibfnamefont {D.-G.}\ \bibnamefont {Lai}},\
  and\ \bibinfo {author} {\bibfnamefont {J.-Q.}\ \bibnamefont {Liao}},\
  }\bibfield  {title} {\bibinfo {title} {Controllable generation of mechanical
  quadrature squeezing via dark-mode engineering in cavity optomechanics},\
  }\href {https://doi.org/10.1103/PhysRevA.108.013516} {\bibfield  {journal}
  {\bibinfo  {journal} {Phys. Rev. A}\ }\textbf {\bibinfo {volume} {108}},\
  \bibinfo {pages} {013516} (\bibinfo {year} {2023}{\natexlab{a}})}\BibitemShut
  {NoStop}%
\bibitem [{\citenamefont {Huang}\ \emph
  {et~al.}(2022{\natexlab{c}})\citenamefont {Huang}, \citenamefont {Lai},\ and\
  \citenamefont {Liao}}]{PhysRevA.106.063506}%
  \BibitemOpen
  \bibfield  {author} {\bibinfo {author} {\bibfnamefont {J.}~\bibnamefont
  {Huang}}, \bibinfo {author} {\bibfnamefont {D.-G.}\ \bibnamefont {Lai}},\
  and\ \bibinfo {author} {\bibfnamefont {J.-Q.}\ \bibnamefont {Liao}},\
  }\bibfield  {title} {\bibinfo {title} {Thermal-noise-resistant optomechanical
  entanglement via general dark-mode control},\ }\href
  {https://doi.org/10.1103/PhysRevA.106.063506} {\bibfield  {journal} {\bibinfo
   {journal} {Phys. Rev. A}\ }\textbf {\bibinfo {volume} {106}},\ \bibinfo
  {pages} {063506} (\bibinfo {year} {2022}{\natexlab{c}})}\BibitemShut
  {NoStop}%
\bibitem [{\citenamefont {Lai}\ \emph {et~al.}(2022)\citenamefont {Lai},
  \citenamefont {Liao}, \citenamefont {Miranowicz},\ and\ \citenamefont
  {Nori}}]{PhysRevLett.129.063602}%
  \BibitemOpen
  \bibfield  {author} {\bibinfo {author} {\bibfnamefont {D.-G.}\ \bibnamefont
  {Lai}}, \bibinfo {author} {\bibfnamefont {J.-Q.}\ \bibnamefont {Liao}},
  \bibinfo {author} {\bibfnamefont {A.}~\bibnamefont {Miranowicz}},\ and\
  \bibinfo {author} {\bibfnamefont {F.}~\bibnamefont {Nori}},\ }\bibfield
  {title} {\bibinfo {title} {Noise-tolerant optomechanical entanglement via
  synthetic magnetism},\ }\href
  {https://doi.org/10.1103/PhysRevLett.129.063602} {\bibfield  {journal}
  {\bibinfo  {journal} {Phys. Rev. Lett.}\ }\textbf {\bibinfo {volume} {129}},\
  \bibinfo {pages} {063602} (\bibinfo {year} {2022})}\BibitemShut {NoStop}%
\bibitem [{\citenamefont {Huang}\ \emph
  {et~al.}(2023{\natexlab{b}})\citenamefont {Huang}, \citenamefont {Liu},
  \citenamefont {Xu},\ and\ \citenamefont {Liao}}]{huang2023dark}%
  \BibitemOpen
  \bibfield  {author} {\bibinfo {author} {\bibfnamefont {J.}~\bibnamefont
  {Huang}}, \bibinfo {author} {\bibfnamefont {C.}~\bibnamefont {Liu}}, \bibinfo
  {author} {\bibfnamefont {X.-W.}\ \bibnamefont {Xu}},\ and\ \bibinfo {author}
  {\bibfnamefont {J.-Q.}\ \bibnamefont {Liao}},\ }\bibfield  {title} {\bibinfo
  {title} {Dark-mode theorems for quantum networks},\ }\bibfield  {journal}
  {\bibinfo  {journal} {arXiv preprint arXiv:2312.06274}\ }\href
  {https://doi.org/10.48550/arXiv.2312.06274} {10.48550/arXiv.2312.06274}
  (\bibinfo {year} {2023}{\natexlab{b}})\BibitemShut {NoStop}%
\bibitem [{\citenamefont {Liu}\ \emph {et~al.}(2025)\citenamefont {Liu},
  \citenamefont {Sun}, \citenamefont {Liu}, \citenamefont {Wu}, \citenamefont
  {Sillanp{\"a}{\"a}},\ and\ \citenamefont {Li}}]{YulongLiu_2025_NC}%
  \BibitemOpen
  \bibfield  {author} {\bibinfo {author} {\bibfnamefont {Y.}~\bibnamefont
  {Liu}}, \bibinfo {author} {\bibfnamefont {H.}~\bibnamefont {Sun}}, \bibinfo
  {author} {\bibfnamefont {Q.}~\bibnamefont {Liu}}, \bibinfo {author}
  {\bibfnamefont {H.}~\bibnamefont {Wu}}, \bibinfo {author} {\bibfnamefont
  {M.~A.}\ \bibnamefont {Sillanp{\"a}{\"a}}},\ and\ \bibinfo {author}
  {\bibfnamefont {T.}~\bibnamefont {Li}},\ }\bibfield  {title} {\bibinfo
  {title} {Degeneracy-breaking and long-lived multimode microwave
  electromechanical systems enabled by cubic silicon-carbide membrane
  crystals},\ }\href {https://doi.org/10.1038/s41467-025-56497-3} {\bibfield
  {journal} {\bibinfo  {journal} {Nature Communications}\ }\textbf {\bibinfo
  {volume} {16}},\ \bibinfo {pages} {1207} (\bibinfo {year}
  {2025})}\BibitemShut {NoStop}%
\bibitem [{\citenamefont {Hoch}\ \emph {et~al.}(2021)\citenamefont {Hoch},
  \citenamefont {Haas}, \citenamefont {Moller}, \citenamefont {Sommer},
  \citenamefont {Soubelet}, \citenamefont {Finley},\ and\ \citenamefont
  {Poot}}]{Menno_2021_Micromechines}%
  \BibitemOpen
  \bibfield  {author} {\bibinfo {author} {\bibfnamefont {D.}~\bibnamefont
  {Hoch}}, \bibinfo {author} {\bibfnamefont {K.-J.}\ \bibnamefont {Haas}},
  \bibinfo {author} {\bibfnamefont {L.}~\bibnamefont {Moller}}, \bibinfo
  {author} {\bibfnamefont {T.}~\bibnamefont {Sommer}}, \bibinfo {author}
  {\bibfnamefont {P.}~\bibnamefont {Soubelet}}, \bibinfo {author}
  {\bibfnamefont {J.~J.}\ \bibnamefont {Finley}},\ and\ \bibinfo {author}
  {\bibfnamefont {M.}~\bibnamefont {Poot}},\ }\bibfield  {title} {\bibinfo
  {title} {Efficient optomechanical mode-shape mapping of micromechanical
  devices},\ }\bibfield  {journal} {\bibinfo  {journal} {Micromachines}\
  }\textbf {\bibinfo {volume} {12}},\ \href
  {https://doi.org/10.3390/mi12080880} {10.3390/mi12080880} (\bibinfo {year}
  {2021})\BibitemShut {NoStop}%
\bibitem [{\citenamefont {Sanz-Jim{\'e}nez}\ \emph {et~al.}(2023)\citenamefont
  {Sanz-Jim{\'e}nez}, \citenamefont {Ruz}, \citenamefont {Gil-Santos},
  \citenamefont {Malvar}, \citenamefont {Garc{\'i}a-L{\'o}pez}, \citenamefont
  {Kosaka}, \citenamefont {Cano}, \citenamefont {Calleja},\ and\ \citenamefont
  {Tamayo}}]{Tamayo_2023_ACS}%
  \BibitemOpen
  \bibfield  {author} {\bibinfo {author} {\bibfnamefont {A.}~\bibnamefont
  {Sanz-Jim{\'e}nez}}, \bibinfo {author} {\bibfnamefont {J.~J.}\ \bibnamefont
  {Ruz}}, \bibinfo {author} {\bibfnamefont {E.}~\bibnamefont {Gil-Santos}},
  \bibinfo {author} {\bibfnamefont {O.}~\bibnamefont {Malvar}}, \bibinfo
  {author} {\bibfnamefont {S.}~\bibnamefont {Garc{\'i}a-L{\'o}pez}}, \bibinfo
  {author} {\bibfnamefont {P.~M.}\ \bibnamefont {Kosaka}}, \bibinfo {author}
  {\bibfnamefont {{\'A}.}~\bibnamefont {Cano}}, \bibinfo {author}
  {\bibfnamefont {M.}~\bibnamefont {Calleja}},\ and\ \bibinfo {author}
  {\bibfnamefont {J.}~\bibnamefont {Tamayo}},\ }\bibfield  {title} {\bibinfo
  {title} {Square membrane resonators supporting degenerate modes of vibration
  for high-throughput mass spectrometry of single bacterial cells},\ }\href
  {https://doi.org/10.1021/acssensors.3c00338} {\bibfield  {journal} {\bibinfo
  {journal} {ACS Sensors}\ }\textbf {\bibinfo {volume} {8}},\ \bibinfo {pages}
  {2060} (\bibinfo {year} {2023})}\BibitemShut {NoStop}%
\bibitem [{\citenamefont {Liu}\ \emph {et~al.}(2015)\citenamefont {Liu},
  \citenamefont {Kim},\ and\ \citenamefont {Lauhon}}]{Lincoln_2015_NL}%
  \BibitemOpen
  \bibfield  {author} {\bibinfo {author} {\bibfnamefont {C.-H.}\ \bibnamefont
  {Liu}}, \bibinfo {author} {\bibfnamefont {I.~S.}\ \bibnamefont {Kim}},\ and\
  \bibinfo {author} {\bibfnamefont {L.~J.}\ \bibnamefont {Lauhon}},\ }\bibfield
   {title} {\bibinfo {title} {Optical control of mechanical mode-coupling
  within a $\rm {MoS}_{2}$ resonator in the strong-coupling regime},\ }\href
  {https://doi.org/10.1021/acs.nanolett.5b02586} {\bibfield  {journal}
  {\bibinfo  {journal} {Nano Letters}\ }\textbf {\bibinfo {volume} {15}},\
  \bibinfo {pages} {6727} (\bibinfo {year} {2015})}\BibitemShut {NoStop}%
\bibitem [{\citenamefont {Shkarin}\ \emph
  {et~al.}(2014{\natexlab{b}})\citenamefont {Shkarin}, \citenamefont
  {Flowers-Jacobs}, \citenamefont {Hoch}, \citenamefont {Kashkanova},
  \citenamefont {Deutsch}, \citenamefont {Reichel},\ and\ \citenamefont
  {Harris}}]{Harris_2014_PRL}%
  \BibitemOpen
  \bibfield  {author} {\bibinfo {author} {\bibfnamefont {A.~B.}\ \bibnamefont
  {Shkarin}}, \bibinfo {author} {\bibfnamefont {N.~E.}\ \bibnamefont
  {Flowers-Jacobs}}, \bibinfo {author} {\bibfnamefont {S.~W.}\ \bibnamefont
  {Hoch}}, \bibinfo {author} {\bibfnamefont {A.~D.}\ \bibnamefont
  {Kashkanova}}, \bibinfo {author} {\bibfnamefont {C.}~\bibnamefont {Deutsch}},
  \bibinfo {author} {\bibfnamefont {J.}~\bibnamefont {Reichel}},\ and\ \bibinfo
  {author} {\bibfnamefont {J.~G.~E.}\ \bibnamefont {Harris}},\ }\bibfield
  {title} {\bibinfo {title} {Optically mediated hybridization between two
  mechanical modes},\ }\href {https://doi.org/10.1103/PhysRevLett.112.013602}
  {\bibfield  {journal} {\bibinfo  {journal} {Phys. Rev. Lett.}\ }\textbf
  {\bibinfo {volume} {112}},\ \bibinfo {pages} {013602} (\bibinfo {year}
  {2014}{\natexlab{b}})}\BibitemShut {NoStop}%
\bibitem [{\citenamefont {Chakram}\ \emph {et~al.}(2014)\citenamefont
  {Chakram}, \citenamefont {Patil}, \citenamefont {Chang},\ and\ \citenamefont
  {Vengalattore}}]{Vengalattore_2014_PRL}%
  \BibitemOpen
  \bibfield  {author} {\bibinfo {author} {\bibfnamefont {S.}~\bibnamefont
  {Chakram}}, \bibinfo {author} {\bibfnamefont {Y.~S.}\ \bibnamefont {Patil}},
  \bibinfo {author} {\bibfnamefont {L.}~\bibnamefont {Chang}},\ and\ \bibinfo
  {author} {\bibfnamefont {M.}~\bibnamefont {Vengalattore}},\ }\bibfield
  {title} {\bibinfo {title} {Dissipation in ultrahigh quality factor {SiN}
  membrane resonators},\ }\href
  {https://doi.org/10.1103/PhysRevLett.112.127201} {\bibfield  {journal}
  {\bibinfo  {journal} {Phys. Rev. Lett.}\ }\textbf {\bibinfo {volume} {112}},\
  \bibinfo {pages} {127201} (\bibinfo {year} {2014})}\BibitemShut {NoStop}%
\bibitem [{\citenamefont {Barg}\ \emph {et~al.}(2016)\citenamefont {Barg},
  \citenamefont {Tsaturyan}, \citenamefont {Belhage}, \citenamefont {Nielsen},
  \citenamefont {M{\o}ller},\ and\ \citenamefont
  {Schliesser}}]{Nielsen_2016_APB}%
  \BibitemOpen
  \bibfield  {author} {\bibinfo {author} {\bibfnamefont {A.}~\bibnamefont
  {Barg}}, \bibinfo {author} {\bibfnamefont {Y.}~\bibnamefont {Tsaturyan}},
  \bibinfo {author} {\bibfnamefont {E.}~\bibnamefont {Belhage}}, \bibinfo
  {author} {\bibfnamefont {W.~H.~P.}\ \bibnamefont {Nielsen}}, \bibinfo
  {author} {\bibfnamefont {C.~B.}\ \bibnamefont {M{\o}ller}},\ and\ \bibinfo
  {author} {\bibfnamefont {A.}~\bibnamefont {Schliesser}},\ }\bibfield  {title}
  {\bibinfo {title} {Measuring and imaging nanomechanical motion with laser
  light},\ }\href {https://doi.org/10.1007/s00340-016-6585-7} {\bibfield
  {journal} {\bibinfo  {journal} {Applied Physics B}\ }\textbf {\bibinfo
  {volume} {123}},\ \bibinfo {pages} {8} (\bibinfo {year} {2016})}\BibitemShut
  {NoStop}%
\bibitem [{\citenamefont {Jakubek}\ and\ \citenamefont
  {Fries}(2020)}]{Jakubek_2020_JRS}%
  \BibitemOpen
  \bibfield  {author} {\bibinfo {author} {\bibfnamefont {R.~S.}\ \bibnamefont
  {Jakubek}}\ and\ \bibinfo {author} {\bibfnamefont {M.~D.}\ \bibnamefont
  {Fries}},\ }\bibfield  {title} {\bibinfo {title} {Calibration of {Raman}
  wavenumber in large {Raman} images using a mercury-argon lamp},\ }\href
  {https://doi.org/https://doi.org/10.1002/jrs.5887} {\bibfield  {journal}
  {\bibinfo  {journal} {Journal of Raman Spectroscopy}\ }\textbf {\bibinfo
  {volume} {51}},\ \bibinfo {pages} {1172} (\bibinfo {year}
  {2020})}\BibitemShut {NoStop}%
\bibitem [{\citenamefont {Kagi;}(2008)}]{Odake_2008_ASpe}%
  \BibitemOpen
  \bibfield  {author} {\bibinfo {author} {\bibfnamefont {S.~O.~F.}\
  \bibnamefont {Kagi;}},\ }\bibfield  {title} {\bibinfo {title} {High precision
  in {Raman} frequency achieved using real-time calibration with a neon
  emission line: Application to three-dimensional stress mapping
  observations},\ }\href {https://doi.org/10.1366/000370208786049169}
  {\bibfield  {journal} {\bibinfo  {journal} {Applied Spectroscopy}\ }\textbf
  {\bibinfo {volume} {62}},\ \bibinfo {pages} {1084} (\bibinfo {year}
  {2008})}\BibitemShut {NoStop}%
\bibitem [{\citenamefont {Long}\ \emph {et~al.}(1999)\citenamefont {Long},
  \citenamefont {Ustin},\ and\ \citenamefont {Ho}}]{Long_1999_JAP}%
  \BibitemOpen
  \bibfield  {author} {\bibinfo {author} {\bibfnamefont {C.}~\bibnamefont
  {Long}}, \bibinfo {author} {\bibfnamefont {S.~A.}\ \bibnamefont {Ustin}},\
  and\ \bibinfo {author} {\bibfnamefont {W.}~\bibnamefont {Ho}},\ }\bibfield
  {title} {\bibinfo {title} {Structural defects in {3C–SiC} grown on {Si} by
  supersonic jet epitaxy},\ }\href {https://doi.org/10.1063/1.371085}
  {\bibfield  {journal} {\bibinfo  {journal} {Journal of Applied Physics}\
  }\textbf {\bibinfo {volume} {86}},\ \bibinfo {pages} {2509} (\bibinfo {year}
  {1999})}\BibitemShut {NoStop}%
\bibitem [{\citenamefont {Ernst}\ and\ \citenamefont
  {Pirouz}(1989)}]{Ernst_1989_JMR}%
  \BibitemOpen
  \bibfield  {author} {\bibinfo {author} {\bibfnamefont {F.}~\bibnamefont
  {Ernst}}\ and\ \bibinfo {author} {\bibfnamefont {P.}~\bibnamefont {Pirouz}},\
  }\bibfield  {title} {\bibinfo {title} {The formation mechanism of planar
  defects in compound semiconductors grown epitaxially on {\{}100{\}} silicon
  substrates},\ }\href {https://doi.org/10.1557/JMR.1989.0834} {\bibfield
  {journal} {\bibinfo  {journal} {Journal of Materials Research}\ }\textbf
  {\bibinfo {volume} {4}},\ \bibinfo {pages} {834} (\bibinfo {year}
  {1989})}\BibitemShut {NoStop}%
\bibitem [{\citenamefont {{La Via}}\ \emph {et~al.}(2018)\citenamefont {{La
  Via}}, \citenamefont {Severino}, \citenamefont {Anzalone}, \citenamefont
  {Bongiorno}, \citenamefont {Litrico}, \citenamefont {Mauceri}, \citenamefont
  {Schoeler}, \citenamefont {Schuh},\ and\ \citenamefont
  {Wellmann}}]{Lavia_2018_MSSP}%
  \BibitemOpen
  \bibfield  {author} {\bibinfo {author} {\bibfnamefont {F.}~\bibnamefont {{La
  Via}}}, \bibinfo {author} {\bibfnamefont {A.}~\bibnamefont {Severino}},
  \bibinfo {author} {\bibfnamefont {R.}~\bibnamefont {Anzalone}}, \bibinfo
  {author} {\bibfnamefont {C.}~\bibnamefont {Bongiorno}}, \bibinfo {author}
  {\bibfnamefont {G.}~\bibnamefont {Litrico}}, \bibinfo {author} {\bibfnamefont
  {M.}~\bibnamefont {Mauceri}}, \bibinfo {author} {\bibfnamefont
  {M.}~\bibnamefont {Schoeler}}, \bibinfo {author} {\bibfnamefont
  {P.}~\bibnamefont {Schuh}},\ and\ \bibinfo {author} {\bibfnamefont
  {P.}~\bibnamefont {Wellmann}},\ }\bibfield  {title} {\bibinfo {title} {From
  thin film to bulk {3C-SiC} growth: Understanding the mechanism of defects
  reduction},\ }\href
  {https://doi.org/https://doi.org/10.1016/j.mssp.2017.12.012} {\bibfield
  {journal} {\bibinfo  {journal} {Materials Science in Semiconductor
  Processing}\ }\textbf {\bibinfo {volume} {78}},\ \bibinfo {pages} {57}
  (\bibinfo {year} {2018})}\BibitemShut {NoStop}%
\bibitem [{\citenamefont {Zimbone}\ \emph {et~al.}(2018)\citenamefont
  {Zimbone}, \citenamefont {Mauceri}, \citenamefont {Litrico}, \citenamefont
  {Barbagiovanni}, \citenamefont {Bongiorno},\ and\ \citenamefont {{La
  Via}}}]{Massimo_2018_JCG}%
  \BibitemOpen
  \bibfield  {author} {\bibinfo {author} {\bibfnamefont {M.}~\bibnamefont
  {Zimbone}}, \bibinfo {author} {\bibfnamefont {M.}~\bibnamefont {Mauceri}},
  \bibinfo {author} {\bibfnamefont {G.}~\bibnamefont {Litrico}}, \bibinfo
  {author} {\bibfnamefont {E.~G.}\ \bibnamefont {Barbagiovanni}}, \bibinfo
  {author} {\bibfnamefont {C.}~\bibnamefont {Bongiorno}},\ and\ \bibinfo
  {author} {\bibfnamefont {F.}~\bibnamefont {{La Via}}},\ }\bibfield  {title}
  {\bibinfo {title} {Protrusions reduction in {3C-SiC} thin film on {Si}},\
  }\href {https://doi.org/https://doi.org/10.1016/j.jcrysgro.2018.06.003}
  {\bibfield  {journal} {\bibinfo  {journal} {Journal of Crystal Growth}\
  }\textbf {\bibinfo {volume} {498}},\ \bibinfo {pages} {248} (\bibinfo {year}
  {2018})}\BibitemShut {NoStop}%
\bibitem [{\citenamefont {Gigan}\ \emph {et~al.}(2006)\citenamefont {Gigan},
  \citenamefont {B{\"o}hm}, \citenamefont {Paternostro}, \citenamefont
  {Blaser}, \citenamefont {Langer}, \citenamefont {Hertzberg}, \citenamefont
  {Schwab}, \citenamefont {B{\"a}uerle}, \citenamefont {Aspelmeyer},\ and\
  \citenamefont {Zeilinger}}]{Zeilinger_2006_Nature}%
  \BibitemOpen
  \bibfield  {author} {\bibinfo {author} {\bibfnamefont {S.}~\bibnamefont
  {Gigan}}, \bibinfo {author} {\bibfnamefont {H.~R.}\ \bibnamefont {B{\"o}hm}},
  \bibinfo {author} {\bibfnamefont {M.}~\bibnamefont {Paternostro}}, \bibinfo
  {author} {\bibfnamefont {F.}~\bibnamefont {Blaser}}, \bibinfo {author}
  {\bibfnamefont {G.}~\bibnamefont {Langer}}, \bibinfo {author} {\bibfnamefont
  {J.~B.}\ \bibnamefont {Hertzberg}}, \bibinfo {author} {\bibfnamefont {K.~C.}\
  \bibnamefont {Schwab}}, \bibinfo {author} {\bibfnamefont {D.}~\bibnamefont
  {B{\"a}uerle}}, \bibinfo {author} {\bibfnamefont {M.}~\bibnamefont
  {Aspelmeyer}},\ and\ \bibinfo {author} {\bibfnamefont {A.}~\bibnamefont
  {Zeilinger}},\ }\bibfield  {title} {\bibinfo {title} {Self-cooling of a
  micromirror by radiation pressure},\ }\href
  {https://doi.org/10.1038/nature05273} {\bibfield  {journal} {\bibinfo
  {journal} {Nature}\ }\textbf {\bibinfo {volume} {444}},\ \bibinfo {pages}
  {67} (\bibinfo {year} {2006})}\BibitemShut {NoStop}%
\bibitem [{\citenamefont {Ren}\ \emph {et~al.}(2023)\citenamefont {Ren},
  \citenamefont {Pan}, \citenamefont {Yan}, \citenamefont {Sheng},
  \citenamefont {Yang}, \citenamefont {Zhang}, \citenamefont {Ma},
  \citenamefont {Wen}, \citenamefont {Huang}, \citenamefont {Wu},\ and\
  \citenamefont {Zeng}}]{HaibinWu_2023_NC}%
  \BibitemOpen
  \bibfield  {author} {\bibinfo {author} {\bibfnamefont {X.}~\bibnamefont
  {Ren}}, \bibinfo {author} {\bibfnamefont {J.}~\bibnamefont {Pan}}, \bibinfo
  {author} {\bibfnamefont {M.}~\bibnamefont {Yan}}, \bibinfo {author}
  {\bibfnamefont {J.}~\bibnamefont {Sheng}}, \bibinfo {author} {\bibfnamefont
  {C.}~\bibnamefont {Yang}}, \bibinfo {author} {\bibfnamefont {Q.}~\bibnamefont
  {Zhang}}, \bibinfo {author} {\bibfnamefont {H.}~\bibnamefont {Ma}}, \bibinfo
  {author} {\bibfnamefont {Z.}~\bibnamefont {Wen}}, \bibinfo {author}
  {\bibfnamefont {K.}~\bibnamefont {Huang}}, \bibinfo {author} {\bibfnamefont
  {H.}~\bibnamefont {Wu}},\ and\ \bibinfo {author} {\bibfnamefont
  {H.}~\bibnamefont {Zeng}},\ }\bibfield  {title} {\bibinfo {title} {Dual-comb
  optomechanical spectroscopy},\ }\href
  {https://doi.org/10.1038/s41467-023-40771-3} {\bibfield  {journal} {\bibinfo
  {journal} {Nature Communications}\ }\textbf {\bibinfo {volume} {14}},\
  \bibinfo {pages} {5037} (\bibinfo {year} {2023})}\BibitemShut {NoStop}%
\bibitem [{\citenamefont {Liu}\ \emph {et~al.}(2023{\natexlab{b}})\citenamefont
  {Liu}, \citenamefont {Zhen}, \citenamefont {Yang}, \citenamefont {Zheng},
  \citenamefont {Guo}, \citenamefont {Zhang}, \citenamefont {Yang},
  \citenamefont {Lv}, \citenamefont {Qiu},\ and\ \citenamefont
  {Liu}}]{Liu_2023_JP}%
  \BibitemOpen
  \bibfield  {author} {\bibinfo {author} {\bibfnamefont {Y.}~\bibnamefont
  {Liu}}, \bibinfo {author} {\bibfnamefont {J.~P.}\ \bibnamefont {Zhen}},
  \bibinfo {author} {\bibfnamefont {W.~X.}\ \bibnamefont {Yang}}, \bibinfo
  {author} {\bibfnamefont {X.~D.}\ \bibnamefont {Zheng}}, \bibinfo {author}
  {\bibfnamefont {S.~L.}\ \bibnamefont {Guo}}, \bibinfo {author} {\bibfnamefont
  {Y.}~\bibnamefont {Zhang}}, \bibinfo {author} {\bibfnamefont
  {P.}~\bibnamefont {Yang}}, \bibinfo {author} {\bibfnamefont {K.~H.}\
  \bibnamefont {Lv}}, \bibinfo {author} {\bibfnamefont {J.}~\bibnamefont
  {Qiu}},\ and\ \bibinfo {author} {\bibfnamefont {G.~J.}\ \bibnamefont {Liu}},\
  }\bibfield  {title} {\bibinfo {title} {Research on dissipation dilution
  mechanism and boundary dissipation suppression technique for high-stress
  graphene nanoelectromechanical resonator},\ }\href
  {https://doi.org/10.1088/1742-6596/2557/1/012064} {\bibfield  {journal}
  {\bibinfo  {journal} {Journal of Physics: Conference Series}\ }\textbf
  {\bibinfo {volume} {2557}},\ \bibinfo {pages} {012064} (\bibinfo {year}
  {2023}{\natexlab{b}})}\BibitemShut {NoStop}%
\bibitem [{\citenamefont {Matsumoto}\ \emph {et~al.}(2001)\citenamefont
  {Matsumoto}, \citenamefont {Nose}, \citenamefont {Nagata}, \citenamefont
  {Kawashima}, \citenamefont {Yamada}, \citenamefont {Nakano},\ and\
  \citenamefont {Nagai}}]{Matsumoto_2001_JMCS}%
  \BibitemOpen
  \bibfield  {author} {\bibinfo {author} {\bibfnamefont {T.}~\bibnamefont
  {Matsumoto}}, \bibinfo {author} {\bibfnamefont {T.}~\bibnamefont {Nose}},
  \bibinfo {author} {\bibfnamefont {Y.}~\bibnamefont {Nagata}}, \bibinfo
  {author} {\bibfnamefont {K.}~\bibnamefont {Kawashima}}, \bibinfo {author}
  {\bibfnamefont {T.}~\bibnamefont {Yamada}}, \bibinfo {author} {\bibfnamefont
  {H.}~\bibnamefont {Nakano}},\ and\ \bibinfo {author} {\bibfnamefont
  {S.}~\bibnamefont {Nagai}},\ }\bibfield  {title} {\bibinfo {title}
  {Measurement of high-temperature elastic properties of ceramics using a laser
  ultrasonic method},\ }\href
  {https://doi.org/https://doi.org/10.1111/j.1151-2916.2001.tb00871.x}
  {\bibfield  {journal} {\bibinfo  {journal} {Journal of the American Ceramic
  Society}\ }\textbf {\bibinfo {volume} {84}},\ \bibinfo {pages} {1521}
  (\bibinfo {year} {2001})}\BibitemShut {NoStop}%
\bibitem [{\citenamefont {Varshney}\ \emph {et~al.}(2015)\citenamefont
  {Varshney}, \citenamefont {Shriya}, \citenamefont {Varshney}, \citenamefont
  {Singh},\ and\ \citenamefont {Khenata}}]{Varshney_2015_JTAP}%
  \BibitemOpen
  \bibfield  {author} {\bibinfo {author} {\bibfnamefont {D.}~\bibnamefont
  {Varshney}}, \bibinfo {author} {\bibfnamefont {S.}~\bibnamefont {Shriya}},
  \bibinfo {author} {\bibfnamefont {M.}~\bibnamefont {Varshney}}, \bibinfo
  {author} {\bibfnamefont {N.}~\bibnamefont {Singh}},\ and\ \bibinfo {author}
  {\bibfnamefont {R.}~\bibnamefont {Khenata}},\ }\bibfield  {title} {\bibinfo
  {title} {Elastic and thermodynamical properties of cubic ({3C}) silicon
  carbide under high pressure and high temperature},\ }\href
  {https://doi.org/10.1007/s40094-015-0183-7} {\bibfield  {journal} {\bibinfo
  {journal} {Journal of Theoretical and Applied Physics}\ }\textbf {\bibinfo
  {volume} {9}},\ \bibinfo {pages} {221} (\bibinfo {year} {2015})}\BibitemShut
  {NoStop}%
\bibitem [{\citenamefont {Stillinger}\ and\ \citenamefont
  {Weber}(1985)}]{Thomas_1985_PRB}%
  \BibitemOpen
  \bibfield  {author} {\bibinfo {author} {\bibfnamefont {F.~H.}\ \bibnamefont
  {Stillinger}}\ and\ \bibinfo {author} {\bibfnamefont {T.~A.}\ \bibnamefont
  {Weber}},\ }\bibfield  {title} {\bibinfo {title} {Computer simulation of
  local order in condensed phases of silicon},\ }\href
  {https://doi.org/10.1103/PhysRevB.31.5262} {\bibfield  {journal} {\bibinfo
  {journal} {Phys. Rev. B}\ }\textbf {\bibinfo {volume} {31}},\ \bibinfo
  {pages} {5262} (\bibinfo {year} {1985})}\BibitemShut {NoStop}%
\bibitem [{\citenamefont {Youssefi}\ \emph {et~al.}(2023)\citenamefont
  {Youssefi}, \citenamefont {Kono}, \citenamefont {Chegnizadeh},\ and\
  \citenamefont {Kippenberg}}]{Kippenberg_2023_NP}%
  \BibitemOpen
  \bibfield  {author} {\bibinfo {author} {\bibfnamefont {A.}~\bibnamefont
  {Youssefi}}, \bibinfo {author} {\bibfnamefont {S.}~\bibnamefont {Kono}},
  \bibinfo {author} {\bibfnamefont {M.}~\bibnamefont {Chegnizadeh}},\ and\
  \bibinfo {author} {\bibfnamefont {T.~J.}\ \bibnamefont {Kippenberg}},\
  }\bibfield  {title} {\bibinfo {title} {A squeezed mechanical oscillator with
  millisecond quantum decoherence},\ }\href
  {https://doi.org/10.1038/s41567-023-02135-y} {\bibfield  {journal} {\bibinfo
  {journal} {Nature Physics}\ }\textbf {\bibinfo {volume} {19}},\ \bibinfo
  {pages} {1697} (\bibinfo {year} {2023})}\BibitemShut {NoStop}%
\bibitem [{\citenamefont {Allan}(1966)}]{Allan_1966_IEEE}%
  \BibitemOpen
  \bibfield  {author} {\bibinfo {author} {\bibfnamefont {D.~W.}\ \bibnamefont
  {Allan}},\ }\bibfield  {title} {\bibinfo {title} {Statistics of atomic
  frequency standards},\ }\href {https://doi.org/10.1109/PROC.1966.4634}
  {\bibfield  {journal} {\bibinfo  {journal} {Proceedings of the IEEE}\
  }\textbf {\bibinfo {volume} {54}},\ \bibinfo {pages} {221} (\bibinfo {year}
  {1966})}\BibitemShut {NoStop}%
\bibitem [{\citenamefont {Sadeghi}\ \emph
  {et~al.}(2020{\natexlab{a}})\citenamefont {Sadeghi}, \citenamefont {Demir},
  \citenamefont {Villanueva}, \citenamefont {K{\"a}hler},\ and\ \citenamefont
  {Schmid}}]{Schimid_2010_PRB}%
  \BibitemOpen
  \bibfield  {author} {\bibinfo {author} {\bibfnamefont {P.}~\bibnamefont
  {Sadeghi}}, \bibinfo {author} {\bibfnamefont {A.}~\bibnamefont {Demir}},
  \bibinfo {author} {\bibfnamefont {L.~G.}\ \bibnamefont {Villanueva}},
  \bibinfo {author} {\bibfnamefont {H.}~\bibnamefont {K{\"a}hler}},\ and\
  \bibinfo {author} {\bibfnamefont {S.}~\bibnamefont {Schmid}},\ }\bibfield
  {title} {\bibinfo {title} {Frequency fluctuations in nanomechanical silicon
  nitride string resonators},\ }\href
  {https://doi.org/10.1103/PhysRevB.102.214106} {\bibfield  {journal} {\bibinfo
   {journal} {Phys. Rev. B}\ }\textbf {\bibinfo {volume} {102}},\ \bibinfo
  {pages} {214106} (\bibinfo {year} {2020}{\natexlab{a}})}\BibitemShut
  {NoStop}%
\bibitem [{\citenamefont {Sadeghi}\ \emph
  {et~al.}(2020{\natexlab{b}})\citenamefont {Sadeghi}, \citenamefont {Demir},
  \citenamefont {Villanueva}, \citenamefont {K\"ahler},\ and\ \citenamefont
  {Schmid}}]{Schmid_2020_PRB}%
  \BibitemOpen
  \bibfield  {author} {\bibinfo {author} {\bibfnamefont {P.}~\bibnamefont
  {Sadeghi}}, \bibinfo {author} {\bibfnamefont {A.}~\bibnamefont {Demir}},
  \bibinfo {author} {\bibfnamefont {L.~G.}\ \bibnamefont {Villanueva}},
  \bibinfo {author} {\bibfnamefont {H.}~\bibnamefont {K\"ahler}},\ and\
  \bibinfo {author} {\bibfnamefont {S.}~\bibnamefont {Schmid}},\ }\bibfield
  {title} {\bibinfo {title} {Frequency fluctuations in nanomechanical silicon
  nitride string resonators},\ }\href
  {https://doi.org/10.1103/PhysRevB.102.214106} {\bibfield  {journal} {\bibinfo
   {journal} {Phys. Rev. B}\ }\textbf {\bibinfo {volume} {102}},\ \bibinfo
  {pages} {214106} (\bibinfo {year} {2020}{\natexlab{b}})}\BibitemShut
  {NoStop}%
\bibitem [{\citenamefont {Kim}\ \emph {et~al.}(2005)\citenamefont {Kim},
  \citenamefont {Candler}, \citenamefont {Hopcroft}, \citenamefont {Agarwal},
  \citenamefont {Park},\ and\ \citenamefont {Kenny}}]{Kim_2005_Conference}%
  \BibitemOpen
  \bibfield  {author} {\bibinfo {author} {\bibfnamefont {B.}~\bibnamefont
  {Kim}}, \bibinfo {author} {\bibfnamefont {R.}~\bibnamefont {Candler}},
  \bibinfo {author} {\bibfnamefont {M.}~\bibnamefont {Hopcroft}}, \bibinfo
  {author} {\bibfnamefont {M.}~\bibnamefont {Agarwal}}, \bibinfo {author}
  {\bibfnamefont {W.-T.}\ \bibnamefont {Park}},\ and\ \bibinfo {author}
  {\bibfnamefont {T.}~\bibnamefont {Kenny}},\ }\bibfield  {title} {\bibinfo
  {title} {Frequency stability of wafer-scale encapsulated mems resonators},\
  }in\ \href {https://doi.org/10.1109/SENSOR.2005.1497485} {\emph {\bibinfo
  {booktitle} {The 13th International Conference on Solid-State Sensors,
  Actuators and Microsystems, 2005. Digest of Technical Papers. TRANSDUCERS
  '05.}}},\ Vol.~\bibinfo {volume} {2}\ (\bibinfo {year} {2005})\ pp.\ \bibinfo
  {pages} {1965--1968 Vol. 2}\BibitemShut {NoStop}%
\bibitem [{\citenamefont {Pandit}\ \emph {et~al.}(2021)\citenamefont {Pandit},
  \citenamefont {Mustafazade}, \citenamefont {Sobreviela}, \citenamefont
  {Zhao}, \citenamefont {Zou},\ and\ \citenamefont {Seshia}}]{Ashwin_2021_JMS}%
  \BibitemOpen
  \bibfield  {author} {\bibinfo {author} {\bibfnamefont {M.}~\bibnamefont
  {Pandit}}, \bibinfo {author} {\bibfnamefont {A.}~\bibnamefont {Mustafazade}},
  \bibinfo {author} {\bibfnamefont {G.}~\bibnamefont {Sobreviela}}, \bibinfo
  {author} {\bibfnamefont {C.}~\bibnamefont {Zhao}}, \bibinfo {author}
  {\bibfnamefont {X.}~\bibnamefont {Zou}},\ and\ \bibinfo {author}
  {\bibfnamefont {A.~A.}\ \bibnamefont {Seshia}},\ }\bibfield  {title}
  {\bibinfo {title} {Experimental observation of temperature and pressure
  induced frequency fluctuations in silicon mems resonators},\ }\href
  {https://doi.org/10.1109/JMEMS.2021.3077633} {\bibfield  {journal} {\bibinfo
  {journal} {Journal of Microelectromechanical Systems}\ }\textbf {\bibinfo
  {volume} {30}},\ \bibinfo {pages} {500} (\bibinfo {year} {2021})}\BibitemShut
  {NoStop}%
\bibitem [{\citenamefont {Antonio}\ \emph {et~al.}(2012)\citenamefont
  {Antonio}, \citenamefont {Zanette},\ and\ \citenamefont
  {L{\'o}pez}}]{Antonio_2012_NC}%
  \BibitemOpen
  \bibfield  {author} {\bibinfo {author} {\bibfnamefont {D.}~\bibnamefont
  {Antonio}}, \bibinfo {author} {\bibfnamefont {D.~H.}\ \bibnamefont
  {Zanette}},\ and\ \bibinfo {author} {\bibfnamefont {D.}~\bibnamefont
  {L{\'o}pez}},\ }\bibfield  {title} {\bibinfo {title} {Frequency stabilization
  in nonlinear micromechanical oscillators},\ }\href
  {https://doi.org/10.1038/ncomms1813} {\bibfield  {journal} {\bibinfo
  {journal} {Nature Communications}\ }\textbf {\bibinfo {volume} {3}},\
  \bibinfo {pages} {806} (\bibinfo {year} {2012})}\BibitemShut {NoStop}%
\bibitem [{\citenamefont {Olcum}\ \emph {et~al.}(2014)\citenamefont {Olcum},
  \citenamefont {Cermak}, \citenamefont {Wasserman}, \citenamefont {Christine},
  \citenamefont {Atsumi}, \citenamefont {Payer}, \citenamefont {Shen},
  \citenamefont {Lee}, \citenamefont {Belcher}, \citenamefont {Bhatia},\ and\
  \citenamefont {Manalis}}]{Selim_2014_PNAS}%
  \BibitemOpen
  \bibfield  {author} {\bibinfo {author} {\bibfnamefont {S.}~\bibnamefont
  {Olcum}}, \bibinfo {author} {\bibfnamefont {N.}~\bibnamefont {Cermak}},
  \bibinfo {author} {\bibfnamefont {S.~C.}\ \bibnamefont {Wasserman}}, \bibinfo
  {author} {\bibfnamefont {K.~S.}\ \bibnamefont {Christine}}, \bibinfo {author}
  {\bibfnamefont {H.}~\bibnamefont {Atsumi}}, \bibinfo {author} {\bibfnamefont
  {K.~R.}\ \bibnamefont {Payer}}, \bibinfo {author} {\bibfnamefont
  {W.}~\bibnamefont {Shen}}, \bibinfo {author} {\bibfnamefont {J.}~\bibnamefont
  {Lee}}, \bibinfo {author} {\bibfnamefont {A.~M.}\ \bibnamefont {Belcher}},
  \bibinfo {author} {\bibfnamefont {S.~N.}\ \bibnamefont {Bhatia}},\ and\
  \bibinfo {author} {\bibfnamefont {S.~R.}\ \bibnamefont {Manalis}},\
  }\bibfield  {title} {\bibinfo {title} {Weighing nanoparticles in solution at
  the attogram scale},\ }\href {https://doi.org/10.1073/pnas.1318602111}
  {\bibfield  {journal} {\bibinfo  {journal} {Proceedings of the National
  Academy of Sciences}\ }\textbf {\bibinfo {volume} {111}},\ \bibinfo {pages}
  {1310} (\bibinfo {year} {2014})}\BibitemShut {NoStop}%
\bibitem [{\citenamefont {Feng}\ \emph {et~al.}(2007)\citenamefont {Feng},
  \citenamefont {He}, \citenamefont {Yang},\ and\ \citenamefont
  {Roukes}}]{XL_Feng_2007_Confernce}%
  \BibitemOpen
  \bibfield  {author} {\bibinfo {author} {\bibfnamefont {X.}~\bibnamefont
  {Feng}}, \bibinfo {author} {\bibfnamefont {R.}~\bibnamefont {He}}, \bibinfo
  {author} {\bibfnamefont {P.}~\bibnamefont {Yang}},\ and\ \bibinfo {author}
  {\bibfnamefont {M.}~\bibnamefont {Roukes}},\ }\bibfield  {title} {\bibinfo
  {title} {Phase noise and frequency stability of very-high frequency silicon
  nanowire nanomechanical resonators},\ }in\ \href
  {https://doi.org/10.1109/SENSOR.2007.4300134} {\emph {\bibinfo {booktitle}
  {TRANSDUCERS 2007 - 2007 International Solid-State Sensors, Actuators and
  Microsystems Conference}}}\ (\bibinfo {year} {2007})\ pp.\ \bibinfo {pages}
  {327--330}\BibitemShut {NoStop}%
\bibitem [{\citenamefont {Asano}\ \emph {et~al.}(2024)\citenamefont {Asano},
  \citenamefont {Yamaguchi},\ and\ \citenamefont {Okamoto}}]{Asano_2024_Arxiv}%
  \BibitemOpen
  \bibfield  {author} {\bibinfo {author} {\bibfnamefont {M.}~\bibnamefont
  {Asano}}, \bibinfo {author} {\bibfnamefont {H.}~\bibnamefont {Yamaguchi}},\
  and\ \bibinfo {author} {\bibfnamefont {H.}~\bibnamefont {Okamoto}},\
  }\bibfield  {title} {\bibinfo {title} {Cavity optomechanical liquid level
  meter using a twin-microbottle resonator},\ }\bibfield  {journal} {\bibinfo
  {journal} {arXiv preprint arXiv:2401.12529}\ }\href
  {https://doi.org/10.48550/arXiv.2401.12529} {10.48550/arXiv.2401.12529}
  (\bibinfo {year} {2024})\BibitemShut {NoStop}%
\bibitem [{\citenamefont {Watkins}\ \emph {et~al.}(2025)\citenamefont
  {Watkins}, \citenamefont {Lee}, \citenamefont {McCandless}, \citenamefont
  {Hall},\ and\ \citenamefont {Philip}}]{XL_Feng_2025_JMS}%
  \BibitemOpen
  \bibfield  {author} {\bibinfo {author} {\bibfnamefont {C.~A.}\ \bibnamefont
  {Watkins}}, \bibinfo {author} {\bibfnamefont {J.}~\bibnamefont {Lee}},
  \bibinfo {author} {\bibfnamefont {J.~P.}\ \bibnamefont {McCandless}},
  \bibinfo {author} {\bibfnamefont {H.~J.}\ \bibnamefont {Hall}},\ and\
  \bibinfo {author} {\bibfnamefont {X.-L.~F.}\ \bibnamefont {Philip}},\
  }\bibfield  {title} {\bibinfo {title} {Single-crystal silicon
  thermal-piezoresistive resonators as high-stability frequency references},\
  }\href {https://doi.org/10.1109/JMEMS.2024.3515098} {\bibfield  {journal}
  {\bibinfo  {journal} {Journal of Microelectromechanical Systems}\ }\textbf
  {\bibinfo {volume} {34}},\ \bibinfo {pages} {15} (\bibinfo {year}
  {2025})}\BibitemShut {NoStop}%
\bibitem [{\citenamefont {Chaste}\ \emph {et~al.}(2012)\citenamefont {Chaste},
  \citenamefont {Eichler}, \citenamefont {Moser}, \citenamefont {Ceballos},
  \citenamefont {Rurali},\ and\ \citenamefont {Bachtold}}]{Chaste_2012_NN}%
  \BibitemOpen
  \bibfield  {author} {\bibinfo {author} {\bibfnamefont {J.}~\bibnamefont
  {Chaste}}, \bibinfo {author} {\bibfnamefont {A.}~\bibnamefont {Eichler}},
  \bibinfo {author} {\bibfnamefont {J.}~\bibnamefont {Moser}}, \bibinfo
  {author} {\bibfnamefont {G.}~\bibnamefont {Ceballos}}, \bibinfo {author}
  {\bibfnamefont {R.}~\bibnamefont {Rurali}},\ and\ \bibinfo {author}
  {\bibfnamefont {A.}~\bibnamefont {Bachtold}},\ }\bibfield  {title} {\bibinfo
  {title} {A nanomechanical mass sensor with yoctogram resolution},\ }\href
  {https://doi.org/10.1038/nnano.2012.42} {\bibfield  {journal} {\bibinfo
  {journal} {Nature Nanotechnology}\ }\textbf {\bibinfo {volume} {7}},\
  \bibinfo {pages} {301} (\bibinfo {year} {2012})}\BibitemShut {NoStop}%
\bibitem [{\citenamefont {Schwender}\ \emph {et~al.}(2018)\citenamefont
  {Schwender}, \citenamefont {Tsioutsios}, \citenamefont {Tavernarakis},
  \citenamefont {Dong}, \citenamefont {Jin}, \citenamefont {Staufer},\ and\
  \citenamefont {Bachtold}}]{Adrian_2018_APL}%
  \BibitemOpen
  \bibfield  {author} {\bibinfo {author} {\bibfnamefont {J.}~\bibnamefont
  {Schwender}}, \bibinfo {author} {\bibfnamefont {I.}~\bibnamefont
  {Tsioutsios}}, \bibinfo {author} {\bibfnamefont {A.}~\bibnamefont
  {Tavernarakis}}, \bibinfo {author} {\bibfnamefont {Q.}~\bibnamefont {Dong}},
  \bibinfo {author} {\bibfnamefont {Y.}~\bibnamefont {Jin}}, \bibinfo {author}
  {\bibfnamefont {U.}~\bibnamefont {Staufer}},\ and\ \bibinfo {author}
  {\bibfnamefont {A.}~\bibnamefont {Bachtold}},\ }\bibfield  {title} {\bibinfo
  {title} {Improving the read-out of the resonance frequency of nanotube
  mechanical resonators},\ }\href {https://doi.org/10.1063/1.5045309}
  {\bibfield  {journal} {\bibinfo  {journal} {Applied Physics Letters}\
  }\textbf {\bibinfo {volume} {113}},\ \bibinfo {pages} {063104} (\bibinfo
  {year} {2018})}\BibitemShut {NoStop}%
\bibitem [{\citenamefont {Moser}\ \emph {et~al.}(2014)\citenamefont {Moser},
  \citenamefont {Eichler}, \citenamefont {G{\"u}ttinger}, \citenamefont
  {Dykman},\ and\ \citenamefont {Bachtold}}]{Moser_2014_NN}%
  \BibitemOpen
  \bibfield  {author} {\bibinfo {author} {\bibfnamefont {J.}~\bibnamefont
  {Moser}}, \bibinfo {author} {\bibfnamefont {A.}~\bibnamefont {Eichler}},
  \bibinfo {author} {\bibfnamefont {J.}~\bibnamefont {G{\"u}ttinger}}, \bibinfo
  {author} {\bibfnamefont {M.~I.}\ \bibnamefont {Dykman}},\ and\ \bibinfo
  {author} {\bibfnamefont {A.}~\bibnamefont {Bachtold}},\ }\bibfield  {title}
  {\bibinfo {title} {Nanotube mechanical resonators with quality factors of up
  to 5 million},\ }\href {https://doi.org/10.1038/nnano.2014.234} {\bibfield
  {journal} {\bibinfo  {journal} {Nature Nanotechnology}\ }\textbf {\bibinfo
  {volume} {9}},\ \bibinfo {pages} {1007} (\bibinfo {year} {2014})}\BibitemShut
  {NoStop}%
\bibitem [{\citenamefont {Kaisar}\ \emph {et~al.}(2023)\citenamefont {Kaisar},
  \citenamefont {Yousuf}, \citenamefont {Lee}, \citenamefont {Qamar},
  \citenamefont {Rais-Zadeh}, \citenamefont {Mandal},\ and\ \citenamefont
  {Feng}}]{XL_Feng_2023_IEEE}%
  \BibitemOpen
  \bibfield  {author} {\bibinfo {author} {\bibfnamefont {T.}~\bibnamefont
  {Kaisar}}, \bibinfo {author} {\bibfnamefont {S.~M. E.~H.}\ \bibnamefont
  {Yousuf}}, \bibinfo {author} {\bibfnamefont {J.}~\bibnamefont {Lee}},
  \bibinfo {author} {\bibfnamefont {A.}~\bibnamefont {Qamar}}, \bibinfo
  {author} {\bibfnamefont {M.}~\bibnamefont {Rais-Zadeh}}, \bibinfo {author}
  {\bibfnamefont {S.}~\bibnamefont {Mandal}},\ and\ \bibinfo {author}
  {\bibfnamefont {P.~X.-L.}\ \bibnamefont {Feng}},\ }\bibfield  {title}
  {\bibinfo {title} {Five low-noise stable oscillators referenced to the same
  multimode {AlN/Si} mems resonator},\ }\href
  {https://doi.org/10.1109/TUFFC.2023.3312159} {\bibfield  {journal} {\bibinfo
  {journal} {IEEE Transactions on Ultrasonics, Ferroelectrics, and Frequency
  Control}\ }\textbf {\bibinfo {volume} {70}},\ \bibinfo {pages} {1213}
  (\bibinfo {year} {2023})}\BibitemShut {NoStop}%
\bibitem [{\citenamefont {Song}\ \emph {et~al.}(2017)\citenamefont {Song},
  \citenamefont {Du}, \citenamefont {Qi}, \citenamefont {Li}, \citenamefont
  {Li},\ and\ \citenamefont {Li}}]{Song_2017_JMM}%
  \BibitemOpen
  \bibfield  {author} {\bibinfo {author} {\bibfnamefont {C.}~\bibnamefont
  {Song}}, \bibinfo {author} {\bibfnamefont {L.}~\bibnamefont {Du}}, \bibinfo
  {author} {\bibfnamefont {L.}~\bibnamefont {Qi}}, \bibinfo {author}
  {\bibfnamefont {Y.}~\bibnamefont {Li}}, \bibinfo {author} {\bibfnamefont
  {X.}~\bibnamefont {Li}},\ and\ \bibinfo {author} {\bibfnamefont
  {Y.}~\bibnamefont {Li}},\ }\bibfield  {title} {\bibinfo {title} {Residual
  stress measurement in a metal microdevice by micro {Raman} spectroscopy},\
  }\href {https://doi.org/10.1088/1361-6439/aa8912} {\bibfield  {journal}
  {\bibinfo  {journal} {Journal of Micromechanics and Microengineering}\
  }\textbf {\bibinfo {volume} {27}},\ \bibinfo {pages} {105014} (\bibinfo
  {year} {2017})}\BibitemShut {NoStop}%
\bibitem [{\citenamefont {Xiao}\ \emph {et~al.}(2023)\citenamefont {Xiao},
  \citenamefont {Han}, \citenamefont {Zhu},\ and\ \citenamefont
  {Wu}}]{Guoqiang_Wu_2023_APL}%
  \BibitemOpen
  \bibfield  {author} {\bibinfo {author} {\bibfnamefont {Y.}~\bibnamefont
  {Xiao}}, \bibinfo {author} {\bibfnamefont {J.}~\bibnamefont {Han}}, \bibinfo
  {author} {\bibfnamefont {K.}~\bibnamefont {Zhu}},\ and\ \bibinfo {author}
  {\bibfnamefont {G.}~\bibnamefont {Wu}},\ }\bibfield  {title} {\bibinfo
  {title} {A piezoelectric mechanically coupled lamé mode resonator with
  ultra-high {Q}},\ }\href {https://doi.org/10.1063/5.0141778} {\bibfield
  {journal} {\bibinfo  {journal} {Applied Physics Letters}\ }\textbf {\bibinfo
  {volume} {122}},\ \bibinfo {pages} {114101} (\bibinfo {year}
  {2023})}\BibitemShut {NoStop}%
\bibitem [{\citenamefont {Zhang}\ and\ \citenamefont
  {St-Gelais}(2023)}]{ChangZhang_2023_APL}%
  \BibitemOpen
  \bibfield  {author} {\bibinfo {author} {\bibfnamefont {C.}~\bibnamefont
  {Zhang}}\ and\ \bibinfo {author} {\bibfnamefont {R.}~\bibnamefont
  {St-Gelais}},\ }\bibfield  {title} {\bibinfo {title} {Demonstration of
  frequency stability limited by thermal fluctuation noise in silicon nitride
  nanomechanical resonators},\ }\href {https://doi.org/10.1063/5.0145780}
  {\bibfield  {journal} {\bibinfo  {journal} {Applied Physics Letters}\
  }\textbf {\bibinfo {volume} {122}},\ \bibinfo {pages} {193501} (\bibinfo
  {year} {2023})}\BibitemShut {NoStop}%
\bibitem [{\citenamefont {Snell}\ \emph {et~al.}(2022)\citenamefont {Snell},
  \citenamefont {Zhang}, \citenamefont {Mu}, \citenamefont {Bouchard},\ and\
  \citenamefont {St-Gelais}}]{Raphael_2022_PRApplied}%
  \BibitemOpen
  \bibfield  {author} {\bibinfo {author} {\bibfnamefont {N.}~\bibnamefont
  {Snell}}, \bibinfo {author} {\bibfnamefont {C.}~\bibnamefont {Zhang}},
  \bibinfo {author} {\bibfnamefont {G.}~\bibnamefont {Mu}}, \bibinfo {author}
  {\bibfnamefont {A.}~\bibnamefont {Bouchard}},\ and\ \bibinfo {author}
  {\bibfnamefont {R.}~\bibnamefont {St-Gelais}},\ }\bibfield  {title} {\bibinfo
  {title} {Heat transport in silicon nitride drum resonators and its influence
  on thermal fluctuation-induced frequency noise},\ }\href
  {https://doi.org/10.1103/PhysRevApplied.17.044019} {\bibfield  {journal}
  {\bibinfo  {journal} {Phys. Rev. Appl.}\ }\textbf {\bibinfo {volume} {17}},\
  \bibinfo {pages} {044019} (\bibinfo {year} {2022})}\BibitemShut {NoStop}%
\bibitem [{\citenamefont {Manzaneque}\ \emph {et~al.}(2023)\citenamefont
  {Manzaneque}, \citenamefont {Ghatkesar}, \citenamefont {Alijani},
  \citenamefont {Xu}, \citenamefont {Norte},\ and\ \citenamefont
  {Steeneken}}]{Peter_20232_PRApplied}%
  \BibitemOpen
  \bibfield  {author} {\bibinfo {author} {\bibfnamefont {T.}~\bibnamefont
  {Manzaneque}}, \bibinfo {author} {\bibfnamefont {M.~K.}\ \bibnamefont
  {Ghatkesar}}, \bibinfo {author} {\bibfnamefont {F.}~\bibnamefont {Alijani}},
  \bibinfo {author} {\bibfnamefont {M.}~\bibnamefont {Xu}}, \bibinfo {author}
  {\bibfnamefont {R.~A.}\ \bibnamefont {Norte}},\ and\ \bibinfo {author}
  {\bibfnamefont {P.~G.}\ \bibnamefont {Steeneken}},\ }\bibfield  {title}
  {\bibinfo {title} {Resolution limits of resonant sensors},\ }\href
  {https://doi.org/10.1103/PhysRevApplied.19.054074} {\bibfield  {journal}
  {\bibinfo  {journal} {Phys. Rev. Appl.}\ }\textbf {\bibinfo {volume} {19}},\
  \bibinfo {pages} {054074} (\bibinfo {year} {2023})}\BibitemShut {NoStop}%
\bibitem [{\citenamefont {Be\ifmmode \check{s}\else
  \v{s}\fi{}i\ifmmode~\acute{c}\else \'{c}\fi{}}\ \emph
  {et~al.}(2023)\citenamefont {Be\ifmmode \check{s}\else
  \v{s}\fi{}i\ifmmode~\acute{c}\else \'{c}\fi{}}, \citenamefont {Demir},
  \citenamefont {Steurer}, \citenamefont {Luhmann},\ and\ \citenamefont
  {Schmid}}]{Silvan_2023_PRApplied}%
  \BibitemOpen
  \bibfield  {author} {\bibinfo {author} {\bibfnamefont {H.}~\bibnamefont
  {Be\ifmmode \check{s}\else \v{s}\fi{}i\ifmmode~\acute{c}\else \'{c}\fi{}}},
  \bibinfo {author} {\bibfnamefont {A.}~\bibnamefont {Demir}}, \bibinfo
  {author} {\bibfnamefont {J.}~\bibnamefont {Steurer}}, \bibinfo {author}
  {\bibfnamefont {N.}~\bibnamefont {Luhmann}},\ and\ \bibinfo {author}
  {\bibfnamefont {S.}~\bibnamefont {Schmid}},\ }\bibfield  {title} {\bibinfo
  {title} {Schemes for tracking resonance frequency for micro- and
  nanomechanical resonators},\ }\href
  {https://doi.org/10.1103/PhysRevApplied.20.024023} {\bibfield  {journal}
  {\bibinfo  {journal} {Phys. Rev. Appl.}\ }\textbf {\bibinfo {volume} {20}},\
  \bibinfo {pages} {024023} (\bibinfo {year} {2023})}\BibitemShut {NoStop}%
\bibitem [{\citenamefont {Gavartin}\ \emph {et~al.}(2013)\citenamefont
  {Gavartin}, \citenamefont {Verlot},\ and\ \citenamefont
  {Kippenberg}}]{Gavartin_2013_NC}%
  \BibitemOpen
  \bibfield  {author} {\bibinfo {author} {\bibfnamefont {E.}~\bibnamefont
  {Gavartin}}, \bibinfo {author} {\bibfnamefont {P.}~\bibnamefont {Verlot}},\
  and\ \bibinfo {author} {\bibfnamefont {T.~J.}\ \bibnamefont {Kippenberg}},\
  }\bibfield  {title} {\bibinfo {title} {Stabilization of a linear
  nanomechanical oscillator to its thermodynamic limit},\ }\href
  {https://doi.org/10.1038/ncomms3860} {\bibfield  {journal} {\bibinfo
  {journal} {Nature Communications}\ }\textbf {\bibinfo {volume} {4}},\
  \bibinfo {pages} {2860} (\bibinfo {year} {2013})}\BibitemShut {NoStop}%
\bibitem [{\citenamefont {Sadeghi}\ \emph
  {et~al.}(2020{\natexlab{c}})\citenamefont {Sadeghi}, \citenamefont {Demir},
  \citenamefont {Villanueva}, \citenamefont {K\"ahler},\ and\ \citenamefont
  {Schmid}}]{Silvan_2020_PRB}%
  \BibitemOpen
  \bibfield  {author} {\bibinfo {author} {\bibfnamefont {P.}~\bibnamefont
  {Sadeghi}}, \bibinfo {author} {\bibfnamefont {A.}~\bibnamefont {Demir}},
  \bibinfo {author} {\bibfnamefont {L.~G.}\ \bibnamefont {Villanueva}},
  \bibinfo {author} {\bibfnamefont {H.}~\bibnamefont {K\"ahler}},\ and\
  \bibinfo {author} {\bibfnamefont {S.}~\bibnamefont {Schmid}},\ }\bibfield
  {title} {\bibinfo {title} {Frequency fluctuations in nanomechanical silicon
  nitride string resonators},\ }\href
  {https://doi.org/10.1103/PhysRevB.102.214106} {\bibfield  {journal} {\bibinfo
   {journal} {Phys. Rev. B}\ }\textbf {\bibinfo {volume} {102}},\ \bibinfo
  {pages} {214106} (\bibinfo {year} {2020}{\natexlab{c}})}\BibitemShut
  {NoStop}%
\bibitem [{\citenamefont {Feng}\ \emph {et~al.}(2008)\citenamefont {Feng},
  \citenamefont {White}, \citenamefont {Hajimiri},\ and\ \citenamefont
  {Roukes}}]{Feng_2008_NN}%
  \BibitemOpen
  \bibfield  {author} {\bibinfo {author} {\bibfnamefont {X.~L.}\ \bibnamefont
  {Feng}}, \bibinfo {author} {\bibfnamefont {C.~J.}\ \bibnamefont {White}},
  \bibinfo {author} {\bibfnamefont {A.}~\bibnamefont {Hajimiri}},\ and\
  \bibinfo {author} {\bibfnamefont {M.~L.}\ \bibnamefont {Roukes}},\ }\bibfield
   {title} {\bibinfo {title} {A self-sustaining ultrahigh-frequency
  nanoelectromechanical oscillator},\ }\href
  {https://doi.org/10.1038/nnano.2008.125} {\bibfield  {journal} {\bibinfo
  {journal} {Nature Nanotechnology}\ }\textbf {\bibinfo {volume} {3}},\
  \bibinfo {pages} {342} (\bibinfo {year} {2008})}\BibitemShut {NoStop}%
\bibitem [{\citenamefont {Fedoseev}\ \emph {et~al.}(2021)\citenamefont
  {Fedoseev}, \citenamefont {Luna}, \citenamefont {Hedgepeth}, \citenamefont
  {L\"offler},\ and\ \citenamefont {Bouwmeester}}]{Dirk_2021_PRL}%
  \BibitemOpen
  \bibfield  {author} {\bibinfo {author} {\bibfnamefont {V.}~\bibnamefont
  {Fedoseev}}, \bibinfo {author} {\bibfnamefont {F.}~\bibnamefont {Luna}},
  \bibinfo {author} {\bibfnamefont {I.}~\bibnamefont {Hedgepeth}}, \bibinfo
  {author} {\bibfnamefont {W.}~\bibnamefont {L\"offler}},\ and\ \bibinfo
  {author} {\bibfnamefont {D.}~\bibnamefont {Bouwmeester}},\ }\bibfield
  {title} {\bibinfo {title} {Stimulated {Raman} adiabatic passage in
  optomechanics},\ }\href {https://doi.org/10.1103/PhysRevLett.126.113601}
  {\bibfield  {journal} {\bibinfo  {journal} {Phys. Rev. Lett.}\ }\textbf
  {\bibinfo {volume} {126}},\ \bibinfo {pages} {113601} (\bibinfo {year}
  {2021})}\BibitemShut {NoStop}%
\bibitem [{\citenamefont {Ferrari}\ and\ \citenamefont
  {Lutterotti}(1994)}]{Luca_1994_JAP}%
  \BibitemOpen
  \bibfield  {author} {\bibinfo {author} {\bibfnamefont {M.}~\bibnamefont
  {Ferrari}}\ and\ \bibinfo {author} {\bibfnamefont {L.}~\bibnamefont
  {Lutterotti}},\ }\bibfield  {title} {\bibinfo {title} {Method for the
  simultaneous determination of anisotropic residual stresses and texture by
  {x-ray} diffraction},\ }\href {https://doi.org/10.1063/1.358006} {\bibfield
  {journal} {\bibinfo  {journal} {Journal of Applied Physics}\ }\textbf
  {\bibinfo {volume} {76}},\ \bibinfo {pages} {7246} (\bibinfo {year}
  {1994})}\BibitemShut {NoStop}%
\bibitem [{\citenamefont {Rossini}\ \emph {et~al.}(2012)\citenamefont
  {Rossini}, \citenamefont {Dassisti}, \citenamefont {Benyounis},\ and\
  \citenamefont {Olabi}}]{Olabi_2012_MD}%
  \BibitemOpen
  \bibfield  {author} {\bibinfo {author} {\bibfnamefont {N.}~\bibnamefont
  {Rossini}}, \bibinfo {author} {\bibfnamefont {M.}~\bibnamefont {Dassisti}},
  \bibinfo {author} {\bibfnamefont {K.}~\bibnamefont {Benyounis}},\ and\
  \bibinfo {author} {\bibfnamefont {A.}~\bibnamefont {Olabi}},\ }\bibfield
  {title} {\bibinfo {title} {Methods of measuring residual stresses in
  components},\ }\href
  {https://doi.org/https://doi.org/10.1016/j.matdes.2011.08.022} {\bibfield
  {journal} {\bibinfo  {journal} {Materials \& Design}\ }\textbf {\bibinfo
  {volume} {35}},\ \bibinfo {pages} {572} (\bibinfo {year} {2012})}\BibitemShut
  {NoStop}%
\bibitem [{\citenamefont {Gries}\ \emph {et~al.}(2008)\citenamefont {Gries},
  \citenamefont {Vandenbulcke}, \citenamefont {Simon},\ and\ \citenamefont
  {Canizares}}]{Gries_2008_JAP}%
  \BibitemOpen
  \bibfield  {author} {\bibinfo {author} {\bibfnamefont {T.}~\bibnamefont
  {Gries}}, \bibinfo {author} {\bibfnamefont {L.}~\bibnamefont {Vandenbulcke}},
  \bibinfo {author} {\bibfnamefont {P.}~\bibnamefont {Simon}},\ and\ \bibinfo
  {author} {\bibfnamefont {A.}~\bibnamefont {Canizares}},\ }\bibfield  {title}
  {\bibinfo {title} {Anisotropic biaxial stresses in diamond films by polarized
  {Raman} spectroscopy of cubic polycrystals},\ }\href
  {https://doi.org/10.1063/1.2959338} {\bibfield  {journal} {\bibinfo
  {journal} {Journal of Applied Physics}\ }\textbf {\bibinfo {volume} {104}},\
  \bibinfo {pages} {023524} (\bibinfo {year} {2008})}\BibitemShut {NoStop}%
\bibitem [{\citenamefont {Kneipp}(2007)}]{Katrin_2007_PT}%
  \BibitemOpen
  \bibfield  {author} {\bibinfo {author} {\bibfnamefont {K.}~\bibnamefont
  {Kneipp}},\ }\bibfield  {title} {\bibinfo {title} {Surface-enhanced {Raman}
  scattering},\ }\href {https://doi.org/10.1063/1.2812122} {\bibfield
  {journal} {\bibinfo  {journal} {Physics Today}\ }\textbf {\bibinfo {volume}
  {60}},\ \bibinfo {pages} {40} (\bibinfo {year} {2007})}\BibitemShut {NoStop}%
\bibitem [{\citenamefont {Langer}\ \emph {et~al.}(2020)\citenamefont {Langer},
  \citenamefont {Jimenez~de Aberasturi}, \citenamefont {Aizpurua},
  \citenamefont {Alvarez-Puebla}, \citenamefont {Augui{\'e}}, \citenamefont
  {Baumberg}, \citenamefont {Bazan}, \citenamefont {Bell}, \citenamefont
  {Boisen}, \citenamefont {Brolo}, \citenamefont {Choo}, \citenamefont
  {Cialla-May}, \citenamefont {Deckert}, \citenamefont {Fabris}, \citenamefont
  {Faulds}, \citenamefont {García~de Abajo}, \citenamefont {Goodacre},
  \citenamefont {Graham}, \citenamefont {Haes}, \citenamefont {Haynes},
  \citenamefont {Huck}, \citenamefont {Itoh}, \citenamefont {Käll},
  \citenamefont {Kneipp}, \citenamefont {Kotov}, \citenamefont {Kuang},
  \citenamefont {Le~Ru}, \citenamefont {Lee}, \citenamefont {Li}, \citenamefont
  {Ling}, \citenamefont {Maier}, \citenamefont {Mayerh{\"o}fer}, \citenamefont
  {Moskovits}, \citenamefont {Murakoshi}, \citenamefont {Nam}, \citenamefont
  {Nie}, \citenamefont {Ozaki}, \citenamefont {Pastoriza-Santos}, \citenamefont
  {Perez-Juste}, \citenamefont {Popp}, \citenamefont {Pucci}, \citenamefont
  {Reich}, \citenamefont {Ren}, \citenamefont {Schatz}, \citenamefont {Shegai},
  \citenamefont {Schl{\"u}cker}, \citenamefont {Tay}, \citenamefont {Thomas},
  \citenamefont {Tian}, \citenamefont {Van~Duyne}, \citenamefont {Vo-Dinh},
  \citenamefont {Wang}, \citenamefont {Willets}, \citenamefont {Xu},
  \citenamefont {Xu}, \citenamefont {Xu}, \citenamefont {Yamamoto},
  \citenamefont {Zhao},\ and\ \citenamefont
  {Liz-Marzán}}]{Langer_2020_ACSNano}%
  \BibitemOpen
  \bibfield  {author} {\bibinfo {author} {\bibfnamefont {J.}~\bibnamefont
  {Langer}}, \bibinfo {author} {\bibfnamefont {D.}~\bibnamefont {Jimenez~de
  Aberasturi}}, \bibinfo {author} {\bibfnamefont {J.}~\bibnamefont {Aizpurua}},
  \bibinfo {author} {\bibfnamefont {R.~A.}\ \bibnamefont {Alvarez-Puebla}},
  \bibinfo {author} {\bibfnamefont {B.}~\bibnamefont {Augui{\'e}}}, \bibinfo
  {author} {\bibfnamefont {J.~J.}\ \bibnamefont {Baumberg}}, \bibinfo {author}
  {\bibfnamefont {G.~C.}\ \bibnamefont {Bazan}}, \bibinfo {author}
  {\bibfnamefont {S.~E.~J.}\ \bibnamefont {Bell}}, \bibinfo {author}
  {\bibfnamefont {A.}~\bibnamefont {Boisen}}, \bibinfo {author} {\bibfnamefont
  {A.~G.}\ \bibnamefont {Brolo}}, \bibinfo {author} {\bibfnamefont
  {J.}~\bibnamefont {Choo}}, \bibinfo {author} {\bibfnamefont {D.}~\bibnamefont
  {Cialla-May}}, \bibinfo {author} {\bibfnamefont {V.}~\bibnamefont {Deckert}},
  \bibinfo {author} {\bibfnamefont {L.}~\bibnamefont {Fabris}}, \bibinfo
  {author} {\bibfnamefont {K.}~\bibnamefont {Faulds}}, \bibinfo {author}
  {\bibfnamefont {F.~J.}\ \bibnamefont {García~de Abajo}}, \bibinfo {author}
  {\bibfnamefont {R.}~\bibnamefont {Goodacre}}, \bibinfo {author}
  {\bibfnamefont {D.}~\bibnamefont {Graham}}, \bibinfo {author} {\bibfnamefont
  {A.~J.}\ \bibnamefont {Haes}}, \bibinfo {author} {\bibfnamefont {C.~L.}\
  \bibnamefont {Haynes}}, \bibinfo {author} {\bibfnamefont {C.}~\bibnamefont
  {Huck}}, \bibinfo {author} {\bibfnamefont {T.}~\bibnamefont {Itoh}}, \bibinfo
  {author} {\bibfnamefont {M.}~\bibnamefont {Käll}}, \bibinfo {author}
  {\bibfnamefont {J.}~\bibnamefont {Kneipp}}, \bibinfo {author} {\bibfnamefont
  {N.~A.}\ \bibnamefont {Kotov}}, \bibinfo {author} {\bibfnamefont
  {H.}~\bibnamefont {Kuang}}, \bibinfo {author} {\bibfnamefont {E.~C.}\
  \bibnamefont {Le~Ru}}, \bibinfo {author} {\bibfnamefont {H.~K.}\ \bibnamefont
  {Lee}}, \bibinfo {author} {\bibfnamefont {J.-F.}\ \bibnamefont {Li}},
  \bibinfo {author} {\bibfnamefont {X.~Y.}\ \bibnamefont {Ling}}, \bibinfo
  {author} {\bibfnamefont {S.~A.}\ \bibnamefont {Maier}}, \bibinfo {author}
  {\bibfnamefont {T.}~\bibnamefont {Mayerh{\"o}fer}}, \bibinfo {author}
  {\bibfnamefont {M.}~\bibnamefont {Moskovits}}, \bibinfo {author}
  {\bibfnamefont {K.}~\bibnamefont {Murakoshi}}, \bibinfo {author}
  {\bibfnamefont {J.-M.}\ \bibnamefont {Nam}}, \bibinfo {author} {\bibfnamefont
  {S.}~\bibnamefont {Nie}}, \bibinfo {author} {\bibfnamefont {Y.}~\bibnamefont
  {Ozaki}}, \bibinfo {author} {\bibfnamefont {I.}~\bibnamefont
  {Pastoriza-Santos}}, \bibinfo {author} {\bibfnamefont {J.}~\bibnamefont
  {Perez-Juste}}, \bibinfo {author} {\bibfnamefont {J.}~\bibnamefont {Popp}},
  \bibinfo {author} {\bibfnamefont {A.}~\bibnamefont {Pucci}}, \bibinfo
  {author} {\bibfnamefont {S.}~\bibnamefont {Reich}}, \bibinfo {author}
  {\bibfnamefont {B.}~\bibnamefont {Ren}}, \bibinfo {author} {\bibfnamefont
  {G.~C.}\ \bibnamefont {Schatz}}, \bibinfo {author} {\bibfnamefont
  {T.}~\bibnamefont {Shegai}}, \bibinfo {author} {\bibfnamefont
  {S.}~\bibnamefont {Schl{\"u}cker}}, \bibinfo {author} {\bibfnamefont {L.-L.}\
  \bibnamefont {Tay}}, \bibinfo {author} {\bibfnamefont {K.~G.}\ \bibnamefont
  {Thomas}}, \bibinfo {author} {\bibfnamefont {Z.-Q.}\ \bibnamefont {Tian}},
  \bibinfo {author} {\bibfnamefont {R.~P.}\ \bibnamefont {Van~Duyne}}, \bibinfo
  {author} {\bibfnamefont {T.}~\bibnamefont {Vo-Dinh}}, \bibinfo {author}
  {\bibfnamefont {Y.}~\bibnamefont {Wang}}, \bibinfo {author} {\bibfnamefont
  {K.~A.}\ \bibnamefont {Willets}}, \bibinfo {author} {\bibfnamefont
  {C.}~\bibnamefont {Xu}}, \bibinfo {author} {\bibfnamefont {H.}~\bibnamefont
  {Xu}}, \bibinfo {author} {\bibfnamefont {Y.}~\bibnamefont {Xu}}, \bibinfo
  {author} {\bibfnamefont {Y.~S.}\ \bibnamefont {Yamamoto}}, \bibinfo {author}
  {\bibfnamefont {B.}~\bibnamefont {Zhao}},\ and\ \bibinfo {author}
  {\bibfnamefont {L.~M.}\ \bibnamefont {Liz-Marzán}},\ }\bibfield  {title}
  {\bibinfo {title} {Present and future of surface-enhanced {Raman}
  scattering},\ }\href {https://doi.org/10.1021/acsnano.9b04224} {\bibfield
  {journal} {\bibinfo  {journal} {ACS Nano}\ }\textbf {\bibinfo {volume}
  {14}},\ \bibinfo {pages} {28} (\bibinfo {year} {2020})}\BibitemShut {NoStop}%
\bibitem [{\citenamefont {Kuzyk}\ and\ \citenamefont
  {Wang}(2018)}]{HailinWang_2018_PRX}%
  \BibitemOpen
  \bibfield  {author} {\bibinfo {author} {\bibfnamefont {M.~C.}\ \bibnamefont
  {Kuzyk}}\ and\ \bibinfo {author} {\bibfnamefont {H.}~\bibnamefont {Wang}},\
  }\bibfield  {title} {\bibinfo {title} {Scaling phononic quantum networks of
  solid-state spins with closed mechanical subsystems},\ }\href
  {https://doi.org/10.1103/PhysRevX.8.041027} {\bibfield  {journal} {\bibinfo
  {journal} {Phys. Rev. X}\ }\textbf {\bibinfo {volume} {8}},\ \bibinfo {pages}
  {041027} (\bibinfo {year} {2018})}\BibitemShut {NoStop}%
\bibitem [{\citenamefont {Chu}\ \emph {et~al.}(2018)\citenamefont {Chu},
  \citenamefont {Kharel}, \citenamefont {Yoon}, \citenamefont {Frunzio},
  \citenamefont {Rakich},\ and\ \citenamefont
  {Schoelkopf}}]{YiwenChu_2018_Nature}%
  \BibitemOpen
  \bibfield  {author} {\bibinfo {author} {\bibfnamefont {Y.}~\bibnamefont
  {Chu}}, \bibinfo {author} {\bibfnamefont {P.}~\bibnamefont {Kharel}},
  \bibinfo {author} {\bibfnamefont {T.}~\bibnamefont {Yoon}}, \bibinfo {author}
  {\bibfnamefont {L.}~\bibnamefont {Frunzio}}, \bibinfo {author} {\bibfnamefont
  {P.~T.}\ \bibnamefont {Rakich}},\ and\ \bibinfo {author} {\bibfnamefont
  {R.~J.}\ \bibnamefont {Schoelkopf}},\ }\bibfield  {title} {\bibinfo {title}
  {Creation and control of multi-phonon fock states in a bulk acoustic-wave
  resonator},\ }\href {https://doi.org/10.1038/s41586-018-0717-7} {\bibfield
  {journal} {\bibinfo  {journal} {Nature}\ }\textbf {\bibinfo {volume} {563}},\
  \bibinfo {pages} {666} (\bibinfo {year} {2018})}\BibitemShut {NoStop}%
\end{thebibliography}%
\bigskip
\noindent\textbf{Acknowledgements}\\
We acknowledge the support of Beijing Natural Science Foundation (Z240007), National Natural Science Foundation of China (No.~92365210, No.~12374325, No.~12304387 and No.~62074091), Young Elite Scientists Sponsorship Program by CAST (Grant No.~2023QNRC001), and Beijing Municipal Science and Technology Commission (Grant No.~Z221100002722011).
\bigskip
\\
\noindent\textbf{Author contributions}\\
H. S., Y. C., and Q. L. performed the room temperature Doppler laser measurements as well as low-temperature radio-frequency measurements. H. W. and Y. W. joined in the device fabrication. H. S., Y. C., Q. L., T. L., and Y. L. analyzed the measurement results and wrote the manuscript. All authors gave their final approval for publication.
\bigskip
\\
\noindent\textbf{Competing interests}\\
The authors declare no competing interests.
\bigskip
\\
\noindent\textbf{Additional information}\\
\textbf{Supplemental information} The online version contains supplemental material available at http://...
\end{document}